\documentclass[size=10.5pt, paper=A4]{scrartcl}

\usepackage[margin=1in]{geometry}
\usepackage{placeins}
\usepackage{float}
\usepackage[doublespacing]{setspace}

\usepackage{graphicx}
\usepackage{transparent}
\usepackage{import}
\usepackage{xr-hyper}
\usepackage{hyperref}
\usepackage{xcolor}
\usepackage{booktabs}
\usepackage{enumitem}
\usepackage{placeins}

\usepackage[
style=nature,
doi=false,
url=false, 
isbn=false,
date=year,
sorting=none,
backend=biber,
]{biblatex}

\usepackage{caption}
\DeclareCaptionLabelSeparator{dot}{. }
\usepackage{amsmath}
\usepackage{stix}
\usepackage{FiraSans}
\usepackage[T1]{fontenc}

\def\imgdir{./img}
\usepackage{etoolbox}
\usepackage{authblk}

\newcommand\CoAuthorMark{\footnotemark[\arabic{footnote}]}

\AtBeginEnvironment{picture}{\sffamily\fontsize{8}{9.6}\selectfont\mathversion{sansmath}}
\AtBeginEnvironment{pgfpicture}{\mathversion{sansmath}}
\AtBeginEnvironment{picture}{\fontsize{8}{9.6}\selectfont{}}

\definecolor{LightPurple}{RGB}{120, 94, 240}

\hypersetup{
	pdfcreator = {},
	pdfproducer = {}
}

\begin{document}
	
\title{Solid-State Lifshitz-van der Waals Repulsion through
		Two-Dimensional Materials}
\author[1]{Tian Tian \footnote{These authors contributed equally to this work.}}
\author[1]{Gianluca Vagli \protect\CoAuthorMark}
\author[1]{Franzisca Naef}
\author[1]{Kemal Celebi}
\author[2, 3]{Yen-Ting Li}
\author[2]{Shu-Wei Chang}
\author[4]{Frank Krumeich}
\author[5,6]{Elton J. G. Santos}
\author[2]{Yu-Cheng Chiu}
\author[1]{Chih-Jen Shih \thanks{Corresponding author. Email: chih-jen.shih@chem.ethz.ch}}
\affil[1]{Institute for Chemical and Bioengineering, ETH Z{\"{u}}rich,  CH-8093 Z{\"{u}}rich, Switzerland}
\affil[2]{Department of Chemical Engineering, National Taiwan University of Science and Technology, Taipei 10607, Taiwan}
\affil[3]{National Synchrotron Radiation Research Center, Hsinchu 30076, Taiwan}
\affil[4]{Laboratory of Inorganic Chemistry, ETH Zürich, 8093, Zürich, Switzerland}
\affil[5]{Institute for Condensed Matter Physics and Complex Systems, School of Physics and Astronomy, The University of Edinburgh, EH9 3FD, UK.}
\affil[6]{Higgs Centre for Theoretical Physics, The University of Edinburgh,  EH9 3FD,  United Kingdom}
\date{} 	
	\maketitle{}
	
	% \linenumbers
	\pagebreak{}
	\doublespacing
	
\begin{quote}
  { \bfseries In the 1960s, Lifshitz et al. predicted that quantum
    fluctuations can change the van der Waals (vdW) interactions from
    attraction to repulsion. However, the vdW repulsion, or its
    long-range counterpart - the Casimir repulsion, has only been
    demonstrated in liquid.  Here we show that the atomic thickness
    and birefringent nature of two-dimensional materials make them a
    versatile medium to tailor the Lifshitz-vdW interactions.  Based
    on our theoretical prediction, we present direct force measurement
    of vdW repulsion on 2D material surfaces without liquid immersion
    and demonstrate their substantial influence on epitaxial
    properties.  For example, heteroepitaxy of gold on a sheet of
    freestanding graphene leads to the growth of ultrathin platelets,
    owing to the vdW repulsion-induced ultrafast diffusion of gold
    clusters.  The creation of repulsive force in nanoscale proximity
    offers technological opportunities such as single-molecule
    actuation and atomic assembly.

}
\end{quote}

When two electroneutral objects, A and B, are brought in proximity in
a polarizable medium, m, the correlations in their temporal
electromagnetic (EM) fluctuations usually lead to an attractive
interaction\autocite{Woods_2016_vdW_review}. At small separations (<
10 nm), this is the vdW force\autocite{Parsegian_2010_vdW}, and at
large separations (> 20 nm) known as the Casimir
force\autocite{Casimir_1948_first,Casimir_1948_second}.  Early vdW
theories\autocite{Keesom_1915_vdW,Maitland_1981_intermolecular,London_1937_vdW}
assumed the total interaction between two objects, each consisting of
many molecules, is simply the sum of intermolecular potentials, which
ignored the fact that the molecular interactions can strongly depend
on surroundings. By applying quantum field theory in statistical
physics, seminal work by Lifshitz et
al. \autocite{Dzyaloshinskii_1961_lifshitz} completely abandoned the
assumption and predicted that quantum fluctuations can lead to
repulsive interactions in both vdW and Casimir regimes. Their
existence was later experimentally verified in a number of fluid-based
systems.
\autocite{Munday_2009_afm,Feiler_2008_superlubri,Zhao_2019_casimir_trap}.

As the interaction potential in the Lifshitz
theory\autocite{Dzyaloshinskii_1961_lifshitz,Parsegian_2010_vdW} is
proportional to the product of effective polarizabilities of A and B
screened by m, the most straightforward approach to generate Casimir
or vdW repulsion is to design a set of materials such
that\autocite{Munday_2009_afm,
  GongCorradoMahbubSheldenMunday+2021+523+536}
\begin{equation}
  \label{eq:eps-ineq}
  (\varepsilon_{\mathrm{A}} - \varepsilon_{\mathrm{m}})(\varepsilon_{\mathrm{B}} - \varepsilon_{\mathrm{m}}) < 0
\end{equation}
where $\varepsilon_{\mathrm{A}}$, $\varepsilon_{\mathrm{B}}$,
$\varepsilon_{\mathrm{m}}$ are the frequency-dependent dielectric
responses for A, B, and m, respectively.  Accordingly, the experiments
demonstrating long-range Casimir repulsion were majorly carried out in
high-refractive-index fluids, i.e., m=fluid, in which
$\varepsilon_{\mathrm{m}}$ is between $\varepsilon_{\mathrm{A}}$ and
$\varepsilon_{\mathrm{B}}$ over a wide range of frequencies to obey
inequality \eqref{eq:eps-ineq}.

However, the examination of vdW repulsion in fluid immersion
\autocite{Munday_2009_afm,Milling_1996_afm,Meurk_1997_afm,Lee_2002_afm}
has two fundamental limitations. First, the high-refractive-index
fluid medium is made by highly polar molecules, and their orientation
and polarity within a small separating gap may disturb the force
\autocite{Munday_2010_review}. Second, more critically, the fluid
dielectric response usually drops rapidly beyond the visible frequency
region, lowering $\varepsilon_{\mathrm{m}}$ below
$\varepsilon_{\mathrm{A}}$ and $\varepsilon_{\mathrm{B}}$ which
results in high-frequency attraction. The long-range repulsive force
observed in fluid arises from the retardation of the high-frequency
contributions \autocite{Bostrom_2012_rep}, but when working at small
separations, the full-spectrum summation may convert the force from
repulsion to attraction \autocite{Bostrom_2012_rep}. In this respect,
the demonstration of vdW repulsion in solid-state systems is clearly
of fundamental and practical interests.

In principle, there is no reason why vdW repulsion cannot exist in
solid-state systems.  In addition to proper selection of materials
fulfilling inequality \eqref{eq:eps-ineq}, the major challenge is to
fabricate an ultrathin medium film (m) sandwiched between two bulk
materials (A and B), such that the vdW repulsion is sufficiently
strong to be observed.  Here we show that 2D materials-mediated
systems could nicely address the challenge.
	
The idea of investigating the 2D materials-mediated repulsive vdW
forces was inspired by recent findings of the wetting
transparency\autocite{Rafiee_2012_trans,Shih_2012_prl,Li_2018_vdw,Liu_2018_screening_ml,Ambrosetti_2018_carbon}
and the remote epitaxy\autocite{Kim_2017_remote,Kong_2018_pol} on
graphene-coated substrates. In these systems, the vdW interactions
exerted by the substrate (A) can be transmitted through the
monolayer-containing medium (m) and greatly influence the
thermodynamic properties on the other side (B), meaning that 2D
materials are highly transparent to vdW interactions. We therefore
predict, if the dielectric response of the 2D material medium is
between $\varepsilon_{\mathrm{A}}$ and $\varepsilon_{\mathrm{B}}$, a
strong vdW repulsion may be generated.
        
\begin{figure}[!htbp]
  \centering
  \includegraphics{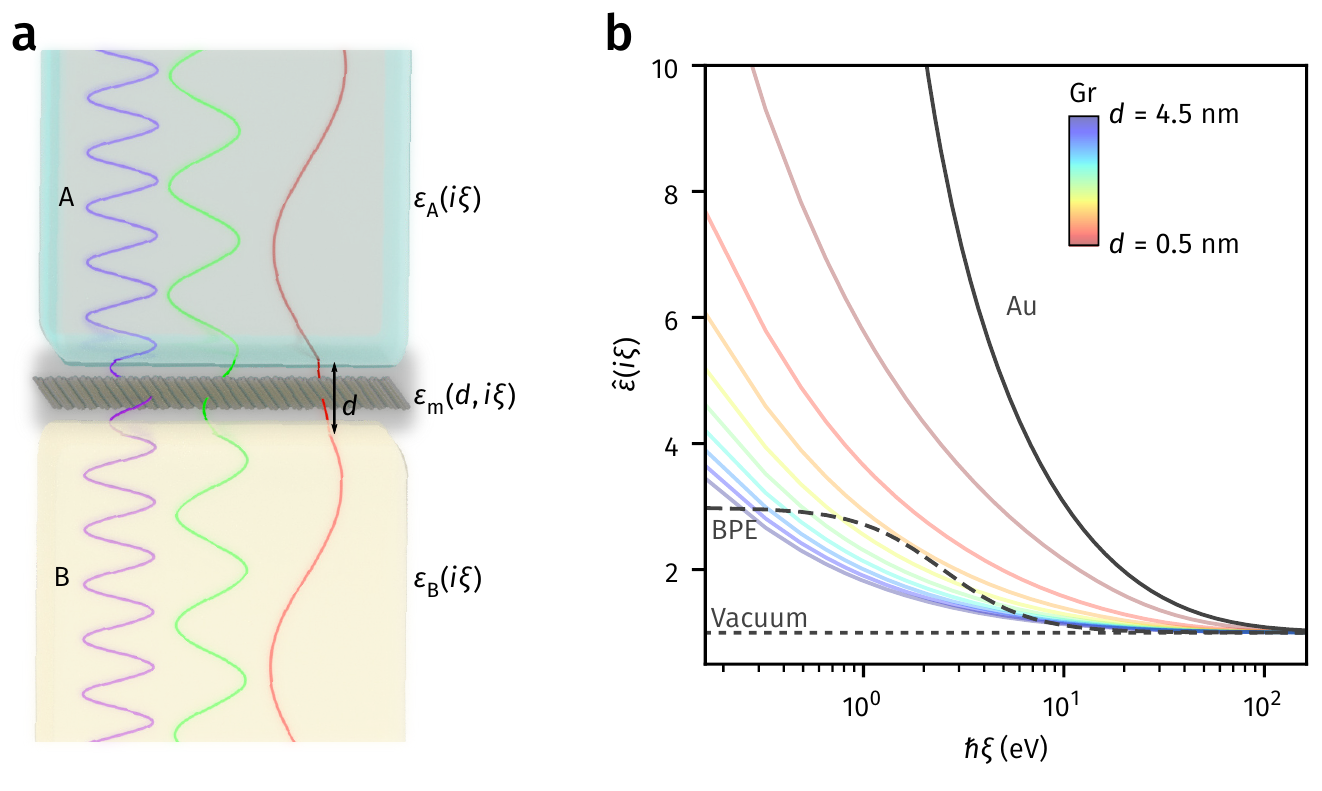}
  \caption{\textbf{vdW repulsion in 2D materials-mediated systems.}
    \textbf{a.} The interaction potential between material A and B
    across a birefringent medium gap m containing a sheet of monolayer
    2D material becomes repulsive when
    $[\varepsilon_{\mathrm{A}} (i\xi)-\hat{\varepsilon}_{\mathrm{m}}
    (d, i\xi)][\varepsilon_{\mathrm{B}} (i\xi) -
    \hat{\varepsilon}_{\mathrm{m}} (d, i\xi)]<0$ at given imaginary
    frequency $i\xi$ and separation $d$.  \textbf{b.} Dielectric
    responses for Au, BPE, Vac, and Gr at different separations as a
    function of electromagnetic energy, $\hbar \xi$, for different
    separations. Accordingly, we predict that vdW repulsion may be
    observed in two sets of materials, A/m/B = Vac/Gr/Au and
    Au/Gr/BPE.}
  \label{fig:1}
\end{figure}

Consider two semi-infinite bulk materials A and B separated by a gap
containing a sheet of monolayer 2D material (Fig. \ref{fig:1}a).  When
the separation $d$ is larger than the vdW thickness of monolayer, we
treat the separating gap containing the monolayer surrounded by vacuum
as an effective birefringent medium with distinct in-plane (IP) and
out-of-plane (OP) dielectric responses
$\varepsilon^{\parallel}_{\mathrm{m}}$ and
$\varepsilon^{\perp}_{\mathrm{m}}$ , which are functions of $d$ and
imaginary frequency $i\xi$, respectively, resulted from the different
IP and OP electronic properties of the monolayer. Using the
polarizability theory of 2D materials\autocite{Tian_2019_nanolett},
$\varepsilon^{\parallel}_{\mathrm{m}}$ and
$\varepsilon^{\perp}_{\mathrm{m}}$ are given by
$\varepsilon_{\mathrm{m}}^{\parallel}(d) = 1 +
\dfrac{\alpha_{\mathrm{2D}}^{\parallel}}{\varepsilon_{0} d}$ and
$\varepsilon_{\mathrm{m}}^{\perp}(d) = \left(1 -
  \dfrac{\alpha_{\mathrm{2D}}^{\perp}}{\varepsilon_{0} d}
\right)^{-1}$ , where $\alpha_{\mathrm{2D}}^{\parallel}$ and
$\alpha_{\mathrm{2D}}^{\perp}$ are the $d$-independent IP and OP
polarizabilities for the 2D material extracted from first principle
calculations, respectively (for details see \textit{Methods}).

The vdW interaction potential between A and B across a birefringent
medium m, $\Phi_{\mathrm{AmB}}^{\mathrm{vdW}}$, is given
by\autocite{Parsegian_2010_vdW} (for details see \textit{Methods}):
\begin{equation}
\label{eq:Phi-aniso}
\Phi_{\mathrm{AmB}}^{\mathrm{vdW}} (d)
= \sum_{n = -\infty}^{\infty} G_{\mathrm{AmB}}(i \xi_{n})
= \sum_{n = -\infty}^{\infty}
\frac{k_{\mathrm{B}} T g_{\mathrm{m}}(i \xi_{n})}{16 \pi d^{2}} 
\left\{
  \int_{\mathcal{r}_{n}}^{\infty} q \ln
  \left[
    1 - \Delta_{\mathrm{Am}}(i \xi_{n}) \Delta_{\mathrm{Bm}}(i \xi_{n}) e^{-q}
  \right] \mathrm{d} \mathcal{q}
\right\}
\end{equation}
where $k_{\mathrm{B}}$ is the Boltzmann constant, $T$ is the absolute
temperature, $\xi_{n}=2\pi n k_{\mathrm{B}} T/ \hbar$ is the n-th
Matsubara frequency, $\hbar$ is the reduced Planck constant,
$\mathcal{r}_{n}=\frac{2d \xi_{n}}{c}
\sqrt{\hat{\varepsilon}_{\mathrm{m}}}$ is the retardation
factor\autocite{Parsegian_2010_vdW}, $c$ is the speed of light in
vacuum and $q$ is a dimensionless auxiliary variable.
$\hat{\varepsilon}_{\mathrm{m}} =
\sqrt{\varepsilon_{\mathrm{m}}^{\parallel}
  \varepsilon_{\mathrm{m}}^{\perp} }$ and
$g_{\mathrm{m}} = \varepsilon_{\mathrm{m}}^{\perp} /
\varepsilon_{\mathrm{m}}^{\parallel}$ are the geometrically-averaged
dielectric function and dielectric
anisotropy\autocite{Tian_2019_nanolett} of m, respectively.  Similar
approach was also used to calculate vdW interactions of layered
materials\autocite{Zhou_2017_lifshitz}.  $\Delta_{\mathrm{Am}}$ and
$\Delta_{\mathrm{Bm}}$ correspond to the dielectric mismatches
following
$ \Delta_{\mathrm{jm}} = \dfrac{ \hat{\varepsilon}_{\mathrm{j}} -
  \hat{\varepsilon}_{\mathrm{m}} }{ \hat{\varepsilon}_{\mathrm{j}} +
  \hat{\varepsilon}_{\mathrm{m}}}$, for j = A, B. Analogous to
inequality \eqref{eq:eps-ineq}, the vdW potential for a given EM mode
$\xi_{n}$ becomes positive when
$\Delta_{\mathrm{Am}} \Delta_{\mathrm{Bm}} < 0$, contributing to vdW
repulsion.
	
Using graphene (Gr), the thinnest carbon-based 2D material, as a model
system, the calculated $\hat{\varepsilon}_{\mathrm{m}}$ as a function
of $\hbar \xi$ for different separations are shown in
Fig. \ref{fig:1}b.  The dielectric responses have the same order of
magnitude with those for high-refractive-index
fluids\autocite{Munday_2009_afm,Feiler_2008_superlubri,Bostrom_2012_rep}
(Supplementary Fig. \ref{si-fig:eps-cascade-full}) but the applicable
separation appears to be significantly smaller.  Indeed, equation
\eqref{eq:Phi-aniso} suggests that the vdW repulsion can be tuned by
the separation $d$ and the dielectric anisotropy $g_{\mathrm{m}}$,
which highlight the versatility of 2D materials-mediated
systems. Fig. \ref{fig:1}b also includes the dielectric responses for
the three bulk materials considered in our experiments later,
including gold (Au),
N,N'-bis(2-phenyl\-ethyl)perylene-3,4,9,10-bis(dicarboximide)
(BPE)\autocite{Mizuguchi_1998_BPE,Ling_2007_BPE} molecular solid, and
vacuum (Vac). It reveals that vdW repulsion may be observed in two
sets of materials, A/m/B = Vac/Gr/Au and Au/Gr/BPE, in which the
former obeys inequality \eqref{eq:eps-ineq} in all separations and
frequencies and the latter for separations $< 2$ nm.

\begin{figure}[!htbp]
  \centering
  \includegraphics[width=0.96\linewidth]{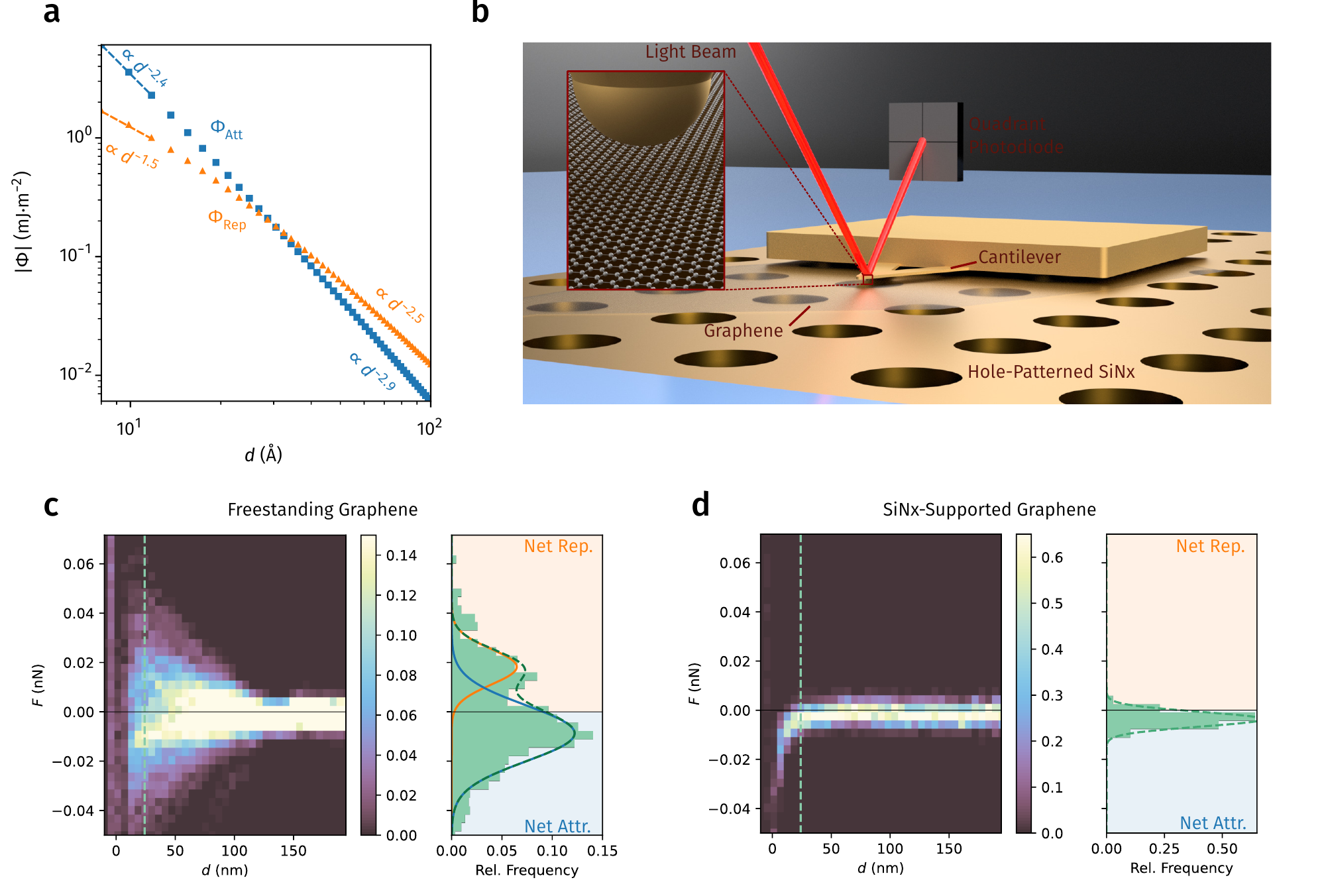}
  \caption{ \textbf{Direct measurement of vdW repulsion on
      freestanding graphene}.  \textbf{a.} Calculated many-body
    repulsive potential,
    $\Phi_{\mathrm{Rep}} = \Phi_{\mathrm{AmB}}^{\mathrm{vdW}}$, and
    two-body attractive potential,
    $\Phi_{\mathrm{Att}}=\Phi_{\mathrm{mB}}^{\mathrm{vdW}}$, as a
    function of $d$. The former is of relatively long-range, scaling
    as $d^{-1.5}$ to $d^{-2.5}$, as compared to the latter scaling as
    $d^{-2.4}$ to $d^{-2.9}$, thereby yielding an energy barrier above
    graphene surface.  \textbf{b.} Experimental setup for measuring
    the force responses between a gold-coated spherical AFM tip and a
    piece of micro-mechanically-exfoliated graphene transferred onto a
    holey SiN$_x$ windows.  \textbf{c, d.} 2D-histograms sampling
    \textbf{c.} 225 measurements over a 3.44 $\mathrm{\mu}$m $\times$
    3.44 $\mathrm{\mu}$m large area on freestanding graphene and
    \textbf{d.} 36 measurements over a 1.37 $\mathrm{\mu}$m $\times$
    1.37 $\mathrm{\mu}$m large area of SiN$_x$-supported graphene. The
    right panels show the force distribution at $d \approx$ 24 nm
    corresponding to the green dashed lines on the histograms. More
    than 90 measurements showed repulsive behavior. The distribution
    on freestanding graphene is bimodal, clearly showing a population
    of repulsive forces.  }
  \label{fig:2-new}
\end{figure}

As pointed out in several theoretical
studies\autocite{Zhou_2017_lifshitz,Ambrosetti_2018_carbon}, for an
A/m/B system where m is a layer of 2D material, the total vdW
potential consists of both attractive and repulsive contributions.
Taking the Vac/Gr/Au system as an example, the gold layer on
freestanding graphene is expected to not only experience a repulsive
potential, $\Phi_{\mathrm{Rep}} = \Phi_{\mathrm{AmB}}^{\mathrm{vdW}}$,
but also an attractive potential, $\Phi_{\mathrm{Att}}$, corresponding
to the two-body vdW potential between gold and graphene,
$\Phi_{\mathrm{mB}}^{\mathrm{vdW}}$.  The total potential acting on
gold, $\Phi_{\mathrm{tot}}=\Phi_{\mathrm{Att}} + \Phi_{\mathrm{Rep}}$,
combines both effects.  Our calculations based on equation
\eqref{eq:Phi-aniso} show that $\Phi_{\mathrm{Rep}}$ is of relatively
longer-range, scaling as $d^{-1.5}$ to $d^{-2.5}$, for separations
from $\sim$1 nm to $\sim$10 nm, as compared to $\Phi_{\mathrm{Att}}$
scaling as $d^{-2.4}$ to $d^{-2.9}$ (Fig. \ref{fig:2-new}a).  As a
result, for $d>3$ nm, the first derivative of $-\Phi_{\mathrm{tot}}$
is positive, thereby yielding a net repulsive force between gold and
vacuum through monolayer graphene.
	
According to our theoretical prediction, we carried out direct
measurement for the vdW force experienced by a gold-coated tip in
atomic force microscopy (AFM) interacting with a sheet of freestanding
graphene (Fig. \ref{fig:2-new}b). The freestanding graphene was
fabricated by transferring a piece of micro-mechanically-exfoliated
graphene to a holey silicon nitride (SiN$_x$) membrane supported by a
silicon chip \autocite{doi:10.1021/nl1008037}, with the hole diameter
of approximately 5 $\mu$m, followed by annealing it in Ar/H$_2$ to
remove contaminants \autocite{Li_2013_contam,Russo_2014_h2}. A
gold-coated AFM tip with a measured radius of 33 nm was chosen for the
force-distance measurements (for details see Supplementary Section
\ref{si-sec:methods}). All measurements were carried out in air.
Supplementary Figs \ref{si-fig:single-f-d}a and
\ref{si-fig:single-f-d}b show representative force-distance responses
for the approach/retraction processes on freestanding and
SiN$_x$-supported graphene, respectively (measurement details see
Supplementary Section \ref{si-sec:F-d-meas}). Note that when
establishing the contact, the force response is quadratic for
freestanding graphene, in contrast to the linear response on supported
region. This is expected considering the mechanical flexibility of
freestanding graphene membrane, which yields elastic response of
higher order \autocite{doi:10.1126/science.1157996}. Indeed, during
the retraction process from a freestanding graphene surface, the tip
remains to adhere to graphene at a large tip displacement, revealing
that both graphene and AFM cantilever were bent before overcoming the
attractive two-body interaction $\Phi_{\mathrm{Att}}$. With the
nonideality in mind, hereafter, we focus on the approach responses
before physical contact with the sample surface.

Figs \ref{fig:2-new}c and \ref{fig:2-new}d compare the two-dimensional
histograms for the force-distance responses extracted from 225 and 36
measurements on freestanding and SiN$_x$-supported graphene,
respectively. The right panels present the force distributions at
$d \approx$ 24 nm corresponding to the green dashed lines on the
two-dimensional histograms. Remarkably, more than 90 measurements
showed repulsive behavior (representative response in Supplementary
Fig. \ref{si-fig:AFM-E-barrier-estimate}). The force distribution on
freestanding graphene (\ref{fig:2-new}c) is bimodal, which can be
nicely decomposed to two Gaussian functions, revealing one population
located at the repulsive regime (mean force $\mu$ = 18.2 pN and
standard deviation $\sigma$ = 7.2 pN). The attractive population
($\mu$ = -8.7 pN and $\sigma$ = 11.3 pN) presumably comes from the
responses characterized on the area contaminated by airborne
adsorbates \autocite{Li_2013_contam}, as well as the impurities
introduced in process history, including transfer and scanning
electron microscopy \autocite{doi:10.1063/1.3062851}. On the other
hand, measurements on supported graphene (Fig. \ref{fig:2-new}d) only
yield attractive responses ($\mu$ = -3.7 pN and $\sigma$ = 3.0 pN).
The theoretical picture presented in Fig. \ref{fig:2-new}a suggests
that the directly measured repulsive force corresponds to the forced
required to overcome the repulsive energy barrier before the
short-range attractive interaction overtakes. Given the tip radius
characterized in SEM (33 nm), we calculated the height of repulsive
energy barrier by integrating the average force response in the
repulsive population with respect to $d$, yielding a value of 19$\pm$5
$\mathrm{\mu}$J$\cdot$m$^{-2}$ (see Supplementary Section
\ref{si-sec:barrier-estimate}), which nicely agrees with our
theoretical prediction (20 $\mathrm{\mu}$J$\cdot$m$^{-2}$; see
Supplementary Fig. \ref{si-fig:lifshitz-barrier}).

\begin{figure}[!htbp]
  \centering
  \includegraphics[width=0.96\linewidth]{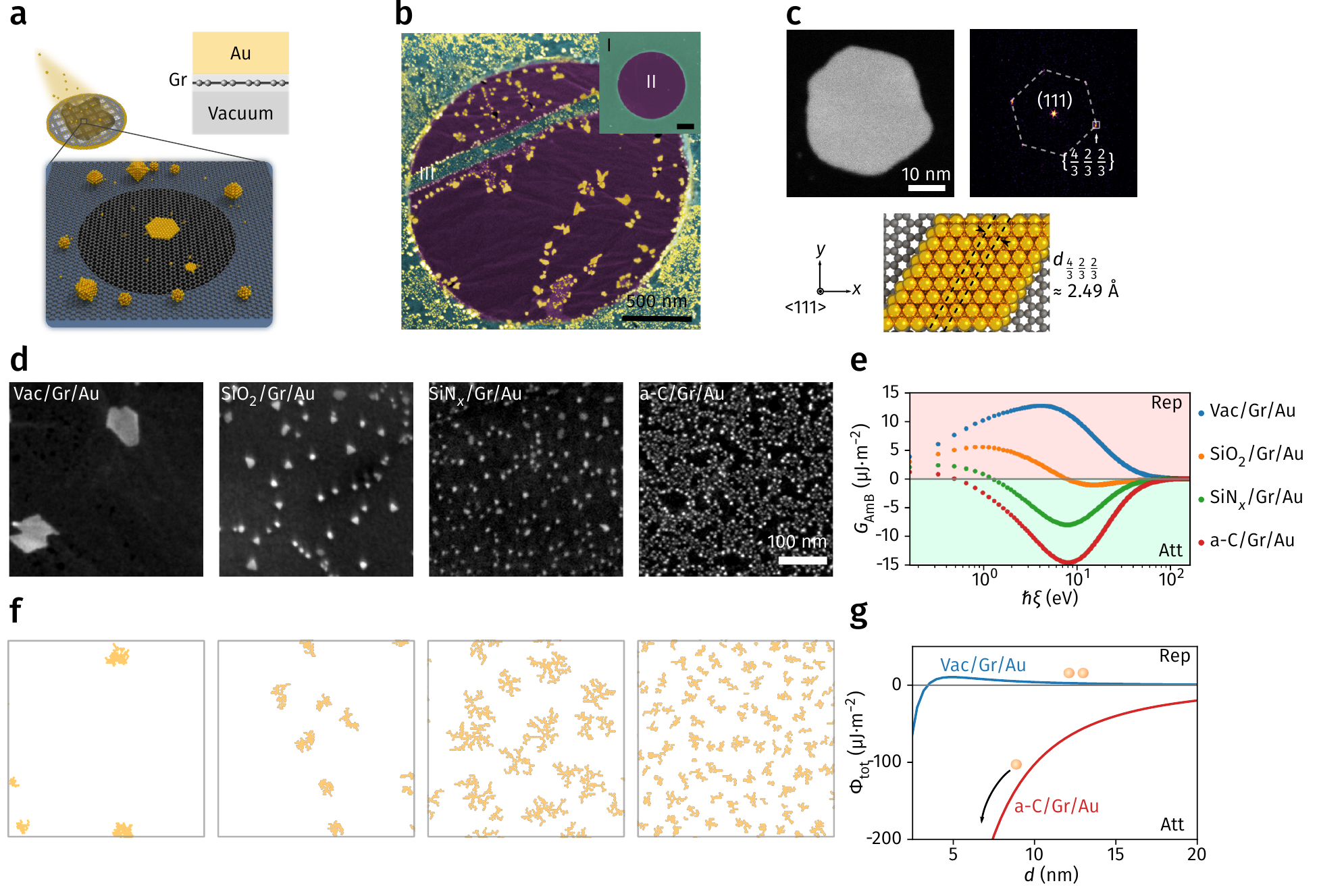}
  \caption{ \textbf{Ultrafast diffusion of gold atoms on freestanding
      graphene at room temperature}.  \textbf{a.} A small amount of
    gold was evaporated in high vacuum and condensed at room
    temperature on a sheet of freestanding graphene transferred to a
    Quantifoil holey carbon grid.  \textbf{b.} False color SEM image
    for gold deposited on graphene near a hole. Inset: SEM image
    before gold deposition that identifies regions I and II
    corresponding to carbon grid-supported and freestanding graphene,
    respectively. Gold on freestanding graphene forms ultrathin
    platelets and leaves a large non-wettable area of up to 1.3
    $\mathrm{\mu}$m$^{2}$.  \textbf{c.} FT-STEM characterization of a
    representative hexagonal platelet (left), revealing a set of Bragg
    diffraction spots (right) corresponding to (4/3 2/3 2/3) lattice
    plane ($d$-spacing of $\sim$2.49 Å).  \textbf{d.} SEM images for
    gold deposited on graphene supported by different substrates (left
    to right: vacuum, SiO$_{2}$, SiN$_{x}$, and amorphous carbon
    (a-C)).  \textbf{e.}  Calculated interaction spectra
    $G_{\mathrm{AmB}}$ as a function of $\hbar \xi$ at $d=0.8$ nm for
    Vac/Gr/Au, SiO$_{2}$/Gr/Au, SiN$_{x}$/Gr/Au and a-C/Gr/Au,
    respectively.  The full-spectrum summation is gradually converted
    from repulsion to attraction, yielding the increase of nucleation
    density in \textbf{D}.  \textbf{f.} KMC simulation snapshots
    modeling 2D growth of gold on a surface by varying the ratio of
    activation energy for diffusion on pristine graphene,
    $\Delta E_{\mathrm{d}}^{0}$, to that for binding,
    $\Delta E_{\mathrm{b}}$ (left to right:
    $\Delta E_{\mathrm{d}}^{0}/\Delta E_{\mathrm{b}}$ = 0.2, 0.5, 1.0,
    1.6).  \textbf{g.} Calculated $\Phi_{\mathrm{tot}}$ profiles for 6
    nm thick gold platelet approaching freestanding (blue) and
    substrate-supported (red) graphene with respect to separation.  }
  \label{fig:2}
\end{figure}

We have found that the vdW repulsion generated in Vac/Gr/Au system is
sufficiently strong to alter the epitaxial properties of Au grown on
freestanding graphene.  As schematically shown in Fig. \ref{fig:2}a,
we evaporated a small amount ($\sim{}3$ ng·mm$^{-2}$) of gold that
condensed on a sheet of freestanding graphene in high vacuum at room
temperature (for details see \textit{Methods}).  Fig. \ref{fig:2}b
shows a representative scanning electron micrograph (SEM) for gold
deposited on graphene.  Two regions, namely amorphous carbon (a-C)
grid-supported (I) and freestanding (II) graphene, can be identified
in the inset SEM image.  The morphology and density for the deposited
gold clusters on regions I and II exhibit substantially different
features.  On region I, as expected, due to a very high surface energy
of gold ($\sim{}1300$ mJ·m$^{-2}$ at room
temperature\autocite{Mills_2006_sur_energy}), fast condensation at
room temperature yields small spherical nano\-clusters with a high
nucleation density.  However, on region II, despite a high degree of
supercooling, the nucleation density is very low, leaving a large
non-wettable area of up to 1.3 $\mathrm{\mu}$m$^{2}$, with a few large
and ultrathin gold platelets grown on the surface.  We performed
crystallographic analysis using the Fourier-transformed scanning
transmission electron microscopy (FT-STEM) on a representative
hexagonal platelet (Fig. \ref{fig:2}c).  A set of Bragg diffraction
spots corresponding to \{$\frac{4}{3} \frac{2}{3} \frac{2}{3}$\}
lattice planes (lattice spacing of $\sim$2.49 Å) was observed, which
was only reported in the atomically-thin face-center-cubic metal
crystals\autocite{Jin_2001_lattice_ag}.  We also notice that gold
deposited on the defective and contaminated domains of freestanding
graphene, e.g., region III, exhibits similar behavior with that on
region I.
	
From a thermodynamic point of view, the growth of ultrathin gold
platelets on freestanding graphene would require the Au-Gr
interactions to be stronger than Au surface energy, or even a negative
interfacial tension \autocite{Israelachvili_2011_intermolecular},
which is unlikely and cannot explain the observed ultralow nucleation
density.  We further transferred graphene onto two other substrates,
silicon oxide (SiO$_{2}$) and silicon nitride (SiN$_{x}$), and
compared the morphologies of gold condensed on top
(Fig. \ref{fig:2}d). Fig. \ref{fig:2}e presents the calculated
many-body vdW interaction spectra as a function of energy,
$G_{\mathrm{AmB}}(i \xi)$ (see equation~\eqref{eq:Phi-aniso}), for the
four systems considered here. The full-spectrum summation indicates
that in contrast to the repulsive Vac/Gr/Au system, for
SiO$_{2}$/Gr/Au, SiN$_{x}$/Gr/Au, and a-C/Gr/Au, the vdW interactions
become increasingly attractive.  Together with Fig. \ref{fig:2}d, it
becomes evident that stronger vdW repulsion would lead to lower
nucleation density. The growth behavior is kinetically controlled
which agrees with morphological statistics of the gold nanoplatelets
based on AFM and SEM analysis (for details see Supplementary Section
\ref{si-sec:frict-diff-au}).  We further exclude the scenario of
remote epitaxy\autocite{Kim_2017_remote} or ``lattice
transparency''\autocite{Chae_2017} of graphene, since the highest
crystalline samples were obtained on freestanding graphene,
contradicting the polarity-dominated mechanism of remote
epitaxy\autocite{Kong_2018_pol}(more discussions see Supplementary
Section \ref{si-sec:effect-latt-transp}).

density $N_{\mathrm{nu}}$ is proportional to $D^{-\frac{1}{3}}$, where
Indeed, the classical nucleation theory suggests that the nucleation
$D$ is the surface diffusivity\autocite{Mo_1991_diffuse}.  According
to the SEM images in Fig. \ref{fig:2}d, we estimate that, by making
graphene freestanding, the surface diffusivity of gold was boosted by
up to approximately 9 orders of magnitude (Supplementary
Fig. \ref{si-fig:support-density}, right $y$-axis), indicating
ultrafast in-plane diffusion.  The platelets grown on freestanding
graphene can be over $10^{2}$ times larger than that on a-C/Gr,
indicating that the platelet growth is dominated by kinetic effect
rather than thermodynamic wettability.  The observation is further
endorsed by our kinetic Monte Carlo (KMC) simulations considering the
competition between Au diffusion on graphene and Au-Au binding
processes on a surface (Fig. \ref{fig:2}f and Supplementary
Fig. \ref{si-fig:kmc-theory-energy}).  Large and sparsely distributed
platelets can only be obtained when the activation energy for
diffusion is negligible compared to that for Au-Au binding (see
Supplementary Section \ref{si-sec:kmc-simulations-au}).

The repulsive potential predicted by the Lifshitz model not only makes
the surface adsorption an energy uphill process at large separation,
but also decreases the vdW potential well depth at the contact
distance by nearly 50\% (Supplementary
Fig. \ref{si-fig:lifshitz-barrier}), which further decrease the
adsorption rate of gold onto freestanding graphene as predicted by
classical aggregation theory\autocite{Fuchs_1934_aggregation}.  As
such, the vdW repulsion-induced energy barrier created above the
surface offers a “highway” for in-plane diffusion (Fig. \ref{fig:2}g),
forming large platelets.

\begin{figure}[!htbp]
  \centering
  \includegraphics[width=0.9\linewidth]{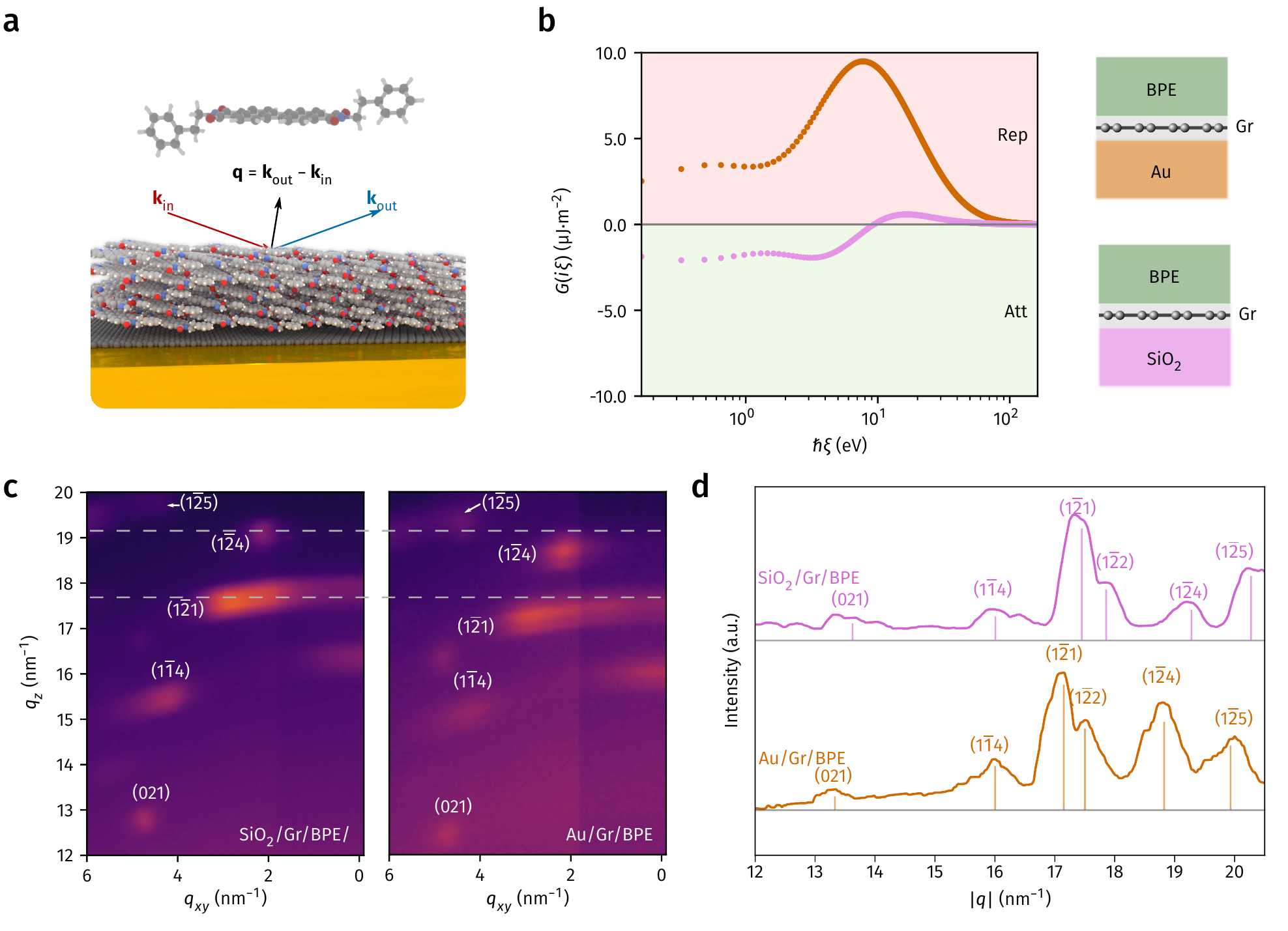}
  \caption{ \textbf{vdW repulsion-induced molecular polymorphism}.
    \textbf{a.} Molecular structure of BPE (top) and schematic of
    molecular solid (bottom) having a layered crystalline structure,
    with the molecular plane oriented in parallel to graphene.  The
    diffraction wave vector $\mathbf{q}$ of a specific lattice plane
    follows the relation
    $\mathbf{q} = \mathbf{k}_{\mathrm{out}} -
    \mathbf{k}_{\mathrm{in}}$, where $\mathbf{k}_{\mathrm{in}}$ and
    $\mathbf{k}_{\mathrm{out}}$ are the wave vectors of incident and
    diffracted light in GIWAXS, respectively.  \textbf{b.} Calculated
    interaction spectra $G_{\mathrm{AmB}}(i\xi)$ as a function of
    $\hbar\xi$ for Au/Gr/BPE (orange dots) and SiO$_{2}$/Gr/BPE
    (purple dots) systems at $d=1$ nm, revealing that the former
    yields a strong vdW repulsion and the latter has a weak
    attraction.  \textbf{c.} Magnified GIWAXS patterns for
    SiO$_{2}$/Gr/BPE (left) and Au/Gr/BPE (right) systems showing the
    Laue spots corresponding to interlayer planes ($1\overline{2}4$)
    and ($1\overline{2}1$) near the $q_{z}$ axis. As compared to
    SiO$_{2}$/Gr/BPE, the $q_{z}$ components for Au/Gr/BPE are
    slightly lower, confirming the interlayer d-spacings are enlarged.
    \textbf{d.} Comparison of integrated line cuts for
    SiO$_{2}$/Gr/BPE, Au/Gr/BPE. The best-fitted peak positions are
    marked as vertical lines.}
  \label{fig:3}
\end{figure}
	
The second set of materials examined here is Au/Gr/BPE
(Fig. \ref{fig:3}a). The BPE molecule has a flat polycyclic aromatic
core, perylene\-tetra\-carboxylic diimide (PTCDI), decorated with two
freely rotatable phenylethyl groups on both ends. Highly ordered BPE
molecular solids are of layered crystalline
structure\autocite{Ling_2007_BPE}, with the interlayer spacing
determined by the sensitive balance between intermolecular $\pi-\pi$
interactions of PTCDI cores and phenylethyl groups conformational
entropy\autocite{Schmidt_2009_ptcdi}.  We therefore hypothesized that,
by orienting the BPE molecular plane parallel to graphene, the force
exerted by the substrate across graphene may alter the interlayer
spacing.
	
The BPE molecules were thermally evaporated onto graphene supported by
gold as well as SiO$_{2}$ for comparison (Fig. \ref{fig:3}b, for
details see \textit{Methods}). Fig. \ref{fig:3}b presents the
calculated $G_{\mathrm{AmB}}(i \xi)$ for the two systems at $d = 0.8$
nm.  Following earlier discussion in Fig. \ref{fig:1}b, for $d<2$ nm,
the vdW potentials between Au and BPE are always repulsive
irrespective of frequency, with the main contribution from the visible
to ultraviolet region.  The full-spectrum summation of
$G_{\mathrm{AmB}}(i \xi)$ according to equation~\eqref{eq:Phi-aniso}
yields a strong vdW repulsion of $\sim{}$1.3 mJ·m$^{-2}$.  On the
other hand, in SiO$_{2}$/Gr/BPE system, the transition from attraction
to repulsion at high frequencies leads to a weakly attractive
potential ($\sim{}$-20 $\mathrm{\mu}$J·m$^{-2}$).

To examine the substrate interactions through graphene, we analyzed
the molecular orientation and crystallographic constants of BPE
molecular solids using synchrotron grazing-incidence wide-angle x-ray
scattering (GIWAXS). In the GIWAXS patterns (Fig. \ref{fig:3}c and
Supplementary Fig. \ref{si-fig:gixd-full-scale}), both systems show
several intense high-angle Laue spots, in particular those
corresponding to the interlayer ($1 \overline{2} 4$) and
($1 \overline{2} 1$) planes (detailed crystallographic analysis see
Supplementary Section \ref{si-sec:molec-epit-bpe}) near the $q_{z}$
axis, confirming the PTCDI plane is preferentially oriented parallel
to the substrate
\autocite{Chiu_2013_PTCDI,Shih_2015_partial}. However, the slightly
shifted $q_{z}$ components reveal polymorphs induced by the substrate
force (Fig. \ref{fig:3}c).  The strongly repulsive substrate, gold,
yielded interlayer spacings of 3.35 and 3.68 Å, respectively, which
are enlarged by $\sim{}$1.8\% as compared to the SiO$_{2}$ control.
The integrated line cuts extracted from GIWAXS patterns
(Fig. \ref{fig:3}d) compares the signals for SiO$_{2}$/Gr/BPE and
Au/Gr/BPE.  The vdW repulsion generated in Au/Gr/BPE appears to
slightly offset the interlayer interactions between BPE
molecules. Consequently, the effect of phenylethyl groups
conformational entropy takes a more active role, increasing the
interlayer spacing for a set of high-$q$ diffraction peaks
($(1 \overline{1} 4)$, $(1 \overline{2} 1)$, $(1 \overline{2} 2)$,
$(1 \overline{2} 4)$ and $(1 \overline{2} 5)$) associated with the
PTCDI basal plane.  Further experimental examination of molecular
packing on graphene-free substrates indicates the breakdown of
classical vdW transparency theory in this system (more details see
Supplementary Sections \ref{si-sec:breakd-wett-transp} and
\ref{si-sec:effect-latt-transp}).
	
We have presented direct force measurement of the vdW repulsion at 2D
materials surfaces and demonstrated that the vdW repulsion can
substantially influence kinetics and thermodynamics of heteroepitaxy.
Our findings imply that the recently reported quantum levitation
\autocite{Munday_2009_afm,Zhao_2019_casimir_trap,Munday_2010_review}
might be even realized without liquid immersion, which give rise to
robust solid-state device miniaturization.  In addition, in view of
the growth of 2D materials family covering an increasingly large range
of properties, they could become versatile surface coatings
selectively repelling objects down to atomic level, which may lead to
new molecular mechanical systems and sensors.  \pagebreak
	
\section*{Author Contributions}
\label{sec:author-contributions}
T.T., G.V. and C.J.S. conceived the idea and designed the
experiments. T.T. and F.N. developed the theoretical
framework. T.T. performed first-principle calculations under guidance
of E.J.G.S.  G.V. carried out AFM force measurement, analyzed the
data, and modeled the force responses.  T.T., G.V. and K.C. fabricated
the freestanding graphene samples and carried out Au deposition. T.T
characterized the freestanding graphene samples.  F.K. performed STEM
measurements.  T.T. carried out KMC simulations. T.T. transferred
graphene onto gold and SiO2 substrates. Y.T.L and S.W.C. deposited BPE
molecules and carried out synchrotron GIWAXS under supervision of
Y.C.C. T.T., Y.T.L. and Y.C.C. analyzed the GIWAXS patterns. T.T.,
G.V. and C.J.S. co-wrote the paper. All authors contributed to this
work, read the manuscript, discussed the results, and agreed to the
contents of the manuscript and supplementary materials.
	
\section*{Acknowledgments}
\label{sec:acknowledgments}
C.J.S. is grateful for financial support from ETH startup funding and
the European Research Council Starting Grant (N849229 CQWLED).
T.T. G.V. and K.C. acknowledge technical support from the Scientific
Center for Optical and Electron Microscopy (ScopeM) and FIRST-Center
for Micro- and Nanoscience of ETH Zurich.  EJGS acknowledges
computational resources through the UK Materials and Molecular
Modeling Hub for access to THOMAS supercluster, which is partially
funded by EPSRC (EP/P020194/1); CIRRUS Tier-2 HPC Service (ec131
Cirrus Project) at EPCC (http://www.cirrus.ac.uk) funded by the
University of Edinburgh and EPSRC (EP/P020267/1); ARCHER UK National
Supercomputing Service (http://www.archer.ac.uk) via d429 Project
code, and the UKCP consortium (Project e89) funded by EPSRC grant ref
EP/P022561/1. EJGS also acknowledges the EPSRC Early Career Fellowship
(EP/T021578/1) and the University of Edinburgh for funding support.
Y.C.C. thanks the financial support by the “Advanced Research Center
for Green Materials Science and Technology” from The Featured Area
Research Center Program within the framework of the Higher Education
Sprout Project by the Ministry of Education (108L9006) and the
Ministry of Science and Technology in Taiwan (MOST 108-3017-F-002-002
and 108-2221-E-011-047).  T.T. thanks Dr. Liqing Zheng for providing
gold substrates.
	
\pagebreak
\section*{Supplementary materials}
\label{sec:suppl}
\begin{itemize}
\item Materials and Methods
\item Supplementary Text
\item Figs. S1 to S34
\item Tables S1 to S4
\item References (S1-S46)
\end{itemize}
% \nolinenumbers
	
\pagebreak{}

\printbibliography{}

\pagebreak{}
		
\end{document}

\title{Supplementary Material for:\\
  Solid-State Lifshitz-van der Waals Repulsion through
  Two-Dimensional Materials}
\author[1]{Tian Tian \footnote{These authors contributed equally to this work.}}
\author[1]{Gianluca Vagli \protect\CoAuthorMark}
\author[1]{Franzisca Naef}
\author[1]{Kemal Celebi}
\author[2, 3]{Yen-Ting Li}
\author[2]{Shu-Wei Chang}
\author[4]{Frank Krumeich}
\author[5,6]{Elton J. G. Santos}
\author[2]{Yu-Cheng Chiu}
\author[1]{Chih-Jen Shih \thanks{Corresponding author. Email: chih-jen.shih@chem.ethz.ch}}
\affil[1]{Institute for Chemical and Bioengineering, ETH Z{\"{u}}rich,  CH-8093 Z{\"{u}}rich, Switzerland}
\affil[2]{Department of Chemical Engineering, National Taiwan University of Science and Technology, Taipei 10607, Taiwan}
\affil[3]{National Synchrotron Radiation Research Center, Hsinchu 30076, Taiwan}
\affil[4]{Laboratory of Inorganic Chemistry, ETH Zürich, 8093, Zürich, Switzerland}
\affil[5]{Institute for Condensed Matter Physics and Complex Systems, School of Physics and Astronomy, The University of Edinburgh, EH9 3FD, UK.}
\affil[6]{Higgs Centre for Theoretical Physics, The University of Edinburgh,  EH9 3FD,  United Kingdom}
\date{} \date{}                   

\maketitle{}

\pagebreak{}
\section{Materials and Methods}
\label{sec:methods}

  \paragraph{Calculation of vdW interaction spectra}
  The dielectric responses as function of imaginary frequency $\varepsilon(i \xi)$
  was calculated
using the Kramers Kronig relationship\autocite{Parsegian_2010_vdW}:
  \begin{equation}
  \label{eq:vdw-KKR-eps}
  \varepsilon(i\xi) =
  1 + {\displaystyle \frac{2}{\pi}}{\displaystyle   \int_{0}^{\infty}}
  {\displaystyle \frac{\omega \text{Im}[\varepsilon(\omega)]}{\omega^{2} + \xi^{2}}}   \mathrm{d}\omega
\end{equation}
where $\omega$ is the real frequency, and $\text{Im}[\varepsilon(\omega)]$ is the
imaginary part of complex dielectric function $\varepsilon(\omega)$.
Frequency-dependent dielectric functions of
SiO$_{2}$\autocite{Palik_1998_handbook},
SiN$_{x}$\autocite{Palik_1998_handbook}, bromobenzene
(BrPh)\autocite{Munday_2009_afm}, amorphous
carbon\autocite{Gioti_2003_aCarbon}, and
Au\autocite{Palik_1998_handbook} were extracted from experimental
data, respectively.  The dielectric function of BPE is estimated using the
single Lorentz oscillator model following
$\varepsilon_{\mathrm{BPE}}(i \xi) = 1 +
\frac{\xi_{\mathrm{p}}^{2}}{\xi_{\mathrm{g}}^{2} + K \xi + \xi^{2}}$
where $\hbar \xi_{\mathrm{g}}=2.30$ eV, $\hbar \xi_{\mathrm{p}}=3.16$,
$\hbar K = 0.1$ eV, yielding an optical refractive index
$n\approx{}1.7$.  Frequency-dependent 2D polarizabilities
($\alpha_{\mathrm{2D}}^{\mathrm{\parallel}}$,
$\alpha_{\mathrm{2D}}^{\mathrm{\perp}}$) of graphene were obtained by
\textit{ab initio} package GPAW\autocite{Mortensen_2005_gpaw} using
the projector augmented wave
method\autocite{Kresse_1999_PAW}. Dielectric responses were calculated
using random phase approximation on top of the Perdew-Burke-Ernzerhof
exchange-correlation functional\autocite{Perdew_1996} with plane wave
cutoff energy of 500 eV, k-point density of 15 Å$^{-1}$ and truncated
Coulomb kernel to avoid spurious interaction from periodic images.

  \paragraph{Graphene growth}
  Monolayer graphene was synthesized by chemical vapor deposition
  (CVD). Copper (Cu) foil (99.999\%, Alfa Aesar) was first cleaned by
  aceton, followed by isopropanol (IPA) and electro\-chemically
  polishing in a mixture of 2:1:1:0.2 deionized water (DI-H$_{2}$O) :
  ortho\-phosphoric acid : ethanol : IPA under a bias of 5 V. The
  polished Cu foil was then annealed in a quartz tube furnace at 1060 °C
  under 5 Torr and 50 standard cubic centimetres per minute (sccm) of
  hydrogen (H$_{2}$) flow. The growth of graphene was followed by
  flowing 40 sccm of methane (CH$_{4}$) and 15 sccm of H$_{2}$ at
  1000 °C under 5 Torr for 10 minutes. After cooling down, graphene on
  the backside of the Cu foil was etched by oxygen plasma.

  \paragraph{Freestanding graphene on Quantifoil grids}
  The Quantifoil® grids consisting of amorphous carbon film with pore
  opening of $1.2 \sim 5$ $\mathrm{\mu}$m supported by Au meshes were cleaned by
  aceton rinsing before use. The grids were placed onto the Cu foil
  with the holey carbon film facing graphene. By dropping $\sim{}5$ $\mathrm{\mu}$L
  of IPA onto the grid and heating the Cu foil at 120 °C, the grid was
  adhered to graphene/Cu by capillary force during evaporation of IPA.
  The Cu foil was etched at the liquid-air interface of 0.5 M ammonium
  persulfate (APS) solution. The APS residue was rinsed with 
  DI-H$_{2}$O for several times. The grid was finally lifted from the liquid-air
  interface and dried under gentle argon flow.
  
  \paragraph{Exfoliated graphene on silicon nitride chip}
  Graphene flakes exfoliated \autocite{doi:10.1021/acsnano.5b04258}
  from natural graphite and transferred to a silicon nitride (SiN$_{x}$) chip of 1 cm$^2$ size, patterned with an 12$\times$12 hole-matrix, in which both hole-diameter and separation distance is approximately 5 $\mu$m. The window with the hole-matrix was fabricated using electron beam lithography and etching. A wedging transfer method \autocite{doi:10.1021/nl1008037} was chosen to avoid damaging the hole-patterned structure. During transfer the polymer-graphene structure was aligned with a micro manipulator to the hole matrix.

  \paragraph{Substrate-supported graphene}
  Monolayer graphene supported by amorphous carbon film was prepared
  by the same method with that for freestanding graphene on Quantifoil
  grids. Graphene films supported by SiO$_{2}$ and SiN$_{x}$ and gold
  were fabricated using a polymer-assisted transfer method. SiO$_{2}$
  (300 nm thermal oxide on Si, Si-Mat) and SiN$_{x}$ (100 nm LPCVD
  low-stress nitride on Si, University Wafer) wafers were cleaned by Piranha solution (7:3
  H$_{2}$SO$_{4}$ : H$_{2}$O$_{2}$) before use. Ultra\-flat Au
  substrates were fabricated by the template-stripping method on silicon
  wafer\autocite{Hegner_1993_template_strip}.  Poly
  methyl\-methacrylate (PMMA, 4\% solution in anisole) was spin-coated
  on CVD-grown graphene/Cu and baked at 120 °C. The Cu foil was etched
  by 0.5 M APS solution. After exchanging APS solution with
  DI-H$_{2}$O, the PMMA/graphene film was transferred onto the desired
  substrate. The substrate was kept in ambient overnight and baked at
  120 °C to enhance adhesion between substrate and graphene. The PMMA
  was removed by aceton and subsequently cleaned using IPA.

  \paragraph{Au deposition}
  Graphene samples (freestanding and substrate-supported) were all
  annealed under 1:1 mixture of Ar and H$_{2}$ at 600 °C for 2 hours
  to remove airborne contamination and loaded in high-vacuum
  evaporation chamber (Plassys MEB550S). Au was deposited by electron
  beam evaporation at room temperature under a pressure $<10^{-7}$
  mBar. The deposition rate was maintained $\sim{} 5 \times 10^{-3}$
  nm·s$^{-1}$. The amount of gold deposited $m_{\mathrm{Au}}$ (mass
  per area) was calculated using nominal thickness
  $\delta_{\mathrm{Au}}$ as
  $m_{\mathrm{Au}} = \delta_{\mathrm{Au}} \rho_{\mathrm{Au}}$, where
  $\rho_{\mathrm{Au}} = 19.30$ g·cm$^{-3}$ is the density of gold.

  \paragraph{BPE deposition}
  Thin film epitaxy of BPE onto the desired substrates were carried
  out in a home-made physical vapor deposition (PVD) chamber at
  pressure below $10^{-5}$ mBar while the substrates are heated up to
  80 °C to faciliate formation of crystalline structures. The
  deposition rate was kept at 0.1 nm·s$^{-1}$ and final thickness of
  BPE was $\sim$50 nm.

  \paragraph{GIWAXS characterizations}
  GIWAXS analysis was conducted on beam\-line BL13A at the National
  Synchrotron Radiation Research Center of Taiwan. The incidence angle
  and beam energy of the X-ray were 0.12 and 12.16 keV, corresponding
  to a wavelength of 1.02143 Å. All of GIWAXS images were collected in
  reflection mode by MAR165 CCD with a 2D area detector.

  \paragraph{Electron microscopy}
  SEM characterizations were carried out on Zeiss ULTRA plus with 3 kV
  beam voltage and 20 $\mathrm{\mu}$m aperture. Scanning transmission electron
  microscopy (STEM) images were acquired with a high-angle annular
  dark field (HAADF) detector both at cryogenic conditions using a
  liquid-nitrogen-cooled holder (Gatan) on a Hitachi HD 2700 CS
  (operation voltage 200 kV). Fourier transformation was performed
  on the phase contrast images.

  \paragraph{AFM characterizations}
  AFM topographies of graphene samples were performed on Bruker
  Resolve using the PeakForce Tapping mode combined with ScanAsyst Air
  ultra\-sharp probe to overcome noise artifacts on freestanding films
  caused by standard tapping mode
  scanning\autocite{Clark_2013_AFM}. Force setpoint were maintained
  under 500 pN to avoid breaking of graphene sheet.

  \paragraph{KMC simulations}
  2D diffusion of Au atoms on graphene surface was simulated by
  standard n-fold KMC algorithm on a triangular lattice with at least
  200 mesh grids in both $x$- and $y$-directions. The following three
  events were considered: (i) deposition of atoms, (ii) diffusion on
  pristine graphene, (iii) diffusion on defective area and (iv)
  interatomic binding, corresponding to kinetic energy barriers of
  $\Delta E_{\mathrm{e}}$, $\Delta E_{\mathrm{d}}^{0}$,
  $\Delta E_{\mathrm{d}}^{*}$ and $\Delta E_{\mathrm{b}}$,
  respectively. Rate $r_{i}$ of individual event $i$ was calculated
  using
  $r_{i} = \nu_{0} \exp(- \frac{\Delta E_{i}} {k_{\mathrm{B}} T})$,
  where $\nu_{0}$ is the rate pre\-factor, and the probability of
  event $i$, $p_{i}$, follows:
  $p_{i} = r_{i} n_{i} / \sum_{i} r_{i} n_{i}$, where $n_{i}$ is the
  degeneracy of event $i$. More details for the parameters used in the
  simulations see Supplementary Information.

\section{Theoretical Simulations}
\label{sec:theory}

\subsection{Modified Lifshitz theory for anisotropic media}
\label{sec:modif-lifsh-theory}

The derivation of equation \ref{main-eq:Phi-aniso} is
described as follows..  The vdW interaction energy of
$\Phi_{\mathrm{AmB}}^{\mathrm{vdW}}$ corresponding to the total energy summed from all
allowed EM modes\autocite{Li_2005_diele}, is given by:
\begin{equation}
\label{eq:vdw-EM-energy}
\Phi^{\mathrm{vdW}}_{\mathrm{AmB}} =
\frac{k_{\mathrm{B}} T}{2(2 \pi)^{2}}  \sum_{n=-\infty}^{\infty} \int_{r_{n}}^{\infty}
\ln \mathcal{D}(i\xi_{n}, \mathbf{k}) \mathrm{d}^{2} \mathbf{k} 
\end{equation}
where $\mathbf{k}=(k_{x}, k_{y})$ is the in-plane wavevector, and
$\mathcal{D}(i\xi_{n}, \mathbf{k})$ is the dispersion relation
for a given geometry.
For generality, the dielectric tensor of material j has diagonal
components $\varepsilon_{\mathrm{j}}^{xx}$,
$\varepsilon_{\mathrm{j}}^{yy}$ and $\varepsilon_{\mathrm{j}}^{zz}$. 
By transforming
\(\mathbf{k} = (\kappa \cos\mathcal{\vartheta}, \kappa\sin
\mathcal{\vartheta})\), and
\(g_{\mathrm{j}} =
\left[\frac{\varepsilon_{\mathrm{j}}^{xx}}{\varepsilon_{\mathrm{j}}^{zz}}
\cos^{2} \vartheta +
\frac{\varepsilon_{\mathrm{j}}^{yy}}{\varepsilon_{\mathrm{j}}^{zz}}
\sin^{2} \vartheta \right]^{-1}\) where $\kappa$, $\vartheta$ are the
corresponding polar coordinates of $\mathbf{k}$, the dispersion
relation $\mathcal{D}$ of an anisotropic A/m/B layered system follows\autocite{Parsegian_2010_vdW}:
\begin{equation}
\label{eq:vdw-disper-D}
\begin{aligned}
\mathcal{D}
&=
1 - 
\underbrace{\left[
\frac{\hat{\varepsilon}_{\mathrm{A}} - \varepsilon_{\mathrm{m}}^{zz} g_{m}^{1/2}(\vartheta) }{\hat{\varepsilon}_{\mathrm{A}} + \varepsilon_{\mathrm{m}}^{zz} g_{\mathrm{m}}^{1/2}(\vartheta)}
\right]}_{\Delta_{\mathrm{Am}}}
\underbrace{\left[
\frac{\hat{\varepsilon}_{\mathrm{B}} - \varepsilon_{\mathrm{m}}^{zz} g_{\mathrm{m}}^{1/2}(\vartheta) }{\hat{\varepsilon}_{\mathrm{B}} + \varepsilon_{\mathrm{m}}^{zz} g_{\mathrm{m}}^{1/2}(\vartheta)}
\right]}_{\Delta_{\mathrm{Bm}}}
e^{-2 g_{\mathrm{m}}^{1/2}(\vartheta) \kappa d} \\
&= 1 - \Delta_{\mathrm{Am}}(\vartheta) \Delta_{\mathrm{Bm}}(\vartheta) e^{-2 g_{\mathrm{m}}^{1/2}(\vartheta) \kappa d}
\end{aligned}
\end{equation}
By further introducing an auxiliary variable
\(x = 2 g_{\mathrm{m}}^{1/2}(\mathcal{\theta}) \kappa d\), it follows:
\begin{equation}
\label{eq:vdw-lifshitz-aniso-final}
\begin{aligned}
  \Phi_{\mathrm{AmB}}^{\mathrm{vdW}} = \dfrac{k_{\mathrm{B}} T}{32 \pi^{2} d^{2}}
  \sum_{n=-\infty}^{\infty} \int_{0}^{2 \pi}
  g_{\mathrm{m}}(i\xi_{n},\vartheta) \mathrm{d}\vartheta
  {\displaystyle \int_{r_{n}}^{\infty}} x \mathrm{d}x \ln[1 -
  \Delta_{\mathrm{Am}}(i\xi_{n},\vartheta)
  \Delta_{\mathrm{Bm}}(i\xi_{n},\vartheta) e^{-x}]
\end{aligned}
\end{equation}
Accordingly, equation \eqref{main-eq:Phi-aniso} is obtained for
$\varepsilon^{xx}_{\mathrm{m}} = \varepsilon^{yy}_{\mathrm{m}}$, i.e. $g_{\mathrm{m}}$ is
independent of $\vartheta$, which is valid for most 2D materials where
the 2D lattice is hexagonal or square. Moreover, in  equation
\eqref{main-eq:Phi-aniso}, $\hat{\varepsilon}_{\mathrm{A}}$ and
$\hat{\varepsilon}_{\mathrm{B}}$ reduce to $\varepsilon_{\mathrm{A}}$ and
$\varepsilon_{\mathrm{B}}$, respectively, when A and B are isotropic bulk
materials. Time-reversal symmetry
$\varepsilon(i \xi) = \varepsilon(-i \xi)$ is used when magnetic response
of the material is negligible, therefore equation \eqref{main-eq:Phi-aniso} only
needs to be evaluated for $\xi_{n} \geq 0$.
Moreover, our numerical analysis suggests the integral of
$\Phi_{\mathrm{AmB}}^{\mathrm{vdW}}$ is dominated by $x \leq 5$, or
equivalently \(\kappa \leq 2.5 (g_{\mathrm{m}} d)^{-1}\).
When $d$ is in the order of 2 nm, and $g_{\mathrm{m}}=2.5$, the
majority of interaction comes from EM modes with \(\kappa<0.05\)
\AA{}\textsuperscript{-1}.  In other words, evaluating 
equation \eqref{main-eq:Phi-aniso} using the material dielectric
functions at the optical limit ($\mathbf{k} \to 0$) would preserve the
accuracy of calculated $\Phi_{\mathrm{AmB}}$ .

\begin{figure}[!htbp]
  \centering \includegraphics[width=0.55\textwidth]{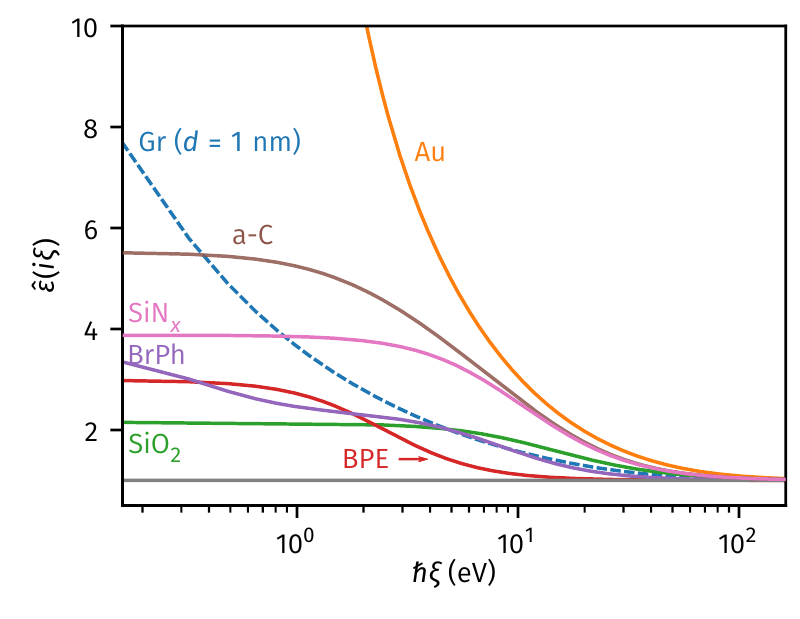}
  \caption{\textbf{Comparison of $\hat{\varepsilon}(i \xi)$
      responses for materials considered in this study}. The effective dielectric function of graphene at $d=1$ nm
    is higher than that of the widely-used high-refractive-index liquid
    bromobenzene (BrPh).}
\label{fig:eps-cascade-full}
\end{figure}

Complementary to  Fig. \ref{main-fig:1}b,
Fig. \ref{fig:eps-cascade-full} compares the dielectric responses of
other materials studied here. Notably, at the order of $d=1$ nm,
$\hat{\varepsilon}_{\mathrm{m}}$ of graphene is comparable to that of
bromobenzene (BrPh), a high-refractive-index liquid commonly used in
experiments demonstrating Casimir
repulsion\autocite{Meurk_1997_afm,Munday_2009_afm}. In this respect, 2D
material appears to be a promising candidate for realizing repulsive vdW
interactions. However, unlike bulk liquid, $\hat{\varepsilon}_{\mathrm{m}}$ of a
2D material strongly depends on the separation $d$, making the
repulsion more pronounced at short distances (the vdW regime).

\subsection{Attractive and repulsive interactions in the Vac/Gr/Au
  system}
\label{sec:thickn-depend-repuls}

\begin{figure}[!htbp]
  \centering
  \includegraphics[width=0.75\textwidth]{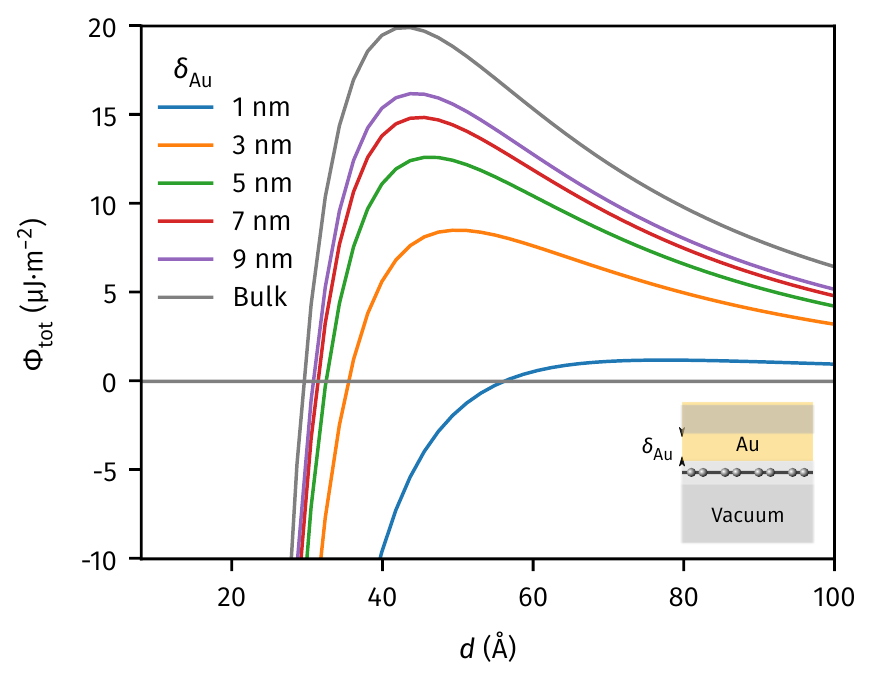}
  \caption{\textbf{Total vdW potential $\Phi_{\mathrm{tot}}$ as a function
      of Au-Gr distance $d$ for different Au layer thickness
      $\delta_{\mathrm{Au}}$ calculated from multi-layer Lifshitz
      approach}. A tiny repulsive potential barrier exists when
    $\delta_{\mathrm{Au}} > 1$ nm, and saturates at $\sim{}20$
    $\mathrm{\mu}$J·m$^{-2}$ for the bulk Au layer, as a result of different power laws
    between attractive and repulsive interactions shown in  Fig. \ref{main-fig:2}b.}
\label{fig:lifshitz-barrier}
\end{figure}

The Lifshitz formalism in equation \ref{main-eq:Phi-aniso} corresponds to the AmB systems when A and B are semi-infinite.
In order to model the interactions of atomically-thin Au platelets,
a multilayer approach was used to calculate
$\Phi_{\mathrm{AmB}}^{\mathrm{vdW}}$ as shown in
Fig. \ref{fig:lifshitz-barrier} inset. The thickness of the Au layer
is $\Delta_{\mathrm{Au}}$, and the effective thickness of graphene is
$\delta_{\mathrm{Gr}}$.

The repulsive interaction $\Phi_{\mathrm{Rep}}$ at separation $d$ is similar to equation \ref{main-eq:Phi-aniso}, with different expression of the dispersion relation\autocite{Parsegian_2010_vdW}:
\begin{equation}
  \label{eq:lifshitz-layer-rep}
  \begin{aligned}[t]
    \Phi_{\mathrm{Rep}}(d, \delta_{\mathrm{Au}}) &= \frac{k_{\mathrm{B}} T}{16 \pi d^{2}} 
 \sum_{n = -\infty}^{\infty}
 \int_{r_{n}}^{\infty} x \ln\left[1 - \Delta^{*}_{\mathrm{L}}(i \xi_{n}) \Delta^{*}_{\mathrm{R}}(i \xi_{n}) e^{-x}\right] \mathrm{d} x \\
 \Delta^{*}_{\mathrm{L}} &= \Delta_{\mathrm{Vac/m}} \\
  \Delta^{*}_{\mathrm{R}} &= \frac{\Delta_{\mathrm{Vac/Au}} e^{-x \frac{\delta_{\mathrm{Au}}}{d}} + \Delta_{\mathrm{Au/m}}}{1 + \Delta_{\mathrm{Vac/Au}}\Delta_{\mathrm{Au/m}} e^{-x \frac{\delta_{\mathrm{Au}}}{d}}}
\end{aligned}
\end{equation}
where the expressions for $\Delta_{\mathrm{Vac/m}}$ and $\Delta_{\mathrm{Au/m}}$ are
analogous to those of $\Delta_{\mathrm{Am}}$ and $\Delta_{\mathrm{Bm}}$ in a
A/m/B system, respectively.

In the platelet system, there is also two-body attractive interaction
between graphene and the Au platelet. The attractive potential between freestanding graphene and Au platelet
$\Phi_{\mathrm{Att}}$ results from the vacuum spacing between the
surfaces of graphene and Au
\autocite{Zhou_2017_lifshitz,Zhou_2018_lifshitz2}. Different from the
repulsive potential, the attractive potential has a shorter
distance $d^{*} = d - \delta_{\mathrm{Gr}}$. The graphene layer is
treated as a dielectric material with the effective dielectric tensor
$\varepsilon_{\mathrm{Gr}}$ same as bulk
graphite and thickness $\delta_{\mathrm{Gr}}$.  Note despite the
breakdown of continuum dielectric function for atomically thin
materials, such effective treatment in Lifshitz theory can still
produce quantitatively correct energy values as compared to other
computationally expensive approaches, such as \textit{ab initio} quantum chemistry simulations. \autocite{Zhou_2017_lifshitz}.
Similar to equation~\eqref{eq:lifshitz-layer-rep},
$\Phi_{\mathrm{Att}}$ of the multilayer configuration is given by:
\begin{equation}
  \label{eq:lifshitz-layer-att}
  \begin{aligned}[t]
    \Phi_{\mathrm{Att}}(d, \delta_{\mathrm{Au}}) &=
    \frac{k_{\mathrm{B}} T}{16 \pi (d^{*})^{2}} \sum_{n =
      -\infty}^{\infty}
    \int_{r_{n}}^{\infty} x \ln\left[1 - \Delta^{*}_{\mathrm{L}}(i \xi_{n}) \Delta^{*}_{\mathrm{R}}(i \xi_{n}) e^{-x}\right] \mathrm{d} x \\
    \Delta^{*}_{\mathrm{L}} &= \frac{\Delta_{\mathrm{Vac/Gr}} e^{-x \frac{\delta_{\mathrm{Gr}}}{d^{*}}} + \Delta_{\mathrm{Gr/Vac}}}{1 + \Delta_{\mathrm{Vac/Gr}}\Delta_{\mathrm{Gr/Vac}} e^{-x \frac{\delta_{\mathrm{Gr}}}{d^{*}}}} \\
    \Delta^{*}_{\mathrm{R}} &= \frac{\Delta_{\mathrm{Vac/Au}} e^{-x
        \frac{\delta_{\mathrm{Au}}}{d^{*}}} +
      \Delta_{\mathrm{Au/Vac}}}{1 +
      \Delta_{\mathrm{Vac/Au}}\Delta_{\mathrm{Au/Vac}}
      e^{-x\frac{\delta_{\mathrm{Au}}}{d^{*}}}}
\end{aligned}
\end{equation}

Equation ~\eqref{eq:lifshitz-layer-att} can be further generalized to
calculate the two-body attraction in the bulk Vac/Gr/Au system, corresponding to
$\delta_{\mathrm{Au}} \to \infty$:
\begin{equation}
  \label{eq:lifshitz-bulk-att}
  \begin{aligned}[t]
    \Phi_{\mathrm{Att}}^{\mathrm{Bulk}}(d) &= \frac{k_{\mathrm{B}}
      T}{16 \pi (d^{*})^{2}} \sum_{n = -\infty}^{\infty}
    \int_{r_{n}}^{\infty} x \ln\left[1 - \Delta^{*}_{\mathrm{L}}(i \xi_{n}) \Delta^{*}_{\mathrm{R}}(i \xi_{n}) e^{-x}\right] \mathrm{d} x \\
    \Delta^{*}_{\mathrm{L}} &= \frac{\Delta_{\mathrm{Vac/Gr}} e^{-x \frac{\delta_{\mathrm{Gr}}}{d^{*}}} + \Delta_{\mathrm{Gr/Vac}}}{1 + \Delta_{\mathrm{Vac/Gr}}\Delta_{\mathrm{Gr/Vac}} e^{-x \frac{\delta_{\mathrm{Gr}}}{d^{*}}}} \\
    \Delta^{*}_{\mathrm{R}} &= \Delta_{\mathrm{Au/Vac}}
\end{aligned}
\end{equation}
Equation \ref{eq:lifshitz-bulk-att} was used to calculate
$\Phi_{\mathrm{Att}}$ in Fig. \ref{main-fig:2}b. For a bulk Vac/Gr/Au
system, when the separation is much larger than
$\delta_{\mathrm{Gr}}$, we have $d \approx d^{*}$. We notice that when
$d \to \infty$, $\Phi_{\mathrm{Att}}$ reduces to the interaction form
between a 2D sheet and semi-infinite bulk material, scaling as
$\Phi_{\mathrm{Att}} \propto d^{-3}$, while $\Phi_{\mathrm{Rep}}$
reduces to the interaction between two semi-infinite bulk materials in
vacuum, yielding $\Phi_{\mathrm{Rep}} \propto d^{-2}$.

The difference in scaling laws between $\Phi_{\mathrm{Att}}$ and
$\Phi_{\mathrm{Rep}}$ results in a repulsive energy barrier beyond
$d>3$ nm, as shown in Fig. \ref{fig:lifshitz-barrier}. As expect, we
observe that the barrier decreases with thinner Au layer. The calculated
magnitude of the barrier ($\sim{}20$ $\mathrm{\mu}$J·m$^{-2}$ in the bulk Vac/Gr/Au
system) is significantly smaller than the surface energy of graphene or gold
and can be overcome by thermal energy at room temperature.
For example, consider a sub-monolayer
of Au atoms on 2D surface with the surface coverage $\lambda$, the thermal
kinetic energy can be estimated using
$E_{\mathrm{kin}} = \lambda k_{\mathrm{B}} T N_{\mathrm{c}} /
S_{\mathrm{c}}$ where $N_{\mathrm{c}}$ and $S_{\mathrm{c}}$ are the
number of atoms and the area of the 2D unit cell, respectively. The
thermal activation energy can be overcome when
$\lambda > 4\times{}10^{-4}$, which is much lower than the
experimental condition. Therefore, we still expect to see nucleation
on the free-standing graphene surface, while the nucleation density is
greatly suppressed due to the existence of such repulsive barrier.

\subsection{Attractive interactions in substrate-supported systems}
\label{sec:attr-inter-substr}

\begin{figure}[!htbp]
  \centering
  \includegraphics[width=0.6\linewidth]{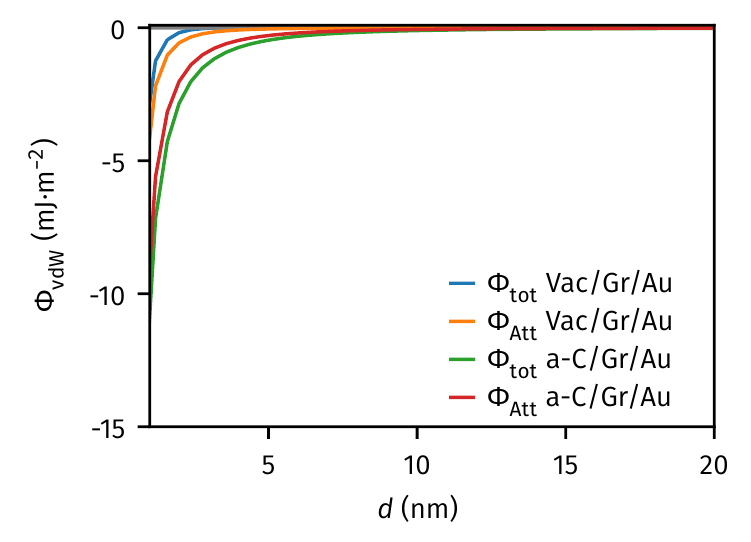}
  \caption{\textbf{$\Phi_{\mathrm{tot}}$ and $\Phi_{\mathrm{Att}}$
      profiles as functions of $d$ for Vac/Gr/Au and a-C/Gr/Au systems
      when $\delta_{\mathrm{Au}}=6$ nm.} The $\Phi_{\mathrm{tot}}-d$
    profiles are the same as those in Fig. \ref{main-fig:2}H while
    with much larger $y$-axis range. In both cases, the attractive
    interaction dominates at short distance.}
  \label{fig:layer-energy-full}
\end{figure}

The analysis in section ~\ref{sec:thickn-depend-repuls} can be easily
extended to model the interactions between substrate (Sub)
-supported graphene and a thin layer of gold.  Assume the substrate has
isotropic dielectric response $\varepsilon_{\mathrm{Sub}}$, the attractive
interaction between a substrate-supported graphene and Au layer,
$\Phi_{\mathrm{Att}}^{\mathrm{ss}}$ (to be distinguished from that of
a freestanding Gr/Au system) is given by:
\begin{equation}
  \label{eq:lifshitz-layer-att-sub}
  \begin{aligned}[t]
    \Phi_{\mathrm{Att}}^{\mathrm{ss}}(d, \delta_{\mathrm{Au}}) &=
    \frac{k_{\mathrm{B}} T}{16 \pi (d^{*})^{2}} \sum_{n =
      -\infty}^{\infty}
    \int_{r_{n}}^{\infty} x \ln\left[1 - \Delta^{*}_{\mathrm{L}}(i \xi_{n}) \Delta^{*}_{\mathrm{R}}(i \xi_{n}) e^{-x}\right] \mathrm{d} x \\
    \Delta^{*}_{\mathrm{L}} &= \frac{\Delta_{\mathrm{Sub/Gr}} e^{-x \frac{\delta_{\mathrm{Gr}}}{d^{*}}} + \Delta_{\mathrm{Gr/Vac}}}{1 + \Delta_{\mathrm{Sub/Gr}}\Delta_{\mathrm{Gr/Vac}} e^{-x \frac{\delta_{\mathrm{Gr}}}{d^{*}}}} \\
    \Delta^{*}_{\mathrm{R}} &= \frac{\Delta_{\mathrm{Vac/Au}} e^{-x
        \frac{\delta_{\mathrm{Au}}}{d^{*}}} +
      \Delta_{\mathrm{Au/Vac}}}{1 +
      \Delta_{\mathrm{Vac/Au}}\Delta_{\mathrm{Au/Vac}}
      e^{-x\frac{\delta_{\mathrm{Au}}}{d^{*}}}}
\end{aligned}
\end{equation}
and subsequently the (two-body) attractive interaction between substrate-supported graphene and bulk gold is given by:
\begin{equation}
  \label{eq:lifshitz-bulk-att-sub}
  \begin{aligned}[t]
    \Phi_{\mathrm{Att}}^{\mathrm{ss,Bulk}}(d) &= \frac{k_{\mathrm{B}}
      T}{16 \pi (d^{*})^{2}} \sum_{n = -\infty}^{\infty}
    \int_{r_{n}}^{\infty} x \ln\left[1 - \Delta^{*}_{\mathrm{L}}(i \xi_{n}) \Delta^{*}_{\mathrm{R}}(i \xi_{n}) e^{-x}\right] \mathrm{d} x \\
    \Delta^{*}_{\mathrm{L}} &= \frac{\Delta_{\mathrm{Sub/Gr}} e^{-x \frac{\delta_{\mathrm{Gr}}}{d^{*}}} + \Delta_{\mathrm{Gr/Vac}}}{1 + \Delta_{\mathrm{Sub/Gr}}\Delta_{\mathrm{Gr/Vac}} e^{-x \frac{\delta_{\mathrm{Gr}}}{d^{*}}}} \\
    \Delta^{*}_{\mathrm{R}} &= \Delta_{\mathrm{Au/Vac}}
\end{aligned}
\end{equation}
where the term $\Delta_{\mathrm{Sub/Gr}}$ corresponds to the
dielectric mismatch between the substrate and graphene interface. Such
method is used to construct the potential-distance curve in
Fig. \ref{main-fig:2}H. As seen from equation
\eqref{eq:lifshitz-bulk-att-sub}, $\Delta_{\mathrm{Sub/Gr}}$ has
considerable influence on the attractive potential, due to the atomic
thickness of graphene (i.e.  $\delta_{\mathrm{Gr}} \ll d^{*}$).  In
other words, $\Phi_{\mathrm{Att}}$ between Sub/Gr and gold platelet
also depends on $\varepsilon_{\mathrm{Sub}}$.  Such effect is clearly
different from the classical vdW description of 2D material
interfaces, where the attractive two-body potential is independent
of the type of substrate\autocite{Rafiee_2012_trans,Shih_2012_prl}.
As an
example, the $\Phi_{\mathrm{tot}}$ and $\Phi_{\mathrm{Att}}$ as
functions of $d$ for Vac/Gr/Au and a-C/Gr/Au systems when
$\delta_{\mathrm{Au}}=6$ nm (corresponding to Fig. \ref{main-fig:2}H)
are shown in Fig. \ref{fig:layer-energy-full}.
We observe the $\Phi_{\mathrm{tot}}$ for a-C/Gr/Au system to be
significantly stronger (more negative) than that for Vac/Gr/Au over
the whole $d$-range.  Therefore, it is clear that only in Vac/Gr/Au
system can an overall repulsive vdW barrier be observed.

\section{Force-distance AFM measurements}
\label{sec:F-d-meas}

\subsection{AFM setup and calibration}
\label{sec:tip-calibr}

\begin{figure}[!htbp]
	\centering
	\includegraphics[width=0.8\textwidth]{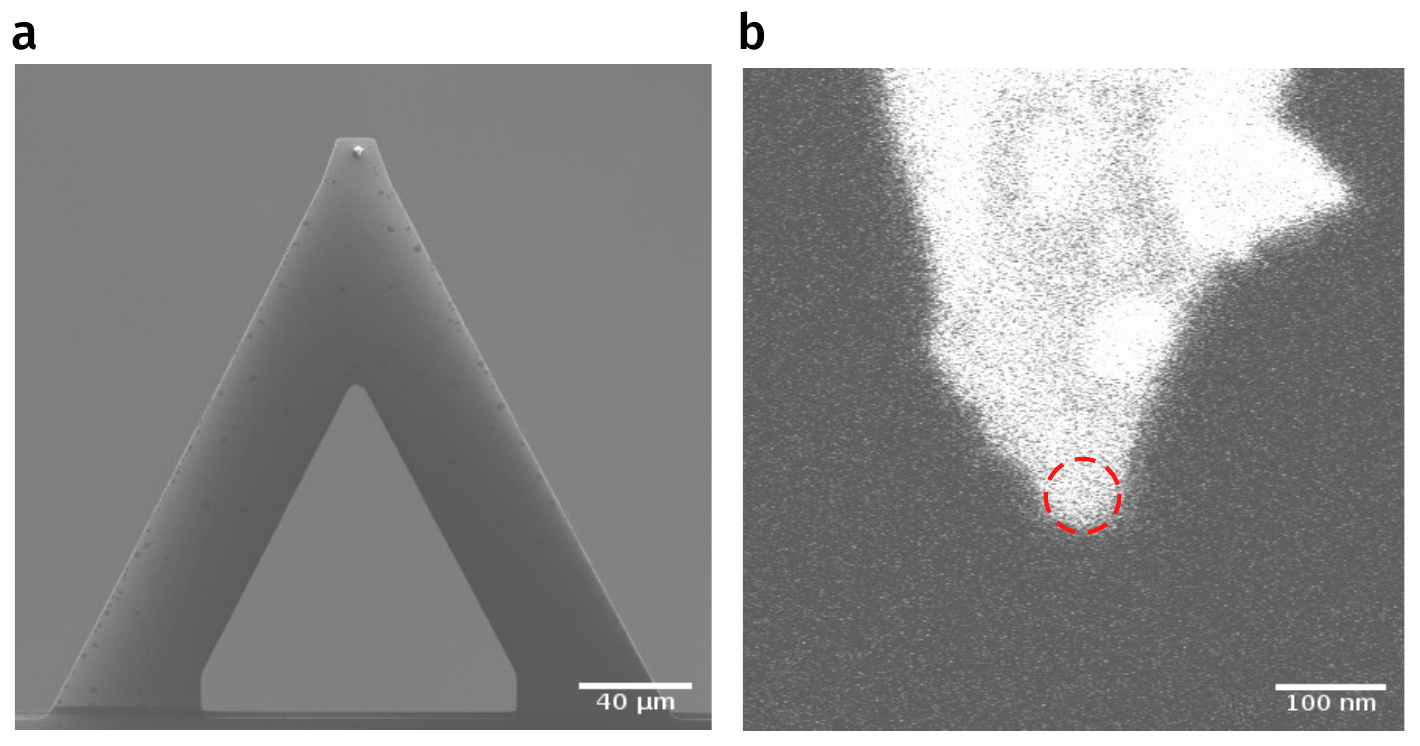}
	\caption{\textbf{SEM images of the AFM probe considered in this study.}
		\textbf{a}. Top-view image of the AFM probe. The cantilever is of triangular shape with a calibrated deflection sensitivity of 69.679 $\frac{\text{nm}}{\text{V}}$ and a spring constant of 0.16136 $\frac{\text{N}}{\text{m}}$.
\textbf{b}. Image of the AFM tip silhouette. The hemispherical shape of the tip is highlighted with a red dashed line and the radius $R_\text{tip}$ was estimated to be approximately 33 nm.						  			}
	\label{fig:AFM-tip}
\end{figure}

The force-distance measurements were carried out with a Bruker BioScope Resolve AFM using the B side of a gold-coated NPG-10 Bruker AFM tip, with a estimated radius $R_\text{tip}$ of approximately 33 nm, as highlighted with a red dashed circle in Fig. \ref{fig:AFM-tip}b. 
In order to obtain the interaction force (N) from the original measured values of the piezo element displacement in (mV), we carried out tip calibration steps to extract the deflection sensitivity $\left(\frac{\text{m}}{\text{V}}\right)$ and spring constant $\left(\frac{\text{N}}{\text{m}}\right)$, by using the PeakForce\texttrademark QNM\texttrademark suite within the NanoScope\textregistered software environment, which determined a deflection sensitivity of 69.679 $\frac{\text{nm}}{\text{V}}$ and a spring constant of 0.16136 $\frac{\text{N}}{\text{m}}$. The spring constant is comparable to the nominal value of 0.12$\frac{\text{N}}{\text{m}}$ provided by the tip vendor. 

\subsection{AFM background noise subtraction}
\label{sec:data-eval}
	
The raw measurement data was processed by subtracting background noise, which was determined by fitting the force-distance measurement data with the following function:
\begin{equation}
	f_{\text{Bkg}} (d) = p_0 + p_1 d + A \sin(\omega d + \theta)
\end{equation} 
within the distance range $d$ between 150 nm to 500 nm, which is far before the tip establishes contact or after it retracts from graphene. The function $f_{\text{Bkg}}$ contains two distinct parts of correction. The linear function $p_0 + p_1 d$, could correct the noise resulting from any slant between the tip and the target. The periodic function $A \sin(\omega d + \theta)$ is to correct the optical interference noise caused by stray light reflected from the supporting material, which is a well-known phenomena taking place during force-distance measurement of reflective surfaces\autocite{PhysRevB.45.11226}, \autocite{doi:10.1063/1.1646767}.  
\begin{figure}[!htbp]
	\centering
	\includegraphics[width=0.6\textwidth]{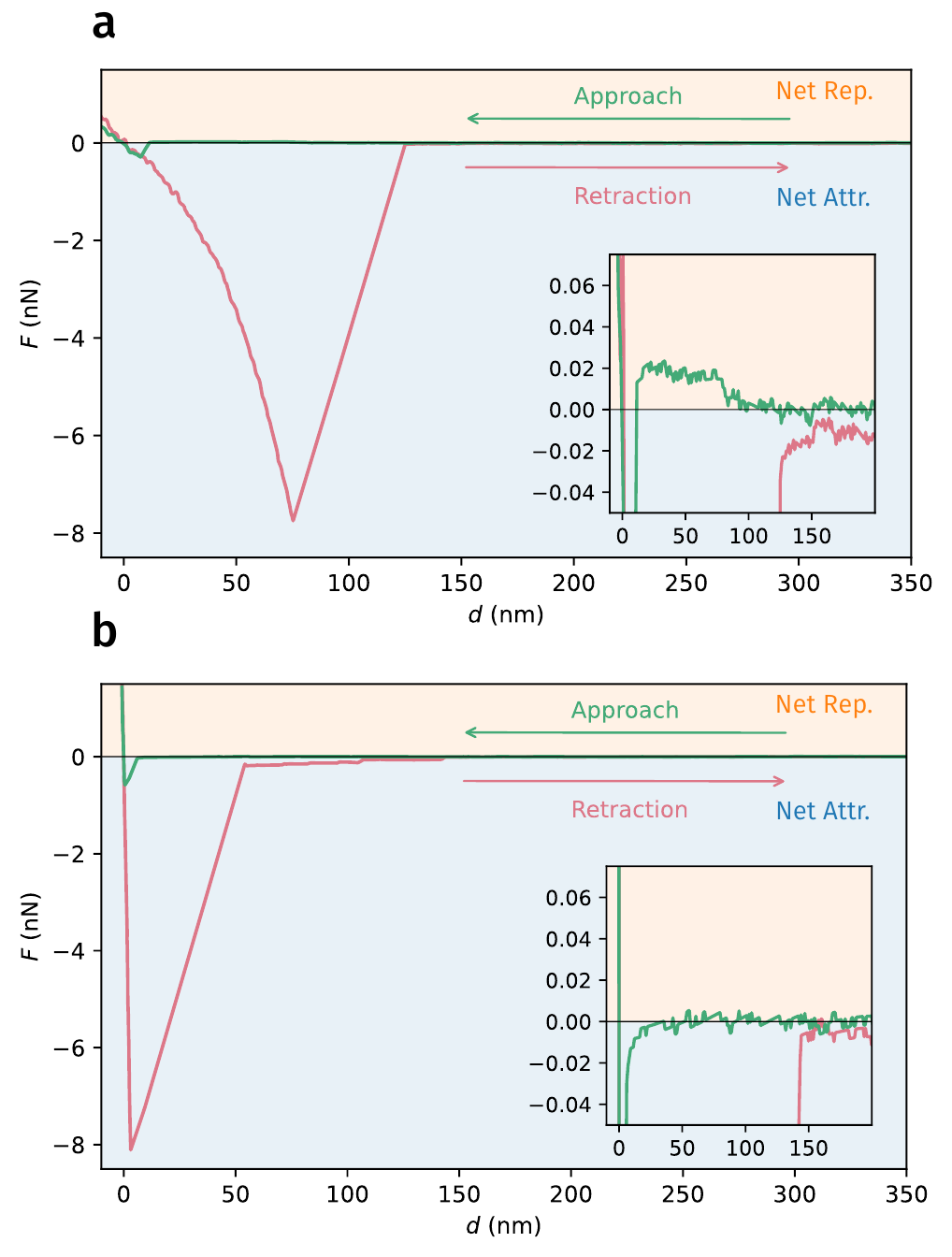}
	\caption{\textbf{Representative force-distance responses after noise substraction.}
		\textbf{a} and \textbf{b} correspond to one force-distance hysteresis measurement on freestanding and SiN$_{x}$ supported graphene, respectively. The insets magnify the force responses within the contact region. Each measurement contains the approach (green) and retraction (red) responses.
	}
	\label{fig:single-f-d}
\end{figure}
This noise subtraction method is applied on both freestanding and SiN$_\text{x}$ supported graphene, which yielded consistent results in terms of a stable baseline for further quantification, regardless of the substrate choice and whether the tip approach/retraction measurements. An example of processed force-distance hysteresis measurement is shown in Fig. \ref{fig:single-f-d}, where Fig. \ref{fig:single-f-d}a and \ref{fig:single-f-d}b correspond to measurements on freestanding and SiN$_\text{x}$ supported graphene, respectively. The distance $d = 0$ nm was set to be the point for which the force $F = 0$ N, after establishing contact. The noise-subtracted force responses exhibit well-defined zero baseline, with all essential information retained within the contact region, such as repulsive and attractive interaction, as well as the mechanical responses on freestanding and supported graphene upon retraction \autocite{doi:10.1126/science.1157996}.

\subsection{Estimation of the repulsive energy barrier}
\label{sec:barrier-estimate}

Consider the approach process for an AFM tip interacting with a surface, the net work experienced by the tip before establishing contact, $W$, is given by:
\begin{equation}\label{eq:net-work}
	W = \int_{\infty}^{d_0} F(d)\text{d}d
\end{equation}
where $F$ is the measured force response as a function of tip displacement $d$, and $d_0$ is the displacement where the contact is established. 

\begin{figure}[!htbp]
	\centering
	\includegraphics[width=0.8\textwidth]{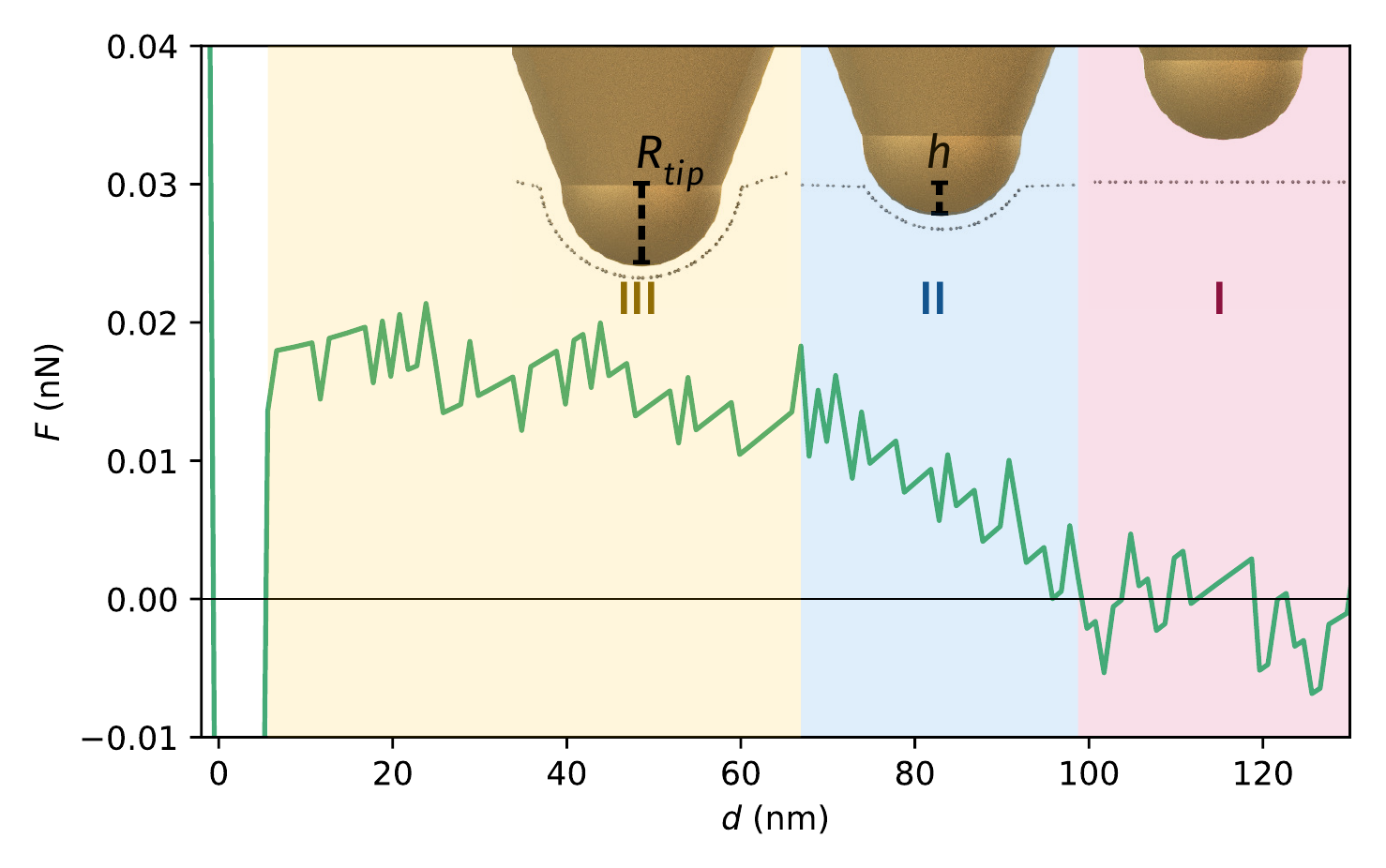}
	\caption{\textbf{Representative force-distance response for the approach process of a gold-coated AFM tip interacting with graphene membrane.}
		The interaction between the AFM tip and freestanding graphene reveals three different regions, describing different force response behavior of the AFM Tip. 
}
	\label{fig:AFM-E-barrier-estimate}
	\end{figure}

Note that the net work $W$ is of unit of energy, and in order to quantitatively compare with the theoretically calculated energy barrier height on freestanding graphene, $\Phi_{\text{b}}$, it would require the estimation of the average area for the interaction between the AFM tip and the freestanding surface, $A_\text{avg}$, following:
\begin{equation}\label{eq:estimate-w-barrier}
	W = A_\text{avg}\Phi_{\text{b}}
\end{equation}
To properly model $A$ in the system of freestanding graphene, we consider a simplified picture of approach process (see Fig. \ref{fig:AFM-E-barrier-estimate}), in which three regimes in the force response are identified. Specifically, upon reducing the separation $d$, the AFM tip starts probing a repulsive force, corresponding to the transition from regime I to II in Fig. \ref{fig:AFM-E-barrier-estimate}, the repulsive force linearly increases with $d$, due to an increase of interaction area upon bending of graphene membrane (regime II in Fig. \ref{fig:AFM-E-barrier-estimate}). After reaching a certain distance, where the depth is approximately equal to $R_\text{tip}$, the slope of force increase is significantly reduced (regime III). We suppose the repulsive force experienced by graphene results in a gradual stretching of graphene membrane. Given the fact that the mechanical behavior of graphene membrane dominates the force-distance response, we estimate the average area between the AFM tip and the freestanding surface throughout the approach process is approximately equal to the surface area of the hemispherical tip interacting with graphene membrane given by:
\begin{equation}\label{eq:avg-area}
	A_{\text{avg}}= 2\pi R_\text{tip}^2
\end{equation}
Given the tip radius characterized in SEM of 33 nm, 
the calculated value of $\Phi_{\text{b}}$ is 19$\pm$5 $\mathrm{\mu}$J$\cdot$m$^{-2}$, which is nicely consistent with that predicted by our theory 20 $\mathrm{\mu}$J$\cdot$m$^{-2}$, as shown in Fig. \ref{fig:lifshitz-barrier}.

\section{Au epitaxy on graphene surfaces}
\label{sec:epitaxy-au}

\subsection{Relation between graphene quality and Au morphology}
\label{sec:quality-gr-au}

\begin{figure}[!htbp]
  \centering
  \includegraphics{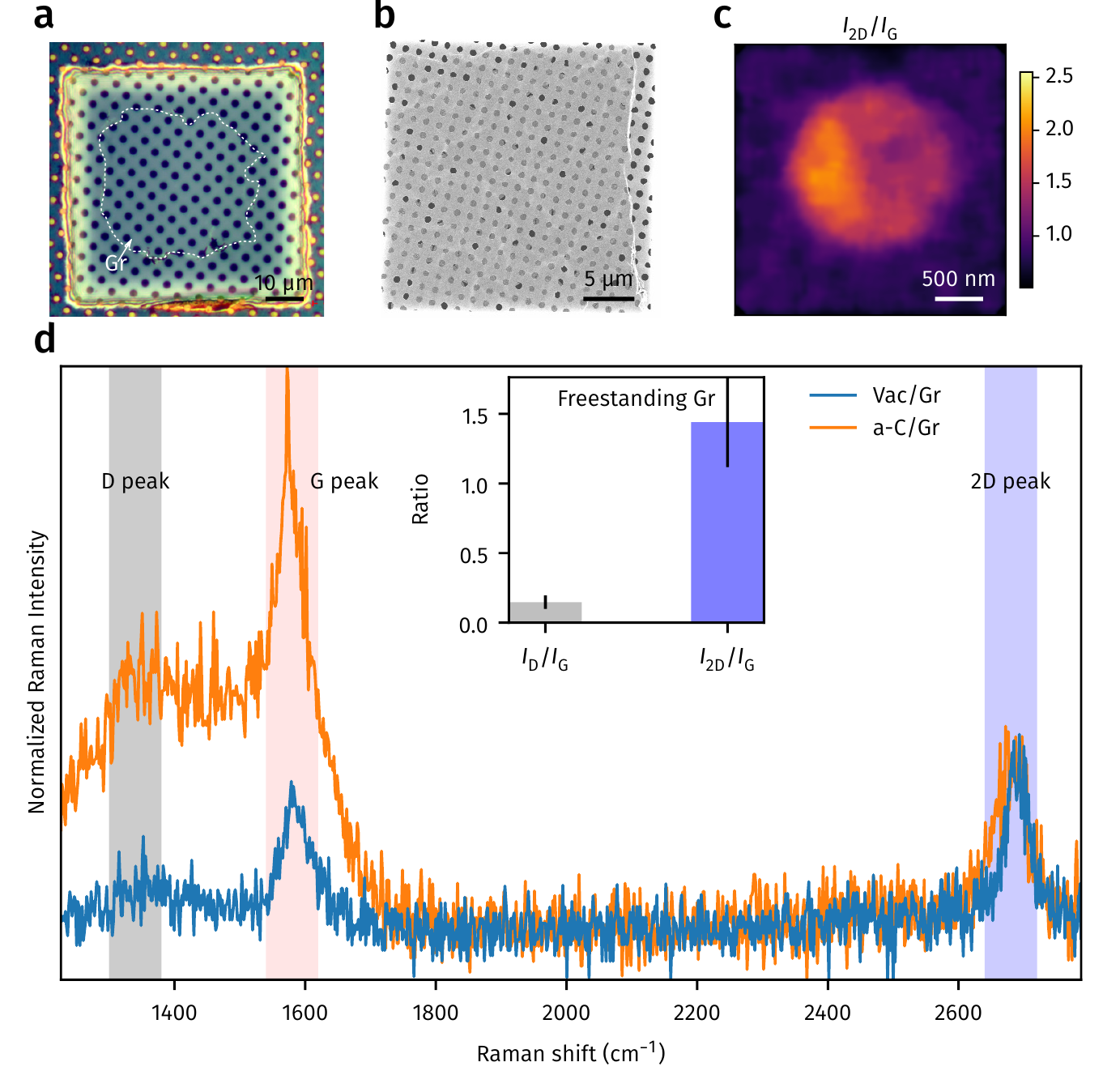}
  \caption{\textbf{Characterization of freestanding graphene
      fabricated by the polymer-free transfer method.}  \textbf{a} Optical micrograph of free-standing graphene covered on
    Quantifoil holes. The boundary of graphene is identified by white
    dashed line.  \textbf{b} SEM image of graphene-covered Quantifoil grid showing high
    yield.  \textbf{c} 2D Raman mapping of the intensity ratio between 2D and
    G peaks near a graphene-covered hole.\textbf{d} Typical single-spot Raman spectra for freestanding
    graphene (blue) and graphene sitting on a-C (orange)  after
    annealing, showing minimal influence of annealing on the
    defect density. The averaged
    $I_{\mathrm{D}} / I_{\mathrm{G}}$ and
    $I_{\mathrm{2D}} / I_{\mathrm{G}}$ ratio values of $>$50 samples
    are shown in the inset.}
  \label{fig:susp-raman}
\end{figure}

We have used multiple characterization techniques to examine the
quality of gold deposited on freestanding graphene. As shown in Figs 
\ref{fig:susp-raman}a and \ref{fig:susp-raman}b, after the wet
transfer process, >75\% pores of the Quantifoil grid are covered by
monolayer graphene. Raman spectroscopy (Renishaw inVia™ confocal
Raman, laser 532 nm) of a typical pore (Fig. \ref{fig:susp-raman}c)
showed that the intensity ratio between the 2D and G resonance peaks
($I_{\mathrm{2D}}/I_{\mathrm{G}}$) is uniformly higher than 1.5 inside
the pore region, indicating the successful transfer of monolayer
graphene.
Annealing of the wet-transferred samples under Ar/H$_{2}$ environment
had minimal influence on the quality of the monolayer graphene with no
significant increase of D peak intensity, as shown in
Fig. \ref{fig:susp-raman}D.

\begin{figure}[!htbp]
  \centering
  \includegraphics{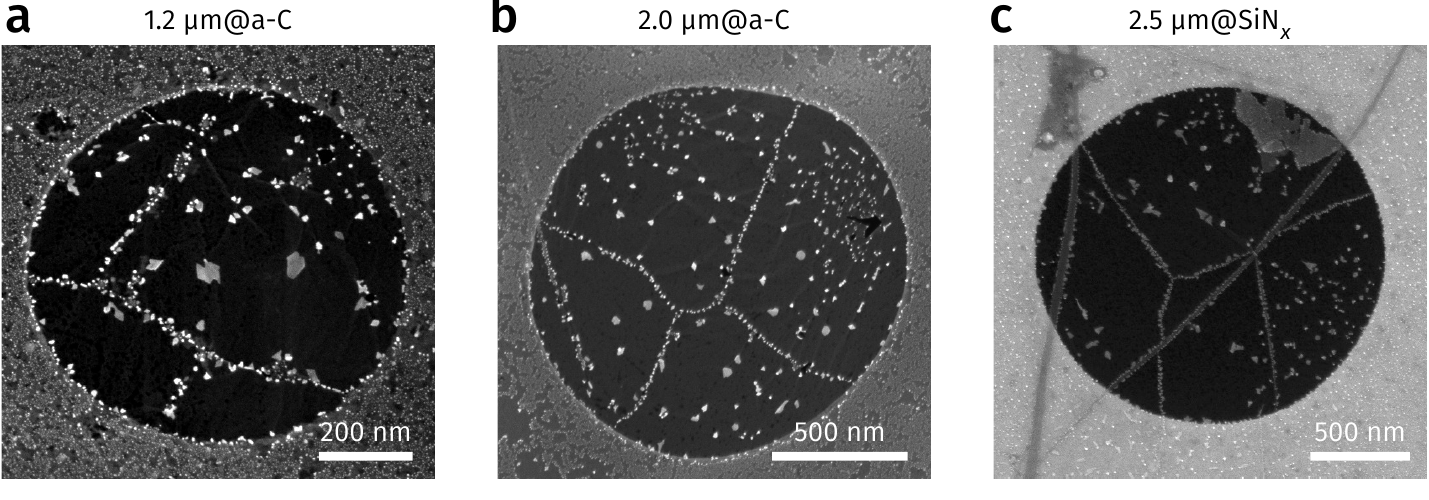}
  \caption{\textbf{SEM images of Au deposited on graphene-covered
      holes on various pore sizes and substrates:} \textbf{a} 1.2 $\mathrm{\mu}$m
    pore on a-C support, \textbf{b} 2.0 $\mathrm{\mu}$m pore on a-C support, and
    \textbf{c} 2.5 $\mathrm{\mu}$m pore on SiN$_{x}$ support.}
\label{fig:different-chip}
\end{figure}

The method in this study can be used to fabricate Au
nanostructures on freestanding graphene samples with varied pore size
and supporting substrate. As shown in Figs \ref{fig:different-chip}a
and \ref{fig:different-chip}b, similar nanostructure morphology was
observed for Vac/Gr/Au with 1.2 $\mathrm{\mu}$m and 2.0 $\mathrm{\mu}$m pores on
amorphous carbon grids, respectively. In addition, we also fabricated
freestanding graphene samples suspended on perforated SiN$_{x}$
chips. The SiN$_{x}$ chips with pore opening ranging from
1.5$\sim$20 $\mathrm{\mu}$m were fabricated according to
Ref. \cite{Celebi_2014_science}. CVD-grown graphene coated with 
Poly(methyl methacrylate)
(PMMA) was transferred onto the holes using standard
wet-etching process\autocite{Liang_2011_clean_graphene}. After
completely drying in ambient, the PMMA/Gr/SiN$_{x}$ stack was directly
annealed under Ar/H$_{2}$ environment at 600 °C to remove
polymer coating. Although the radicals generated during the thermal decomposition
of PMMA might bond with defective graphene
surface\autocite{Lin_2011_PMMA_decomp}, as shown in
Fig. \ref{fig:different-chip}c, at sub-$\mathrm{\mu}$m scale, the Au nucleation
density on freestanding graphene transferred on perforated SiN$_{x}$
substrates is similar to the samples prepared without polymer,
indicating the successful fabrication of clean graphene surfaces using
both methods.

\begin{figure}[!htbp]
  \centering
  \includegraphics{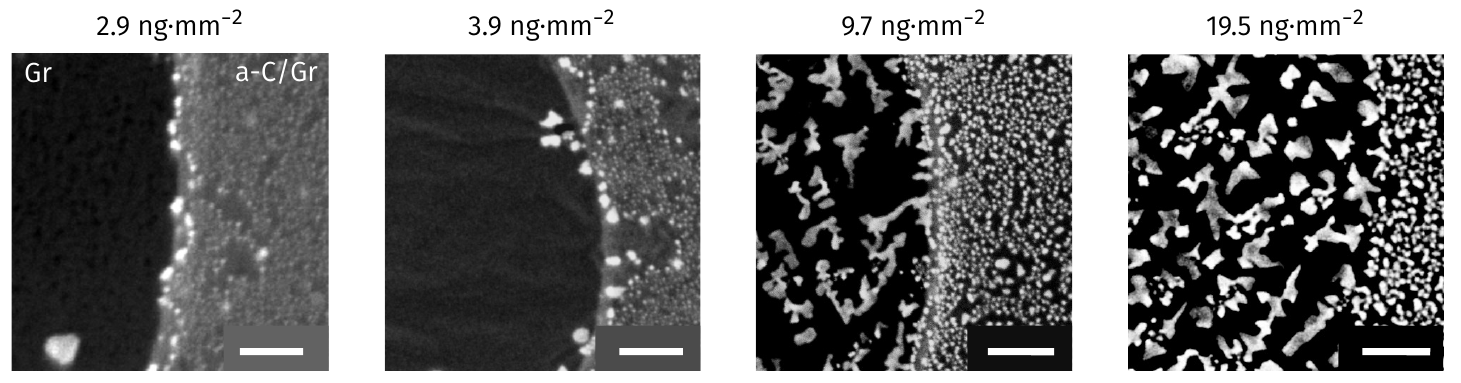}
  \caption{\textbf{Morphology of Au nanostructures by varying amount of
      evaporation}. Scale bars: 100 nm.}
  \label{fig:support-height-au}
\end{figure}

We also investigated the Au morphology change on freestanding
and supported graphene by varying the amount of evaporation. As
shown in Fig. \ref{fig:support-height-au}, when Au evaporation
increased from 2.9 ng·mm$^{-2}$ to 19.5 ng·mm$^{-2}$ (corresponding to
nominal thickness of 0.15 nm to 1 nm, respectively), the nucleation
density on the freestanding graphene remained lower than that on
a-C/Gr.
When $>9.7$ ng·mm$^{-2}$ of Au was evaporated, coalescence of
Au platelets was observed which formed larger nanostructures.  It is
worth noting that these interconnected platelets
substantially differs the dendritic Au patterns
grown on graphite surface at room
temperature\autocite{Cihan_2016_augr,Cihan_2015_Au_graphite}, 
which we will further explain using KMC simulations in section 
\ref{sec:kmc-simulations-au}, as an indirect evidence of
ultra-fast in-plane movement of Au on freestanding graphene.

Despite a low  defect density in the CVD-grown graphene as
indicated by Raman spectroscopy, atomistic defects and contamination
still have strong influence on the Au deposition process.
\begin{figure}[!htbp]
  \centering
  \includegraphics{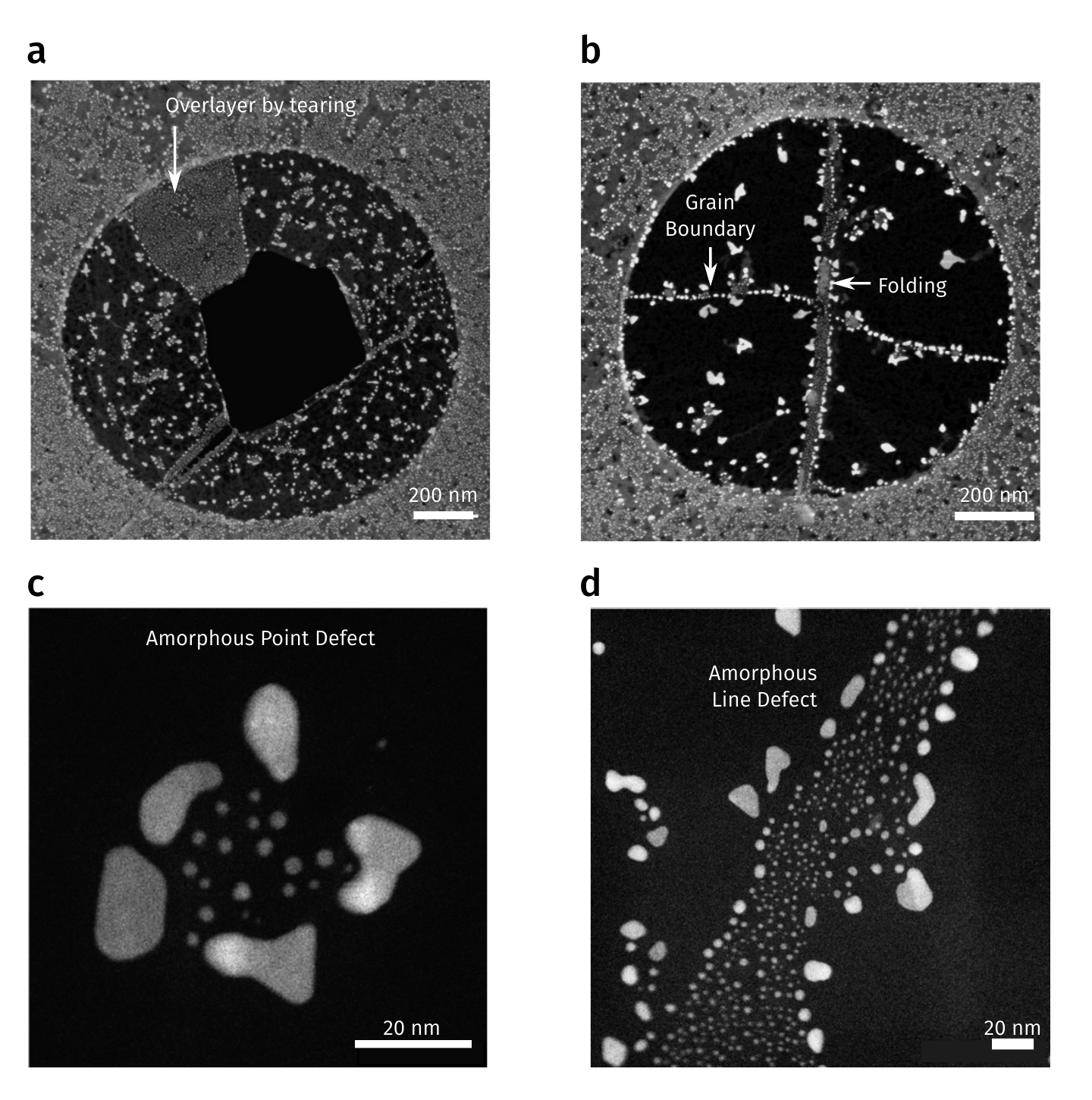}
  \caption{\textbf{Different types of high-nucleation-density areas on
      freestanding graphene due to surface defects and contamination.} \textbf{a} SEM
    image of high-nucleation-density Au deposited on overlayer graphene formed due to tearing. \textbf{b} SEM image of high-nucleation-density
    Au grown near the line defect and folded regions on freestanding graphene. 
\textbf{c} STEM image showing the morphology of Au deposited near
    a point contamination site on freestanding graphene. \textbf{d} STEM
    image of the Au morphology near a line defect.
In both cases, smaller spherical
    particles are formed on the contaminated area, surrounded by
    larger planar structures, as a result of different diffusivities on
    different surfaces. }
\label{fig:defect-type}
\end{figure}
Electron micro\-graphs in Fig. \ref{fig:defect-type} demonstrate
different types of high-nucleation-density Au nanostructures on
defective or contaminated areas of freestanding Gr. As shown in
Fig. \ref{fig:defect-type}a and \ref{fig:defect-type}b, high density
Au particles were observed on mechanical defects of graphene surface, such as (i)
over\-layer (ii) folding and (iii) grain boundary regions, which
agrees with previous observation of metal epitaxy on graphite surface
\autocite{Cross_2007_line}.
The high-nucleation-density areas on graphene can be linked to point
and line defects, as indicated by STEM images in
Fig. \ref{fig:defect-type}c and \ref{fig:defect-type}D,
respectively. Au nanostructures on such defective regions usually
appear as small and spherical particles surrounded by large
nano\-platelets, which agrees with our hypothesis that friction of Au
motion on freestanding Gr is negligible.

\begin{figure}[!htbp]
  \centering
  \includegraphics[width=0.7\textwidth]{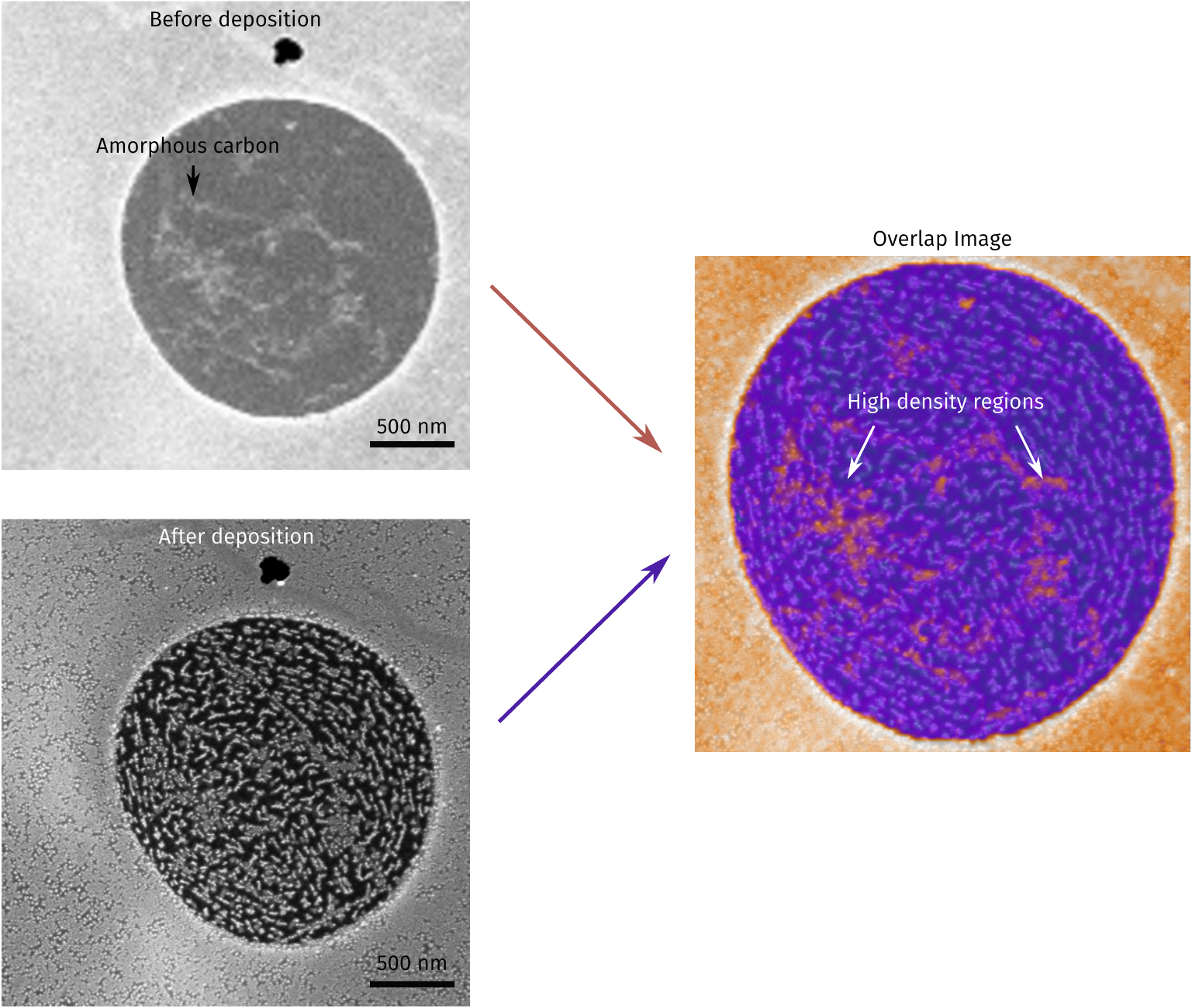}
  \caption{\textbf{Influence of surface contamination on the Au morphology 
   on freestanding graphene.} SEM images show that the brighter regions
    corresponding to amorphous carbon contamination on freestanding graphene before
    deposition (top left panel) coincide with the areas of high
    nucleation density after deposition (bottom left panel), which is
    confirmed by superimposing the SEM images before and after
    deposition (right panel). }
  \label{fig:defect-exsitu}
\end{figure}

The correlation between surface contamination and formation of
spherical Au nanostructures is further elaborated by comparing SEM
images of the same graphene-covered hole in the Quantifoil grid before
and after deposition of Au, as shown in
Fig. \ref{fig:defect-exsitu}. Our results indicate  the cleanness
of freestanding graphene is crucial for the observation of
low-nucleation-density surface, in accordance with previous TEM studies of atomic-scale metal nucleation on graphene
\autocite{Zan_2011_au_gr}.

\begin{figure}[!htbp]
  \centering
  \includegraphics{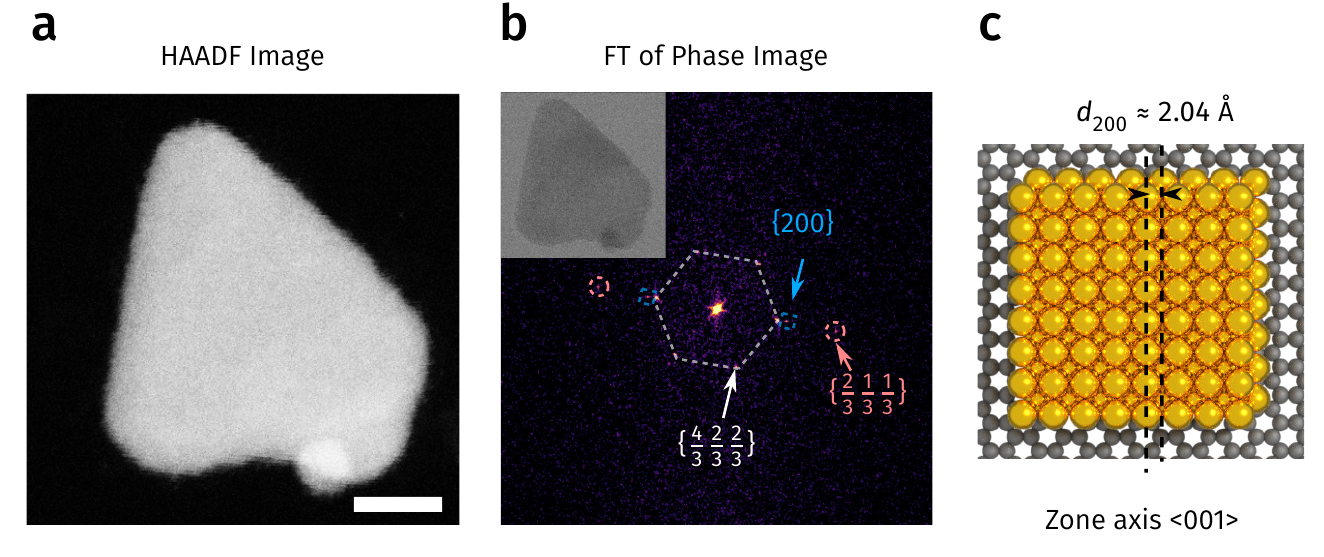}
  \caption{\textbf{Characterization of triangular Au nanostructures.}
    \textbf{a} STEM image of selected triangular Au nanostructure with
    high-angle annular dark field (HAADF) detector.  Scale bar: 5
    nm. \textbf{b} Fourier-transformed (FT) image of the phase signal
    (inset), showing mixed lattice plane sets corresponding to
    \{$200$\} (blue) , \{$\frac{4}{3} \frac{2}{3} \frac{2}{3}$\}
    (white) and \{$\frac{2}{3} \frac{1}{3} \frac{1}{3}$\} (pink,
    double-spacing of \{$\frac{4}{3} \frac{2}{3} \frac{2}{3}$\})
    families. \textbf{c} Scheme of the ($200$) lattice plane seen from
    the <$001$> zone axis of the face-center-cubic (fcc) bulk crystal
    of Au.}
\label{fig:STEM-lattice-thick}
\end{figure}

Unlike spherical clusters observed in a-C/Gr/Au system, Au
platelets in Vac/Gr/Au usually appear more regular geometries, such as
hexagons or triangles. In addition to hexagonal
Au platelets shown in Fig. \ref{main-fig:2}D,
Fig. \ref{fig:STEM-lattice-thick} displays the STEM characteristics of a
triangular Au platelet. From the FT-STEM image in
Fig. \ref{fig:STEM-lattice-thick}b, in addition to the
$\{\frac{4}{3} \frac{2}{3} \frac{2}{3}\}$ (and its double-spacing counterpart
$\{\frac{2}{3} \frac{1}{3} \frac{1}{3}\}$) lattice planes, another set
of $\{200\}$ peaks emerges, corresponding to the <001> zone axis
perpendicular to the graphene surface. The co-existence of two sets of
diffraction planes indicate the Au platelet is thicker than a few
monolayers. Detailed topography profiles of these Au nanostructures
were measured by atomic force microscopy (AFM) in section
\ref{sec:mechanism-gr}.

\subsection{AFM characterizations}
\label{sec:mechanism-gr}

\begin{figure}[!htbp]
  \centering
  \includegraphics{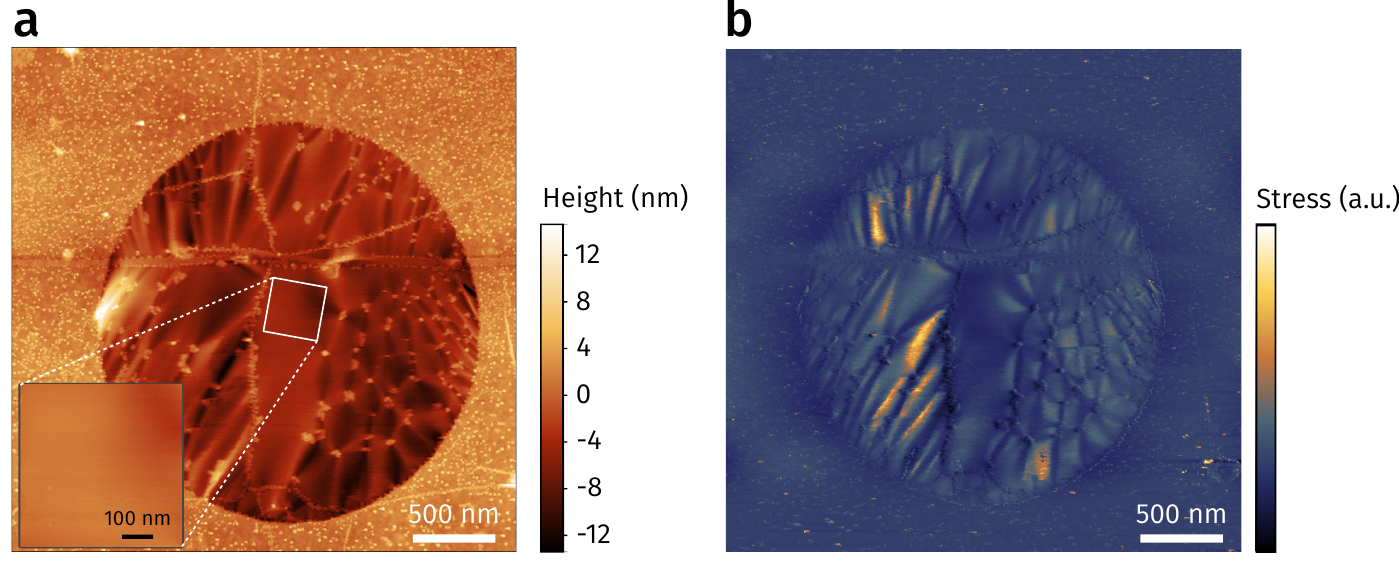}
  \caption{\textbf{AFM topographic and stress images of a
      graphene-covered hole in SiN$_{x}$ substrate after Au
      deposition}. \textbf{a} and \textbf{b} show the same mapped area
    with height and deformation signals, respectively. Inset of
    \textbf{a} shows the topography on a non-wettable region on
    freestanding graphene (color scale normalized for visualization).}
  \label{fig:AFM-peakforce-large}
\end{figure}

To investigate the mechanism of 2D Au nano\-platelets formation on
freestanding graphene, we performed detailed analysis of their height
and size distribution on different graphene surfaces using AFM.
Freestanding graphene on perforated SiN$_{x}$ chips were used for all
AFM experiments to overcome the issue of high surface corrugation of
amorphous carbon substrate after annealing.
The SiN$_{x}$ substrate surface remained flat after annealing. As
shown in Fig. \ref{fig:AFM-peakforce-large}a, the freestanding
graphene surface was highly corrugated, with height prominence at the
order of 20 nm, which is corroborated by deformation mapping of the
same area as shown in Fig. \ref{fig:AFM-peakforce-large}b. Despite the
use of polymer coating during graphene transfer, the annealing process
almost completely removed the PMMA residue, as indicated by the AFM
topography in the inset of Fig. \ref{fig:AFM-peakforce-large}a.

Table
\ref{tab:afm-rough} compares the root mean square (RMS) roughness
$R_{\mathrm{q}}$ and arithmetic roughness $R_{\mathrm{a}}$ for
as-prepared freestanding and substrate-supported graphene
surfaces. The roughness of freestanding graphene is significantly
higher than that of substrate-supported graphene, indicating the
observed low nucleation density on freestanding graphene is not caused
by structural
super\-lubricity\autocite{Cihan_2016_augr,Ozogul_2017_platinum_graphite}
which were observed on ultra\-flat graphite surface.

\FloatBarrier
\begin{table}[h!]
  \centering
  \caption{\textbf{Root mean square (RMS) roughness  $R_{\mathrm{q}}$
      and arithmetic roughness $R_{\mathrm{a}}$ of different graphene
      surfaces after annealing.} The mean roughness of freestanding
    graphene is substantially higher than the substrate-supported
    graphene surfaces due to high degree of winkle and surface
    corrugation.}
  \begin{tabular}[!htbp]{lrrrr}
  \hline{}
  \  & Freestanding Gr & a-C/Gr& SiN$_{x}$/Gr & SiO$_{2}$/Gr \\
  \hline{}
  RMS Roughness $R_{\mathrm{q}}$ (nm) & 2.15$\pm$0.61 &  0.663$\pm$0.029 & 0.409$\pm$0.042 & 0.242$\pm$0.019 \\
  Arithmetic Roughness $R_{\mathrm{a}}$ (nm) & 1.70$\pm$0.45 &  0.450$\pm$0.015 & 0.324$\pm$0.028 & 0.195$\pm$0.011 \\
  \hline{}
\end{tabular}

   \label{tab:afm-rough}
\end{table}

\begin{figure}[!htbp]
  \centering
  \includegraphics{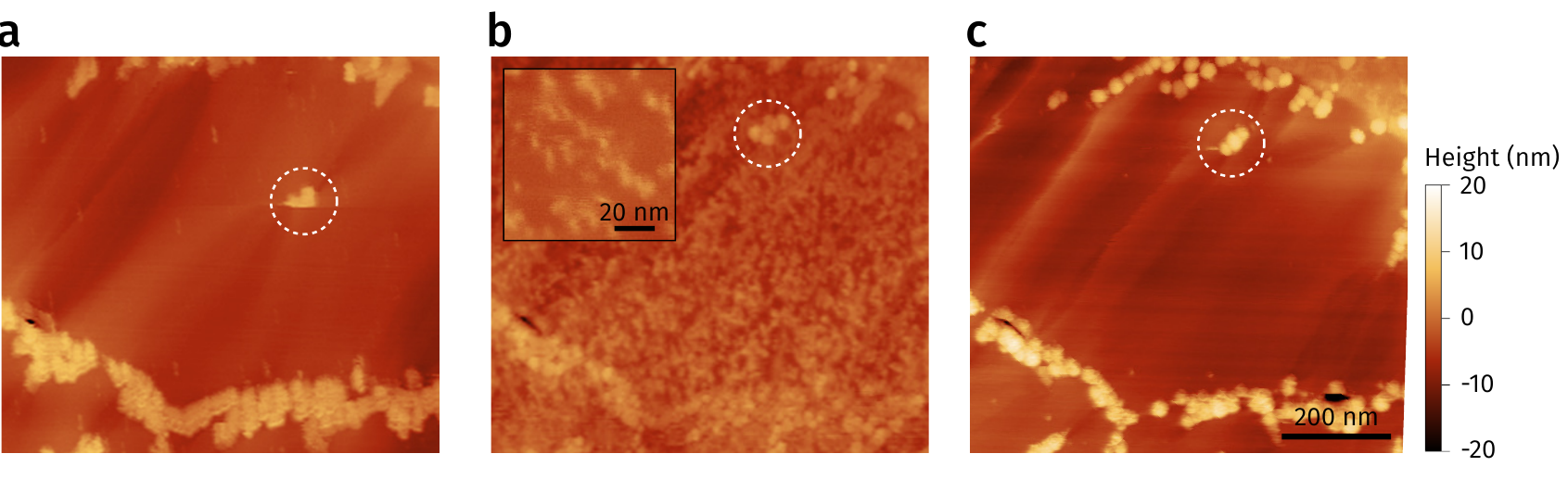}
  \caption{\textbf{Effect of thermal annealing on the quality of
      freestanding graphene and Au nanostructure.} \textbf{a} AFM
    topography of Vac/Gr/Au directly after Au deposition. \textbf{b} AFM
    topography of same area after the sample was kept in ambient for 24
    hours. Airborne contamination can clearly be observed. Inset:
    zoomed-in mapping on the contaminated area. \textbf{c} The same
    area after re-annealing the sample at 600 °C. The
    amorphous contaminants were completely  removed, while the
    average height of Au nanostructures increases. Note the Au
    nanostructure marked in white circle appeared to be moved 
    from \textbf{a} to \textbf{c}. }
  \label{fig:AFM-annealing}
\end{figure}

Using AFM, we were also able to identify the influence of thermal annealing by comparing the
topography of the same region before and after thermal
treatment. Fig. \ref{fig:AFM-annealing}a shows the AFM topography of
Vac/Gr/Au directly after Au evaporation with large non-wettable
area. However, after placing the sample under ambient conditions for 1
day, significant contamination layer with average height of $1\sim{}2$ nm
was observed on the previously clean regions 
(Fig. \ref{fig:AFM-annealing}b and inset). Such airborne contaminants
resembles those in a recent report\autocite{Temiryazev_2019_ambient},
and their average height is distinguishable from the evaporated Au
nanostructures. Interestingly, after re-annealing the sample under
Ar/H$_{2}$ environment, the contaminants completely removed
(Fig. \ref{fig:AFM-annealing}c). The comparison clearly
indicates thermal annealing of graphene sample prior to the Au
evaporation is the key to obtain clean and non-wettable freestanding
graphene surface, as the hydrocarbon contaminants are known to promote
metal nucleation on graphitic
surfaces\autocite{Wayman_1975_depo_au_gr1,Darby_1975_depo_au_gr2,Zan_2011_au_gr}.
We also found that one Au particle on the freestanding graphene
surface (marked in white circle, Fig. \ref{fig:AFM-annealing}a-c)
had apparent lateral displacement before and after the thermal
treatment. On the contrary, although the shape and height changed
after thermal annealing, Au nanostructures near line defects 
and on SiN$_{x}$/Gr surface showed
negligible movement. We therefore conclude that on
freestanding graphene, Au nanostructures with lateral length up to tens of
nanometers could still have considerable high mobility and low friction.

\begin{figure}[!htbp]
  \centering
  \includegraphics{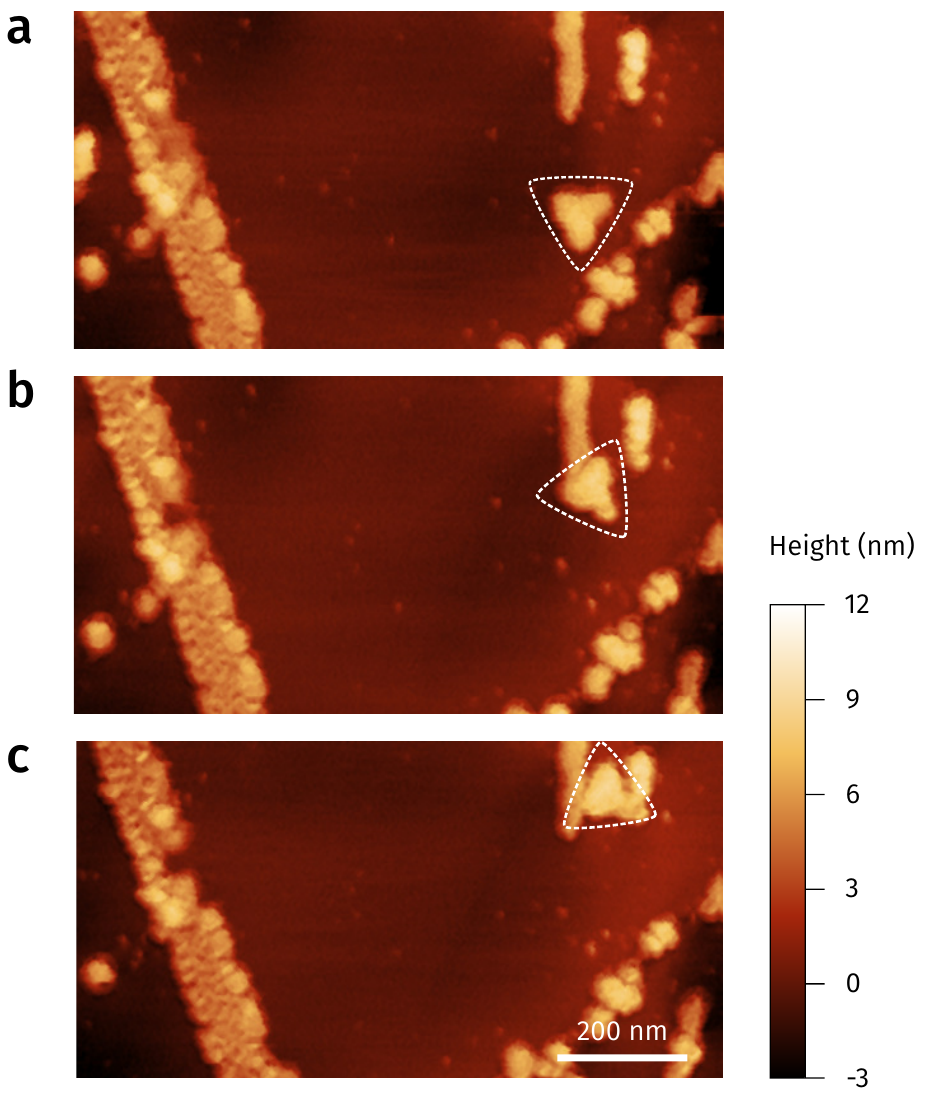}
  \caption{\textbf{Observation of Au nanostructure displacement during
      AFM measurement in ambient}. \textbf{a}, \textbf{b} and \textbf{c} are
    consecutive AFM images of Au nanostructure on freestanding
    graphene surface taken during continuous scanning. A triangular
    nano\-platelet on freestanding graphene appeared to move >200 nm (>2 times the
    lateral size of Au platelet) induced by the motion of AFM tip. On
    the other hand, no apparent change of morphology was observed for
    Au deposited on line defects. The results indicate that the
    friction between freestanding graphene and Au nanostructures with
    lateral size $>50$ nm is still considerably small.}
  \label{fig:ex-site-move-afm}
\end{figure}

The movement of Au  nanostructures could also be induced by external
mechanical force. As shown in Fig. \ref{fig:ex-site-move-afm}, during
continuous AFM scanning in ambient, the triangular Au platelet (marked
in white dashed lines) showed significant lateral displacement ($>2$
times of its size). However, such displacement was not observed for Au
clusters evaporated on line defects or substrate-supported graphene. 
Consider the huge difference
between the lateral size of Au platelet ($>50$ nm) and the apex of AFM
tip (ScanAsys Air, $\sim{}$2 nm), it is unlikely that such
displacement was caused by pickup and re-deposition by the AFM
tip \autocite{Hsieh_2002_AFM_move}.

The tip-induced motion of Au platelets can be explained by the
reduction of equilibrium Gr-Au interaction potential caused by the
many-body repulsive interaction. Following our analysis in section
\ref{sec:attr-inter-substr}, the attractive potential of Vac/Gr/Au
system is significantly reduced compared with the Sub/Gr/Au system
when $d$ is close to the vdW contact distance
(Fig. \ref{fig:layer-energy-full}).  Similarly, on defective area, we
expect the attractive potential to be also large due to the strong
adhesion between defective graphene and
Au\autocite{Zan_2011_au_gr}. As a result, interaction between the AFM
tip and the Au platelet on freestanding graphene can easily overcome
the attractive potential between Au and graphene, leading to
considerable lateral displacement.

\FloatBarrier
\FloatBarrier

\subsection{Discussion about mechanism}
\label{sec:frict-diff-au}

\begin{figure}[!htbp]
  \centering
  \includegraphics{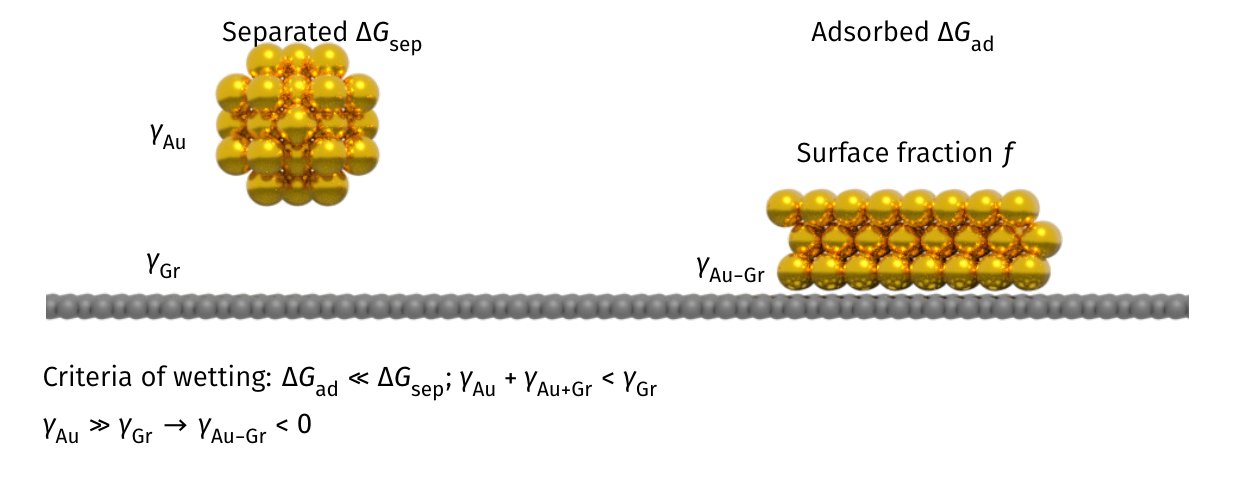}
  \caption{\textbf{Thermodynamic considerations of a 2D platelet formed on
      freestanding graphene}. Wetting of high-surface-energy Au on Gr indicates the
    free energy of adsorbed system $\Delta G_{\mathrm{ad}}$ should be
    smaller than in the separated system $\Delta
    G_{\mathrm{sep}}$, which in turn means the interfacial tension
    $\gamma_{\mathrm{Au-Gr}}$ is negative and adhesion between Au and
    Gr increases. This contradicts to  the experimental observation
    of ultra-low condensation of Au on freestanding graphene.}
\label{fig:hypothesis-2}
\end{figure}

More details about the mechanism of 2D Au platelets formation on
freestanding graphene are discussed as follows. First we rule out the
possibility that larger Au structure formation was due to that Au is
easier to wet freestanding graphene. As shown in
Fig. \ref{fig:hypothesis-2}, if such thermodynamic hypothesis is
correct, it is more energetically favorable for the Au clusters to be
adsorbed on freestanding graphene than the separated system. In other
words, $\Delta G_{\mathrm{ad}} < \Delta G_{\mathrm{sep}}$ is
expected, where $\Delta G_{\mathrm{ad}}$ and $\Delta G_{\mathrm{sep}}$
are the free energy of adsorbed and separated systems,
respectively. Assume the surface area of graphene is
$A_{\mathrm{Gr}}$, the surface area of isolated Au is
$A_{\mathrm{Au}}$, and the fraction of Au-covered graphene surface is $f$,
 we have:
\begin{equation}
  \label{eq:free-energy-Au-Gr}
  \begin{aligned}[t]
    \Delta G_{\mathrm{sep}} &= A_{\mathrm{Gr}} \gamma_{\mathrm{Gr}} + A_{\mathrm{Au}} \gamma_{\mathrm{Au}} + \Delta G_{\mathrm{co}}\\
    \Delta G_{\mathrm{ad}} &= A_{\mathrm{Gr}} (1 - f) \gamma_{\mathrm{Gr}} + A_{\mathrm{Gr}} f \gamma_{\mathrm{Au-Gr}} + A_{\mathrm{Gr}} f \gamma_{\mathrm{Au}} + \Delta G_{\mathrm{co}}
  \end{aligned}
\end{equation}
where $\gamma_{\mathrm{Au}}$ and $\gamma_{\mathrm{Gr}}$ are the
surface energy of Au and freestanding graphene, respectively,
$\gamma_{\mathrm{Au-Gr}}$ is the interfacial energy between Au and
graphene, and $\Delta G_{\mathrm{co}}$ is the free energy of cohesion
(chemical bonding) . Consider the fact
$A_{\mathrm{Gr}} \gg A_{\mathrm{Au}}$, the inequality
$\Delta G_{\mathrm{ad}} < \Delta G_{\mathrm{sep}}$ can be approximated
by
\begin{equation}
  \label{eq:inequality-gamma-Au-Gr}
  \gamma_{\mathrm{Au}} + \gamma_{\mathrm{Au-Gr}} < \gamma_{\mathrm{Gr}}
\end{equation}
Since $\gamma_{\mathrm{Au}} \gg \gamma_{\mathrm{Gr}}$,
$\gamma_{\mathrm{Au-Gr}}$ must be negative. Combine with
$\gamma_{\mathrm{Au-Gr}} = \gamma_{\mathrm{Gr}} + \gamma_{\mathrm{Au}}
- \Delta W_{\mathrm{Au-Gr}}$, where $\Delta W_{\mathrm{Au-Gr}}$ is the
work of adhesion between graphene and Au, $\Delta W_{\mathrm{Au-Gr}}$
should be considerably high to allow the wetting of Au on freestanding
graphene to occur. This is nonphysical since i) attractive vdW
potential is stronger with substrate-supported graphene, following the
discussion in section \ref{sec:attr-inter-substr} and ii) stronger
adhesion leads to higher nucleation density, contradictory to our
experimental findings.

\begin{figure}[!htbp]
  \centering
  \includegraphics[width=0.5\linewidth]{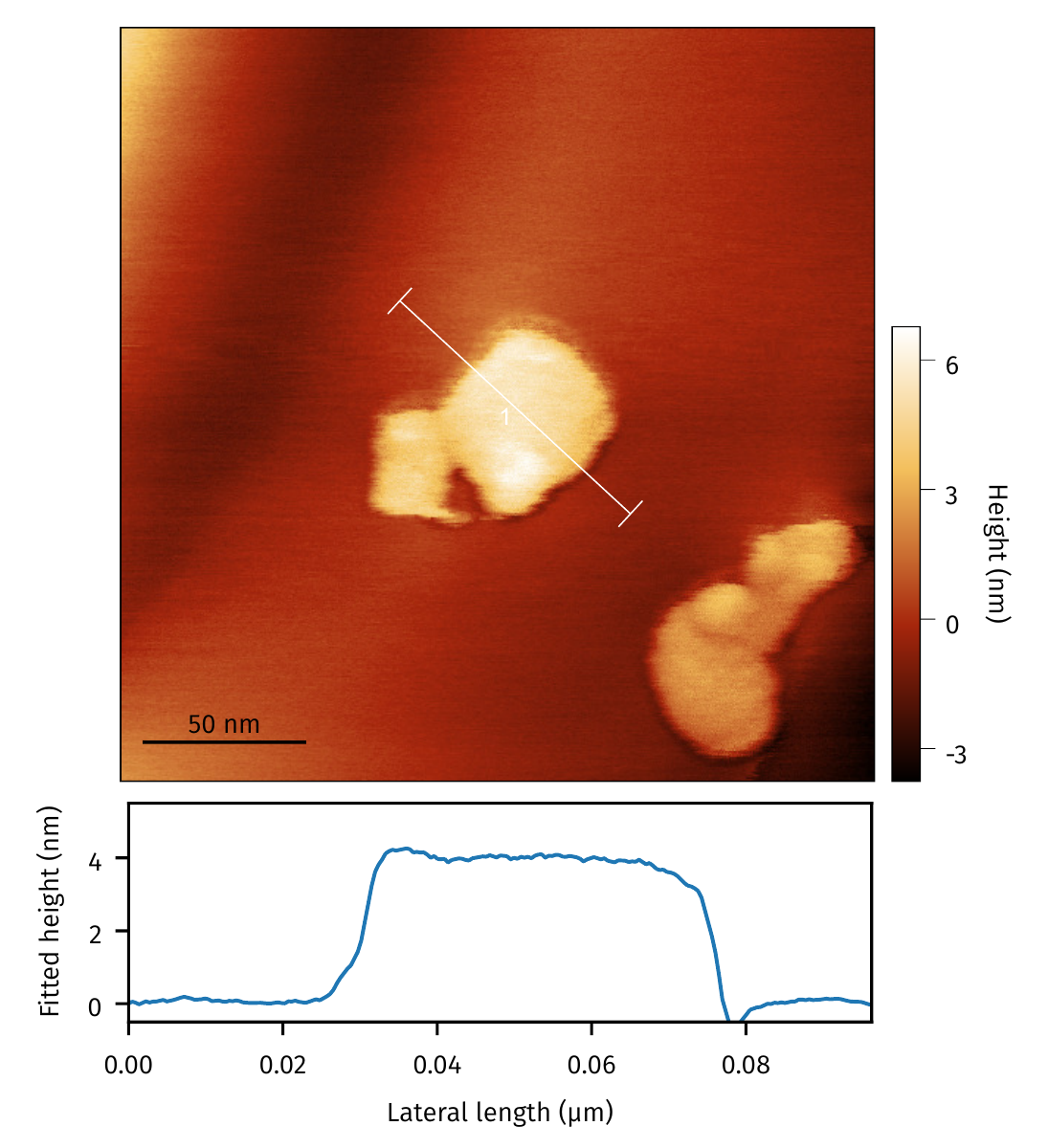}
  \caption{\textbf{Direct AFM height measurement of a typical Au
      platelets on freestanding graphene}. The height profile is
    flattened using polynomial background regression, showing a
    platelet height of 3.90$\pm$0.1 nm.}
\label{fig:afm-direct}
\end{figure}

\begin{figure}[!htbp]
  \centering
  \includegraphics{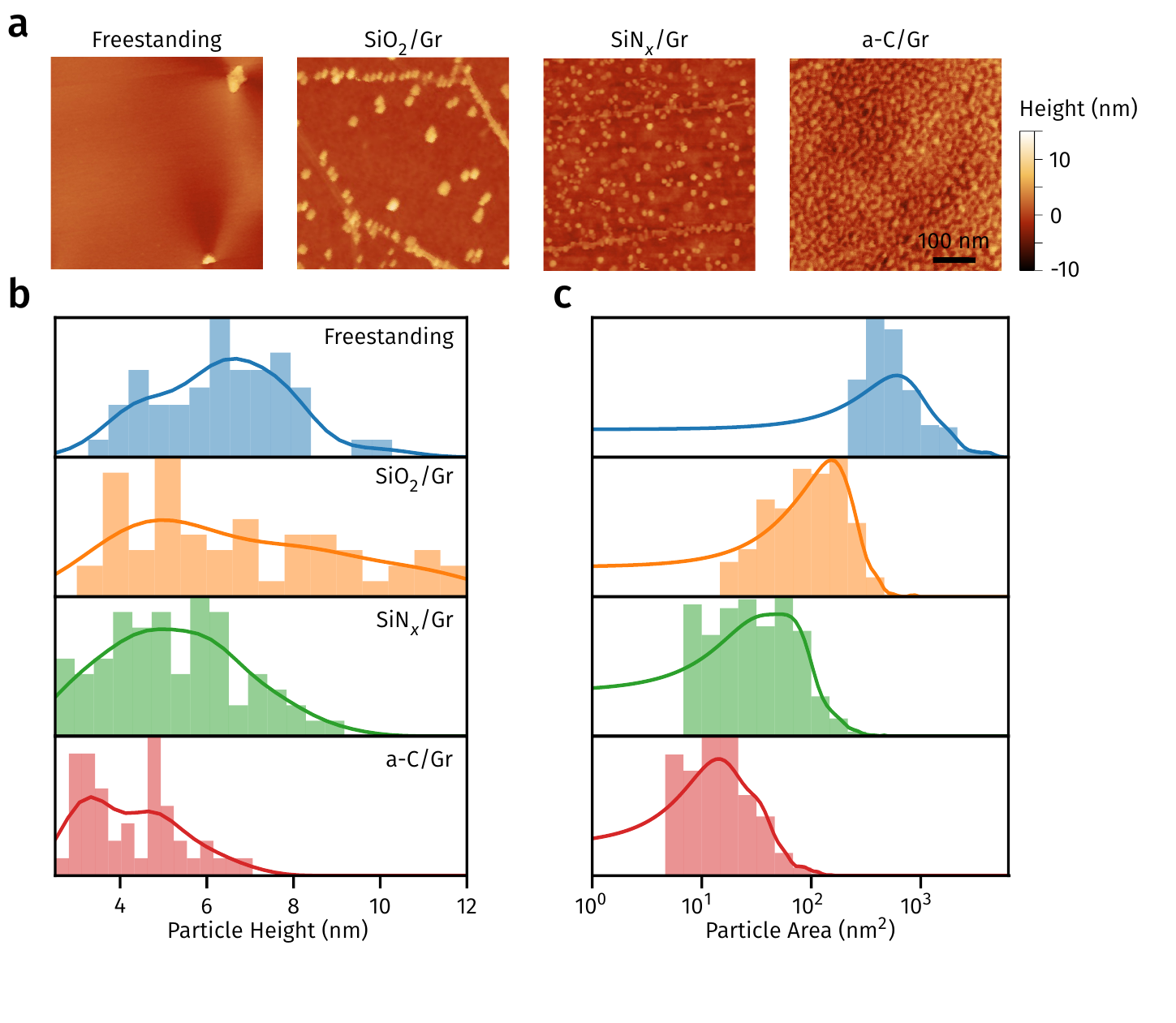}
  \caption{\textbf{Statistics of Au nanostructures grown on
      freestanding or substrate-supported graphene
      surfaces}. \textbf{a}. AFM topography of Au deposited on
    freestanding graphene and graphene supported by SiO$_{2}$,
    SiN$_{x}$ and a-C substrates.  \textbf{b} and \textbf{c}. Height
    and surface area distribution of Au particles on different
    graphene interfaces, respectively. The solid line in \textbf{b}
    and \textbf{c} are the continuous distribution function fitted
    using Gaussian kernel.  Au nanoparticles on freestanding graphene
    are slightly higher while much larger in lateral size compared
    with those on substrate-supported graphene. The statistics were
    based on both AFM and SEM measurements.}
  \label{fig:afm-substrate-au-stat}
\end{figure}

To further study the formation of large 2D platelets on freestanding
graphene, we performed statistic analysis for the Au nanostructures
grown on different graphene surfaces using both AFM topography and SEM
micrograph. As shown in Fig. \ref{fig:afm-substrate-au-stat}, the Au
nucleation density measured by AFM ranks Vac/Gr < SiO$_{2}$/Gr <
SiN$_{x}$/Gr < a-C/Gr, which agrees with our SEM images in
Fig. \ref{main-fig:2}e. Fig. \ref{fig:afm-substrate-au-stat} compares
the  AFM height distribution of Au nanostructures from $>500$
samples points per graphene surface measured. The
fitted distribution curves indicate that the height
of Au nanostructures follows Vac/Gr > SiO$_{2}$/Gr $\approx$ SiN$_{x}$/Gr >
a-C/Gr. In other words, Vac/Gr is actually slightly \textit{less
  wettable} to Au than the substrate-supported graphene surfaces, in
particular a-C/Gr, which agrees with our analysis in
Fig. \ref{fig:layer-energy-full}.
On the other hand, the average
particle area for Vac/Gr/Au increases by $>10^{2}$ times compared with that for
a-C/Gr/Au, as shown in Fig. \ref{fig:afm-substrate-au-stat}c. Notably,
the distribution of particle area for Vac/Gr/Au has an tail towards 
$10^{4}$ nm$^{2}$. In combination with the observation of low
friction interface between Vac/Gr and Au
(Fig. \ref{fig:ex-site-move-afm}), we conclude that the formation of larger
2D crystals on freestanding graphene, can only be ascribed to kinetic
effects.
In fact, such 2D platelet formation may be metastable, as we observed
the height of Au nanostructures on freestanding graphene increased from
$\sim{}6$ nm (as prepared, Fig. \ref{fig:AFM-annealing}a) to
$\sim{}14$ nm after annealing at 600 °C
(Fig. \ref{fig:AFM-annealing}c), further proving that the freestanding
graphene is poorly wettable by Au.

\begin{figure}[!htbp]
  \centering
  \includegraphics{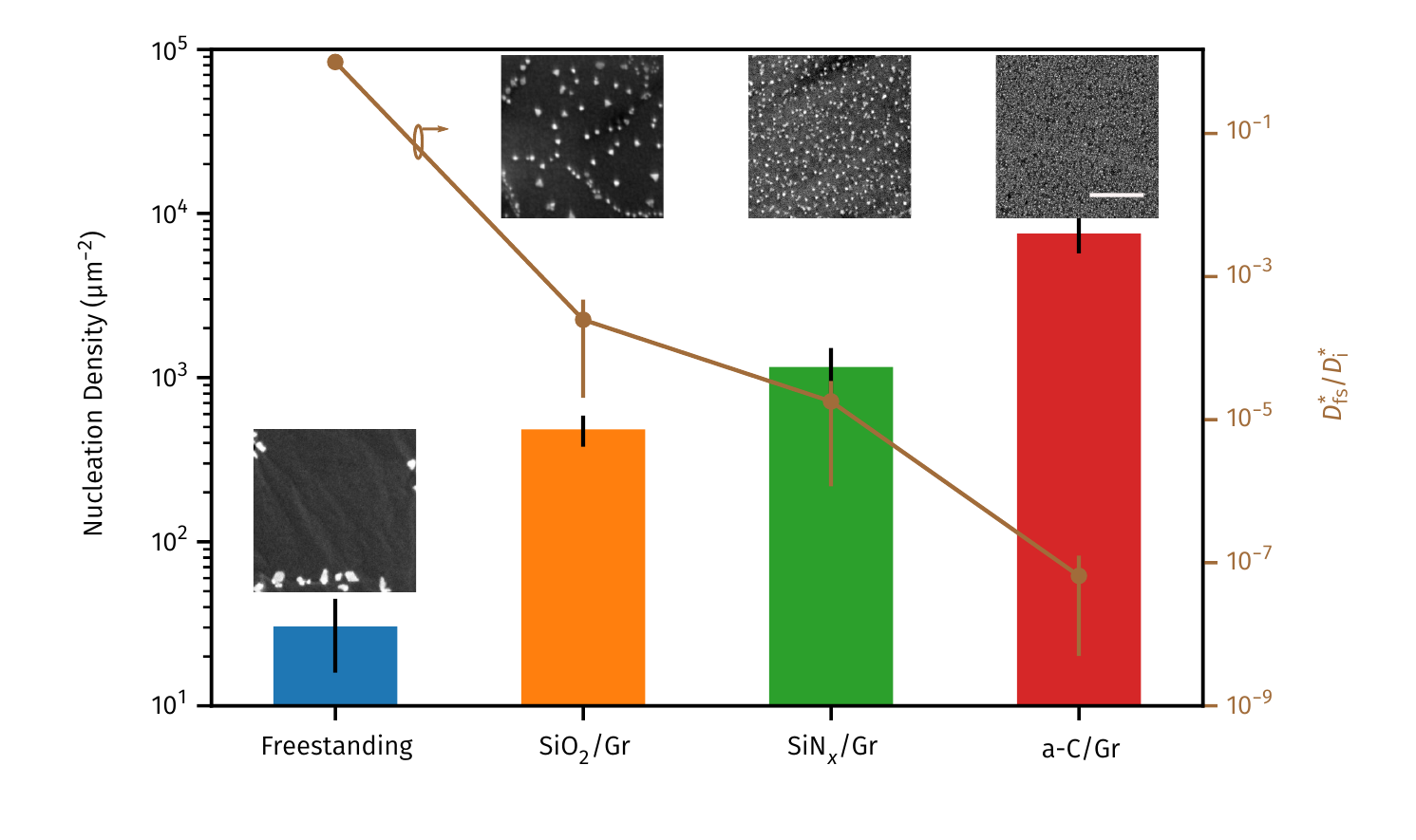}
    \caption{ \textbf{Histograms of nucleation density (left vertical axis) and
    corresponding SEM images of Au morphology on graphene with
    different substrates}. The calculated ratio
    $D^{*}_{\mathrm{fs}} / D^{*}_{\mathrm{i}}$
    between effective surface diffusivities of freestanding graphene
    ($D^{*}_{\mathrm{fs}}$) and graphene supported by substrate i
    ($D^{*}_{\mathrm{i}}$) are shown as the right vertical axis. Scale bars
    of SEM insets: 200 nm.}
\label{fig:support-density}
\end{figure}

To estimate the surface diffusivity $D$ of Au on different graphene
surfaces, we compared the nucleation density $N_{\mathrm{nu}}$
measured by SEM images as shown in Fig. \ref{fig:support-density}
insets. The ratios $D^{*}_{\mathrm{fs}} / D^{*}_{\mathrm{i}}$ between
the effective diffusivities on freestanding graphene
($D^{*}_{\mathrm{fs}}$) and graphene surface i ($D^{*}_{\mathrm{i}}$)
calculated using the relation
$N_{\mathrm{nu}} \propto (D^{*})^{-1/3}$, are also plotted. The
effective diffusivity on freestanding graphene can be up to $10^{9}$
times faster than that on a-C/Gr. Furthermore, although the average
particle area on Vac/Gr/Au is larger than that of substrate-supported
graphene, the low nucleation density leads to less amount of Au
deposited on Vac/Gr surface, as a result of the existence of repulsive
vdW interaction.

\begin{figure}[!htbp]
  \centering
  \includegraphics{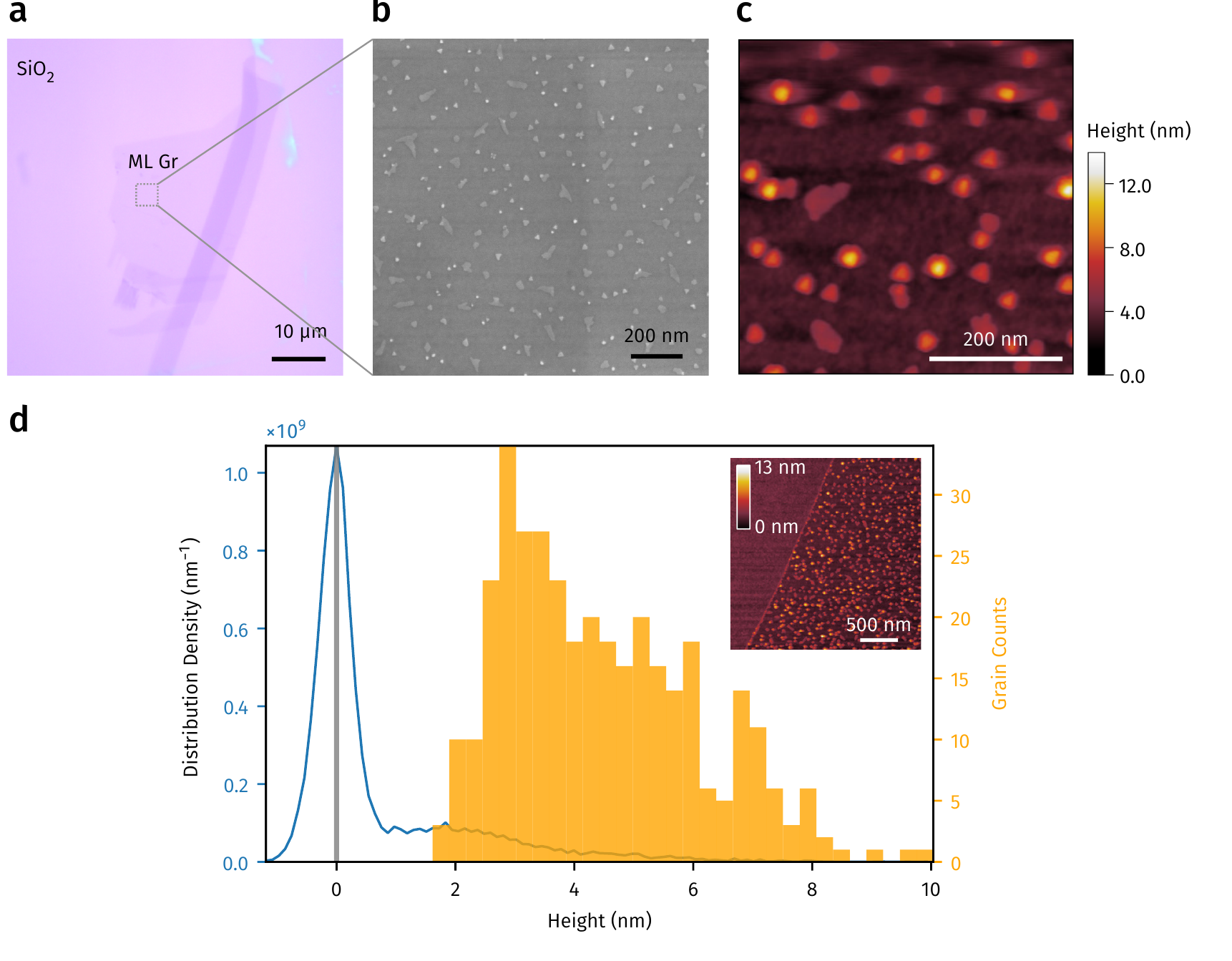}
  \caption{\textbf{Au nanostructures deposited on
      mechanically-exfoliated substrate-supported graphene.}
    \textbf{a} Optical image of a piece of exfoliated graphene
    transferred onto SiO$_{2}$ substrate. Monolayer (ML) regions are
    identified by optical contrast. \textbf{b} and \textbf{c} SEM and
    AFM characterizations of Au nanostructures on freestanding Gr when
    the nominal deposition thickness is 0.1 nm, respectively. Despite
    the clean graphene surface created by mechanical exfoliation,
    nucleation density of Au clusters is still higher than on
    freestanding Gr (Fig. \ref{main-fig:2}D). \textbf{d}. Left axis:
    height distribution density of all pixels in the inset AFM image,
    the reference level of the substrate (0 nm) is taken as the center
    of the first peak. Right axis: histogram of average height per Au
    nano\-platelet in the inset AFM image, showing similar trend with
    Fig. \ref{fig:afm-substrate-au-stat}c.}
\label{fig:sio2-scotch-gr}
\end{figure}

To rule out the possibility that the variation of Au nucleation
density is influenced by the defects of the CVD-grown graphene, we
also measured Au morphology on mechanically-exfoliated
graphene. Monolayer (ML) graphene mechanically exfoliated onto
SiO$_{2}$ substrate was identified by optical micrograph
(Fig. \ref{fig:sio2-scotch-gr}a). As shown in
Fig. \ref{fig:sio2-scotch-gr}b-d, the nucleation density, morphology
and height distributions of Au deposited onto mechanically-exfoliated
graphene are similar to those observed in SiO$_{2}$/Gr/Au samples
prepared using CVD-grown graphene. 

\FloatBarrier

\subsection{KMC simulations of Au diffusion}
\label{sec:kmc-simulations-au}

\begin{figure}[!htbp]
  \centering
  \includegraphics{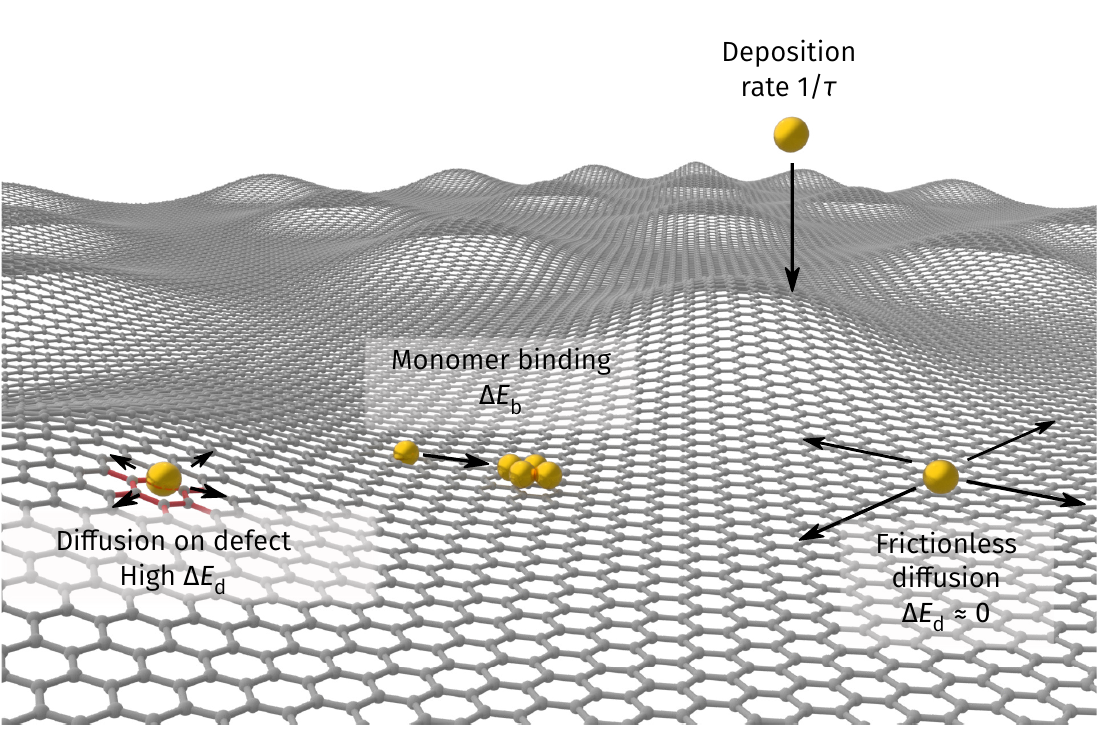}
  \caption{\textbf{Scheme of different kinetic processes for the Au
      deposited on freestanding Gr.} The Au atoms are deposited onto
    the freestanding Gr surface with a rate of $1 / \tau$, where
    $\tau$ is the characteristic time interval of incoming Au
    atoms. On pristine Gr interface, the diffusion energy barrier
    $\Delta E_{\mathrm{d}}$ is nearly negligible, leading to friction-less
    diffusion, while on defective / contaminated area
    $\Delta E_{\mathrm{d}}$ is much higher.  On the other hand, Au
    atoms are bonded to form covalent structures with a energy barrier
    of $\Delta E_{\mathrm{b}}$. The formation of large 2D Au
    structures are enabled when $\Delta E_{\mathrm{d}}$ on pristine
    freestanding Gr is even much lower than $\Delta E_{\mathrm{b}}$.}
\label{fig:kmc-theory-energy}
\end{figure}

\begin{figure}[!htbp]
  \centering
  \includegraphics{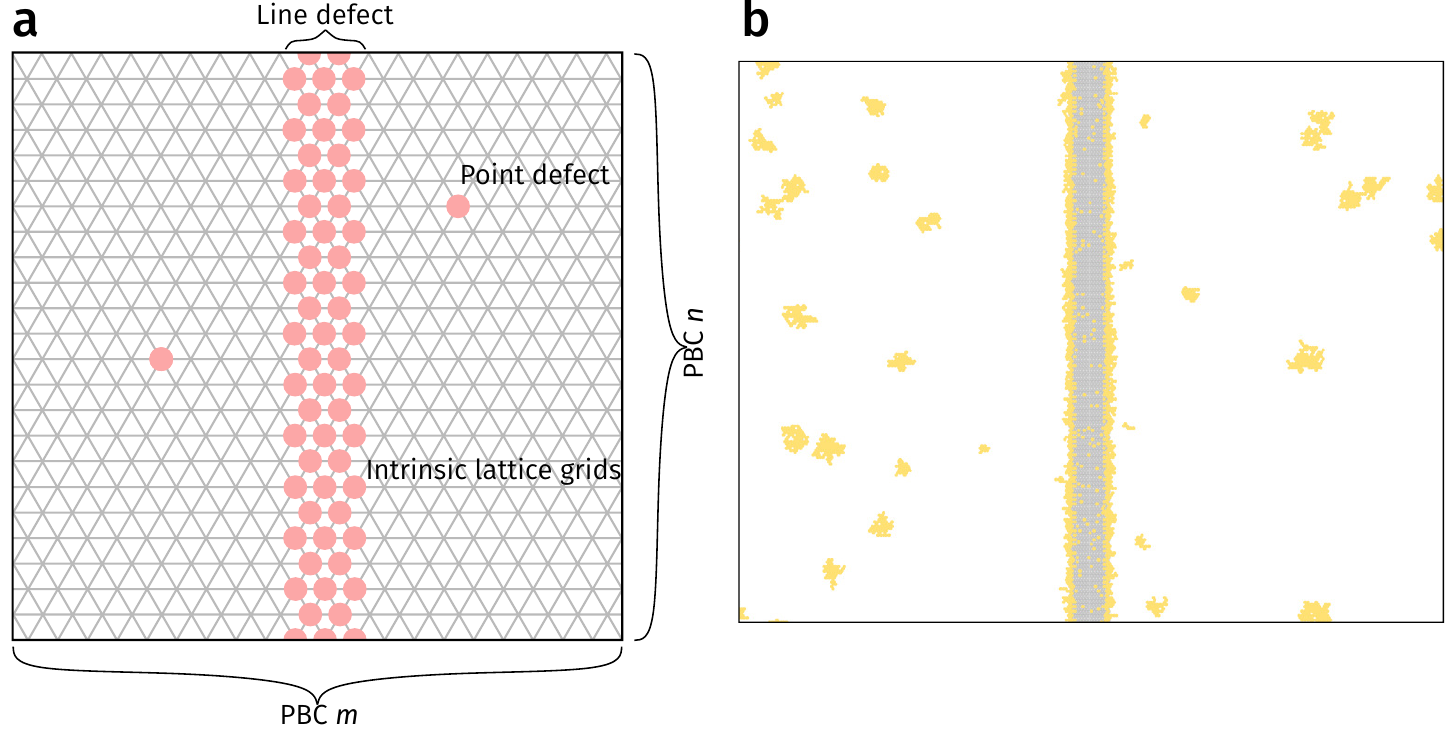}
  \caption{\textbf{Schematic diagram of simulation box used in our 2D
      lattice-KMC}. \textbf{a} The simulation box consists of
    $m\times{}n\times{}2$ triangle grids on which Au atoms are allowed
    to move. Line and point defects (pink dots) are implemented by
    assigning higher $\Delta E_{\mathrm{d}}$ value onto corresponding
    grid points. \textbf{b} Snapshot of a KMC simulation containing
    line defects. Yellow dots represents Au.}
  \label{fig:kmc-mesh}
\end{figure}

The detailed kinetic processes and their corresponding activation energy
barriers in the KMC simulations are schematically shown in
Fig. \ref{fig:kmc-theory-energy}. The KMC simulation box was
implemented on a $m\times{}n\times{}2$ triangular mesh with periodic
boundary conditions (PBC), as shown in Fig. \ref{fig:kmc-mesh}.
To simulate the experimental conditions on the CVD-grown graphene, we
implemented line and point defects on top of the triangular meshes
(Fig. \ref{fig:kmc-mesh}a, pink dots) by assigning higher diffusion
barrier $\Delta E_{\mathrm{d}}^{*}$. Practically, when simulating very fast
Au diffusion on freestanding graphene, such defective regions act as
preferential nucleation sites to promote growth of larger Au
platelets. A typical snapshot of the KMC simulation can be seen in
Fig. \ref{fig:kmc-mesh}b.

The Au evaporation process is modeled as follows. The rate of
deposition is defined as $\nu = 1/ \tau$, where $\tau$ is the characteristic
time scale between two deposition events, which follows:
\begin{equation}
  \label{eq:def-graphene-depo-nu}
  \nu = J_{\mathrm{e}} N_{\mathrm{site}} = \nu_{0} \exp(-\frac{\Delta E_{\mathrm{e}}}{k_{\mathrm{B}} T})
\end{equation}
where $J_{\mathrm{e}}$ is the flux of evaporation (measured in
ML·s$^{-1}$), $N_{\mathrm{site}}$ is the density of lattice sites in the
KMC simulation, $\nu_{0}$ is the rate pre\-factor, and
$\Delta E_{\mathrm{e}}$ is the effective energy barrier for the
evaporation process. For $J_{\mathrm{e}}=0.01$ ML·s$^{-1}$, $m=n=100$
and $\nu_{0}=10^{10}$ s$^{-1}$, we have
$\Delta E_{\mathrm{e}} \approx 15 k_{\mathrm{B}}T$.  We also note in
the current KMC framework, the difference between freestanding and
substrate-supported graphene is reflected by the choice of
$\Delta E_{\mathrm{d}}^{0}$, and surface corrugation is not taken into
account. 

There are several simplifications used in the KMC simulations.
First, we do not specify whether the Au on each mesh grid
is a single atom or a cluster. As a result, the binding between two Au units may be slower than that of surface diffusion.
To simplify the simulation, the Au units are not allowed to move after coalescence occurs.
Moreover, since in this
study we do not care about absolute time scale of cluster formation,
the exact value for $\nu_{0}$ does not affect the final result.

The morphology of Au nanostructures are majorly influenced by the
following factors: i) magnitude of $\Delta E_{\mathrm{e}}$ and ii) competition 
between $\Delta E_{\mathrm{d}}$ and $\Delta E_{\mathrm{b}}$. For all
simulations we fix the diffusion barrier on defects
$\Delta E_{\mathrm{d}}^{*}$ to 15 $k_{\mathrm{B}}T$.

\paragraph{Influence of $\Delta E_{\mathrm{e}}$}

\begin{figure}[!htbp]
  \centering
  \includegraphics{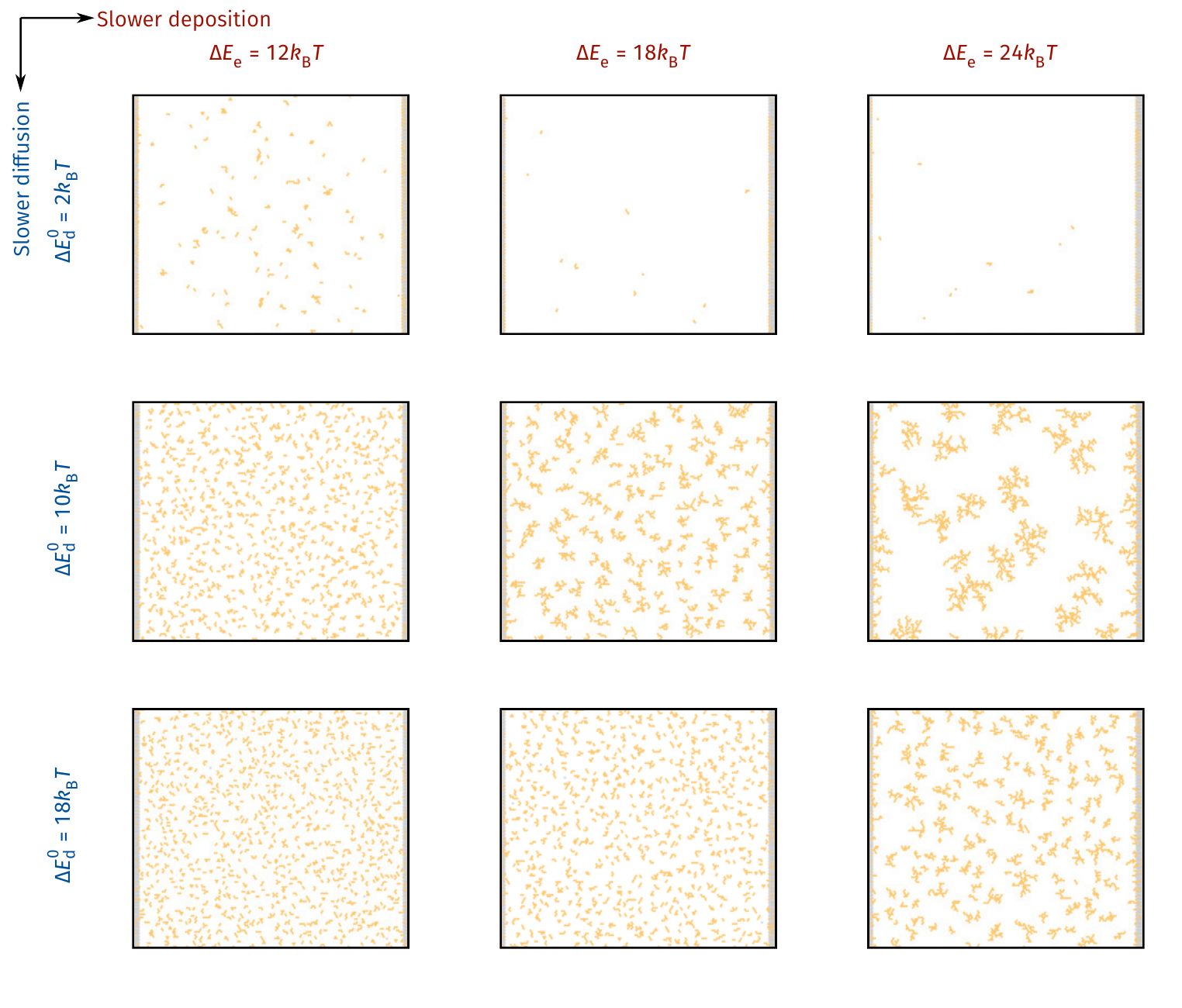}
  \caption{\textbf{Snapshots of Au nanostructure formation at $10^{6}$
      time steps by varying $\Delta E_{\mathrm{e}}$ and
      $\Delta E_{\mathrm{d}}^{0}$}.  Independent of
    $\Delta E_{\mathrm{d}}^{0}$, decreasing $\Delta E_{\mathrm{e}}$
    always leads to higher nucleation density. The value of
    $\Delta E_{\mathrm{b}}$ is $10 k_{\mathrm{B}} T$ for all
    simulations. The line defects are plot at the edge of the
    simulation box for illustration purpose.  Note for very fast
    diffusion cases, the time scale of individual cluster to grow
    appears to be longer than 10$^6$ steps, leading to smaller cluster
    size.  }
  \label{fig:kmc-effec-Ee}
\end{figure}

Fig. \ref{fig:kmc-effec-Ee} compares the different snapshots of Au nanostructures
by varying $\Delta E_{\mathrm{e}}$ (horizontal direction) and
$\Delta E_{\mathrm{d}}^{0}$ (vertical direction). Independent of the
in-plane diffusion rate, by lowering $\Delta E_{\mathrm{e}}$ (faster
deposition), the nucleation density of Au nanostructures always
increases. This is because when increasing the evaporation rate,
 there are more mobile
Au on the graphene surface at the same time, leading to higher
rate of coalescence. In the experiment of Au evaporation, it is
therefore always favorable to use lower deposition rate for the
formation of sparse and large clusters on freestanding graphene.

\paragraph{Influence of $\Delta E_{\mathrm{d}}^{0}$ and $\Delta E_{\mathrm{b}}$}

\begin{figure}[!htbp]
  \centering
  \includegraphics{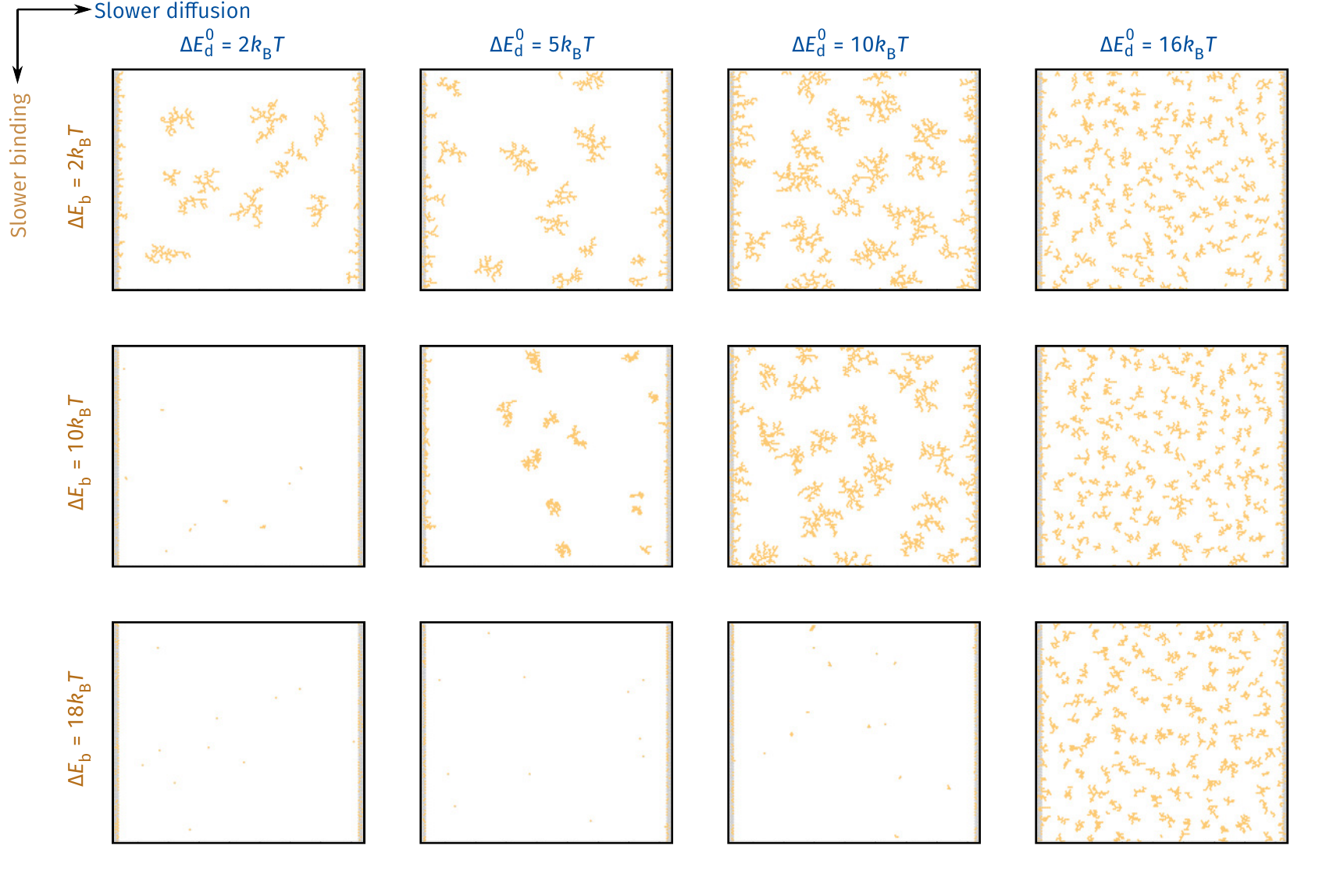}
  \caption{\textbf{Snapshots of Au nanostructure formation at $10^{6}$
      time steps by varying $\Delta E_{\mathrm{b}}$ and
      $\Delta E_{\mathrm{d}}^{0}$}. Different
    $\Delta E_{\mathrm{b}}^{0} / \Delta E_{\mathrm{d}}$ ratios lead to
    distinct morphology and nucleation density of Au nanostructures.
    The value of $\Delta E_{\mathrm{e}}$ is $20 k_{\mathrm{B}} T$ for
    all simulations. The line defects are plot at the edge of the
    simulation box for illustration purpose. Note for very fast
    diffusion cases, the time scale of individual cluster to grow
    appears to be longer than 10$^6$ steps, leading to smaller cluster
    size.}
  \label{fig:kmc-effec-EbEd}
\end{figure}

The influence of $\Delta E_{\mathrm{d}}^{0}$ and
$\Delta E_{\mathrm{b}}$ on the Au nanostructure formation is more
complex. As shown in Fig. \ref{fig:kmc-effec-EbEd}, with a lower
$\Delta E_{\mathrm{d}}^{0} / \Delta E_{\mathrm{b}}$ ratio, the
nucleation density of Au nanostructures decreases. On the other hand,
we also observe that at same
$\Delta E_{\mathrm{d}}^{0} / \Delta E_{\mathrm{b}}$ ratio, increasing
$\Delta E_{\mathrm{d}}^{0}$ leads to formation of higher degree of
nucleation. In both scenarios, the increasing of nucleation density
can be explained by shorter time scale for Au coalescence to occur.

\begin{figure}[!htbp]
  \centering
  \includegraphics{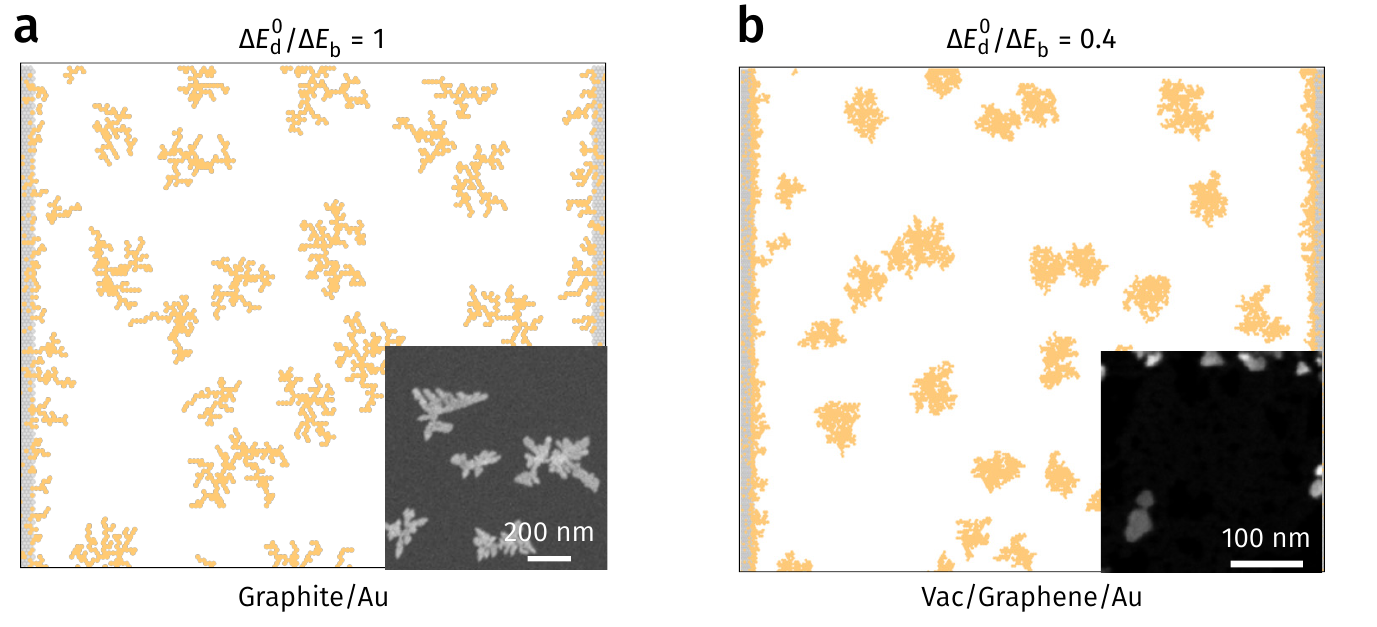}
  \caption{\textbf{Simulated morphology of Au nanostructures on
      bulk graphite a. and freestanding graphene (B) surfaces}. The
    dendritic patterns on graphite are formed as a result of the high
    $\Delta E_{\mathrm{d}}^{0} / \Delta E_{\mathrm{b}}$ ratio, while
    hexagonal and triangular structures are observed on freestanding
    graphene due to lower
    $\Delta E_{\mathrm{d}}^{0} / \Delta E_{\mathrm{b}}$. Experimental
    SEM images for Graphite/Au and Vac/Graphene/Au fabricated from Au
    evaporation at room temperature are shown as insets of \textbf{a}
    and \textbf{b}, respectively.}
  \label{fig:kmc-graphite}
\end{figure}

The $\Delta E_{\mathrm{d}}^{0}/\Delta E_{\mathrm{b}}$ ratio also
influences the morphology of individual Au nanostructure. A general
trend observed in Fig. \ref{fig:kmc-effec-EbEd} is that for higher
$\Delta E_{\mathrm{d}}^{0}/\Delta E_{\mathrm{b}}$ value, the Au
nanostructure tends to dendritic, while for lower
$\Delta E_{\mathrm{d}}^{0}/\Delta E_{\mathrm{b}}$ value the Au
nanostructures becomes more compact. This can be explained by the fact
that when the rate of binding and diffusion are comparable, the
structure formed is not fully relaxed. We use this feature to explain
the distinct Au morphology observed on bulk graphite and freestanding
graphene surfaces, as shown in Fig. \ref{fig:kmc-graphite}. Although
the nucleation density of Graphite/Au (Fig. \ref{fig:kmc-graphite}a)
at room temperature is comparable with that of Vac/Graphene/au
(Fig. \ref{fig:kmc-graphite}b), the dendritic pattern formation on
Graphite/Au indicates the in-plane diffusion of Au is faster on
freestanding graphene, given the fact that the Au-Au coalescence rate
is almost independent on the substrate
\autocite{Lewis_2000_Au_diff,Bardotti_1996_SbAu}.  On the other hand,
previous studies showed that hexagonal/triangular Au nanostructures on
graphite could only be formed during high-temperature
deposition\autocite{Wayman_1975_depo_au_gr1,Darby_1975_depo_au_gr2,Cihan_2015_Au_graphite},
indicating that $\Delta E_{\mathrm{d}}^{0}$ on freestanding graphene
is lower than of bulk graphite.  Taking the surface corrugation of
freestanding graphene into account, the fast in-plane Au diffusion on
freestanding graphene beyond the structural super\-lubricity, can only
be explained by the existence of many\-body repulsive vdW
interactions.

\FloatBarrier

\section{Molecular epitaxy of BPE molecules}
\label{sec:molec-epit-bpe}

\subsection{Practical considerations}

\begin{figure}[!htbp]
  \centering
  \includegraphics{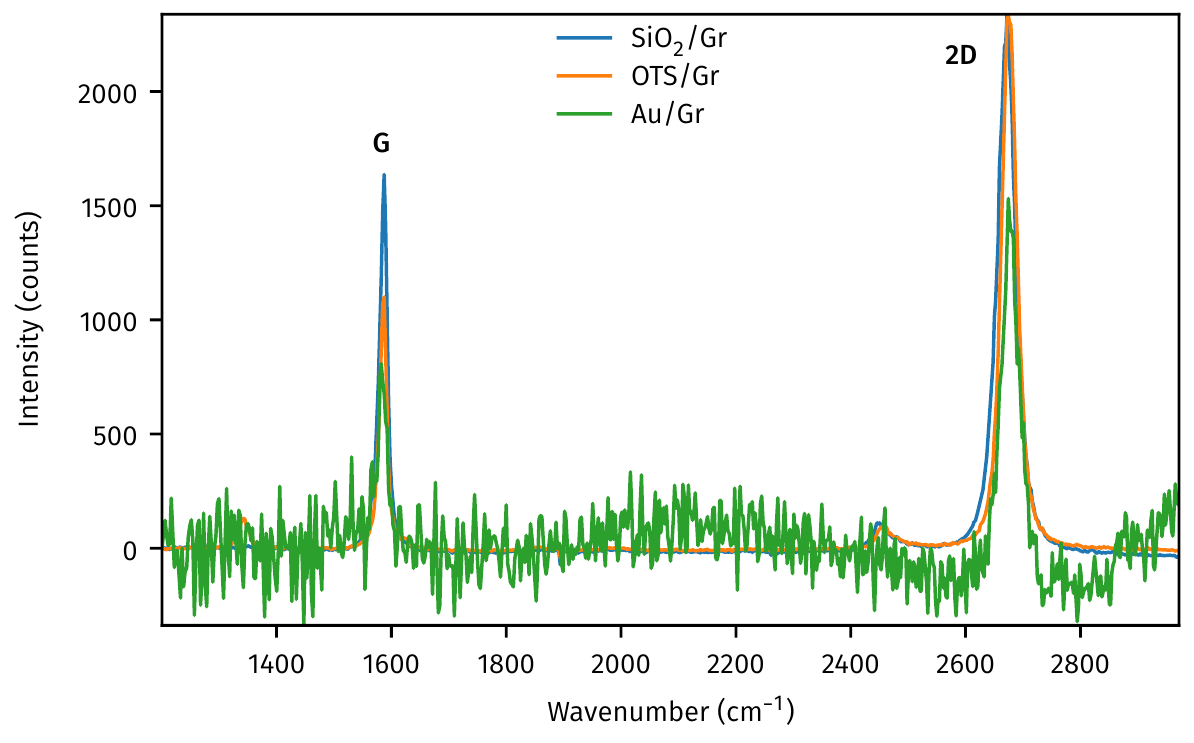}
  \caption{\textbf{Typical Raman spectra of monolayer graphene transferred
      onto as-cleaned SiO$_{2}$ (SiO$_{2}$/Gr), OTS-treated SiO$_{2}$
      (OTS/Gr) and gold (Au/Gr) substrates.
    }
Raman spectra were taken using 532 nm laser. In all cases the 2D
    peak remains higher than G peak indicating monolayer graphene. The
    positions of 2D and G peaks do not have apparent shift, similar
    doping densities in these substrates. }
\label{fig:raman-avg}
\end{figure}

Molecular epitaxy on graphene surface is known to be influenced by the
doping state of graphene \autocite{Tian_2017_doping}. To rule out the
possibility that observed polymorphism of BPE was due to doping, Raman
spectroscopy was used to monitor the charge density of graphene
transferred on as-cleaned SiO$_{2}$ (SiO$_{2}$/Gr), self-assembled
octadecyl\-trichloro\-silane (OTS) on SiO$_{2}$ (OTS/Gr) and gold
(Au/Gr).
The OTS-SiO$_{2}$ substrate was used as a reference for suppressing
potential substrate-induced doping\autocite{Yan_2011_gr}.
As shown in Fig. \ref{fig:raman-avg}, on all substrates the
2D peak is higher than the G peak, and D peak intensity is
negligible. As summarized in Table \ref{tab:raman-sub}, the 2D and G
peak positions of SiO$_{2}$/Gr and Au/Gr samples are statistically
similar.
From literature\autocite{Das_2008_gr_doping}, the positions of the
G-peak for SiO$_{2}$/Gr and Au/Gr indicates both samples have doping
density less than $2\times{}10^{12}$ \textit{e}·cm$^{-2}$.

\begin{table}[h!]
  \centering
  \caption{\textbf{Peak position and full width at half-maximum (FWHM)
      of G and 2D peaks in the Raman spectra in
      Fig. \ref{fig:raman-avg}.}  Statistical average and deviation
    are taken from over 80 data points. The peak parameters are fitted
    using Gaussian function.  }
  \begin{tabular}[!htbp]{lrrrr}
    \hline{}
    Substrate & Position G (cm$^{-1}$) & FWHM G (cm$^{-1}$) & Position 2D (cm$^{-1}$) & FWHM 2D (cm$^{-1}$) \\
    \hline{}
    SiO$_{2}$/Gr & $1585.69\pm2.69$ & $8.33\pm2.30$ & $2675.62\pm4.15$ & $17.75\pm2.52$ \\
    OTS/Gr & $1584.76\pm1.07$ & $8.07\pm1.48$ & $2676.42\pm2.57$ & $16.05\pm3.16$ \\
    Au/Gr & $1582.50\pm2.48$ & $10.31\pm2.93$ & $2676.25\pm3.43$ & $18.80\pm2.05$ \\
    \hline{}
\end{tabular}
   \label{tab:raman-sub}
\end{table}
\FloatBarrier

\subsection{Determining crystallographic parameters}

\begin{figure}[!htbp]
  \centering
  \includegraphics{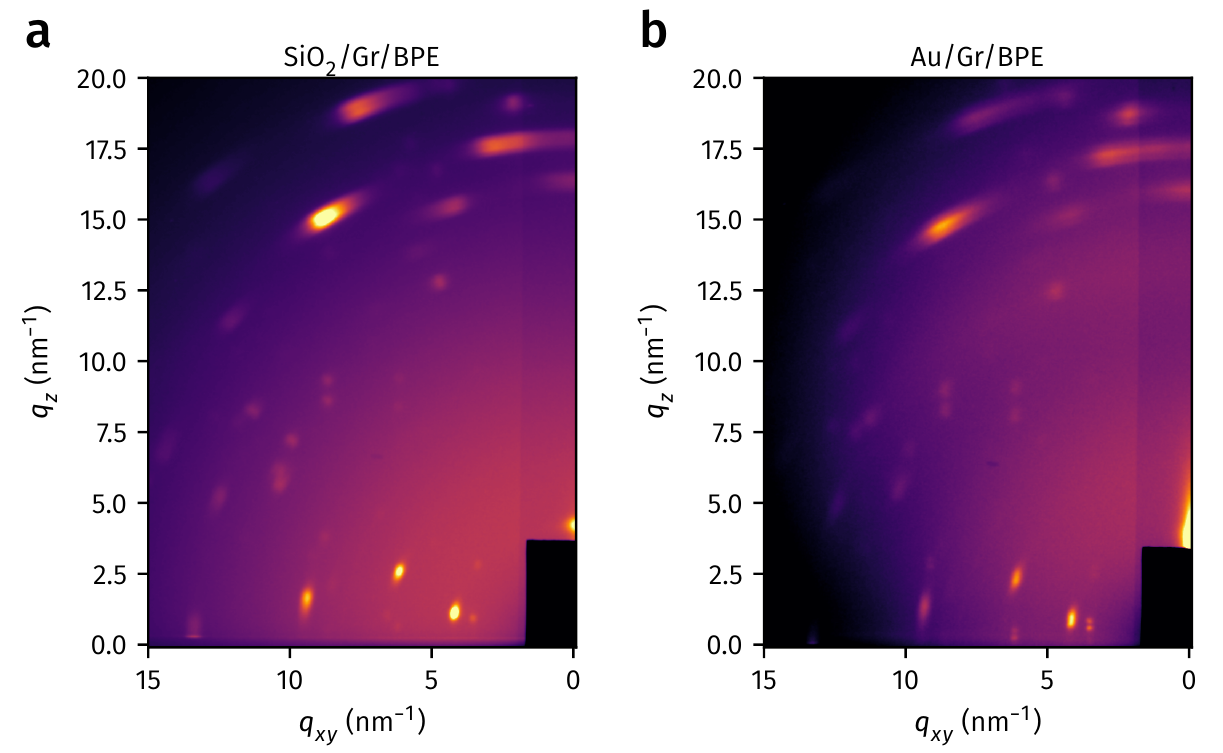}
  \caption{\textbf{Full-range GIWAXS spectra of SiO$_{2}$/Gr/BPE a. and
      Au/Gr/BPE (B) systems corresponding to Fig. \ref{main-fig:3}c}
}
\label{fig:gixd-full-scale}
\end{figure}

The processes for determining the crystallographic parameters (crystal
lattice constant and interplanar distances) are described as follows.
The full-range GIWAXS spectra for SiO$_{2}$/Gr/BPE and Au/Gr/BPE are
shown in Fig. \ref{fig:gixd-full-scale}. A rectangular beam stop was
used to reduced the scattering from the substrate.  The 1D diffraction
profile was generated by first subtracting the
background from the 2D GIWAXS spectra using rolling-ball algorithm\autocite{Schindelin_2012_FIJI}, and calculated using:
\begin{equation}
  \label{eq:1D-extraction}
  I_{\mathrm{1D}}(q) = \int_{0}^{\pi/2} I_{\mathrm{2D}}(q, \theta) q \mathrm{d}q
\end{equation}
where $I_{\mathrm{2D}}$ and $I_{\mathrm{1D}}$ are the 2D and 1D X-ray
diffraction intensity, $q = | \mathbf{q} |$ is magnitude of the
diffraction wave vector, and
$\theta = \arctan \Bigg|\frac{q_{z}}{q_{xy}}\Bigg|$.

\begin{figure}[!htbp]
  \centering
  \includegraphics[width=0.8\linewidth]{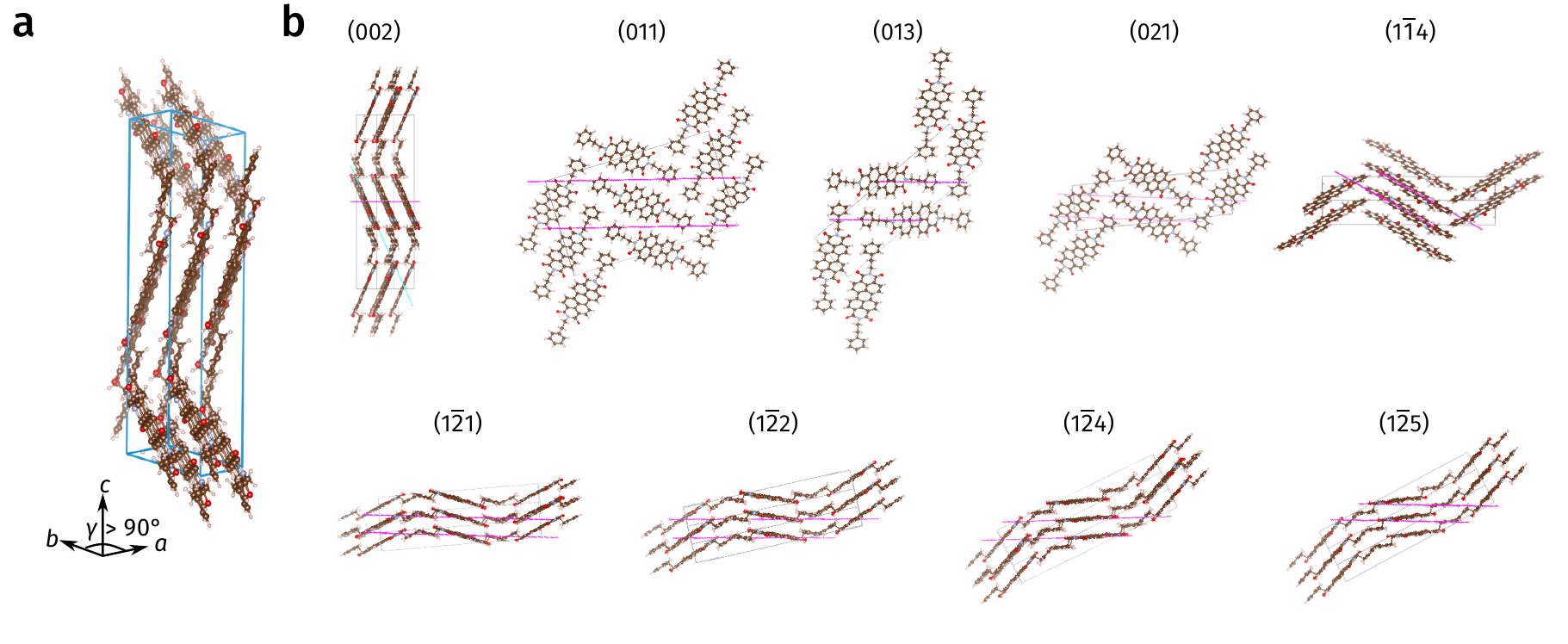}
  \caption{\textbf{Lattice structure of BPE.} \textbf{a}. Schematic of
    BPE unit cell with monoclinic symmetry. \textbf{b}. Typical
    lattice planes in BPE single crystal with corresponding $(hkl)$
    indices.}
  \label{fig:bpe-lattice}
\end{figure}

\begin{figure}[!htbp]
  \centering
  \includegraphics{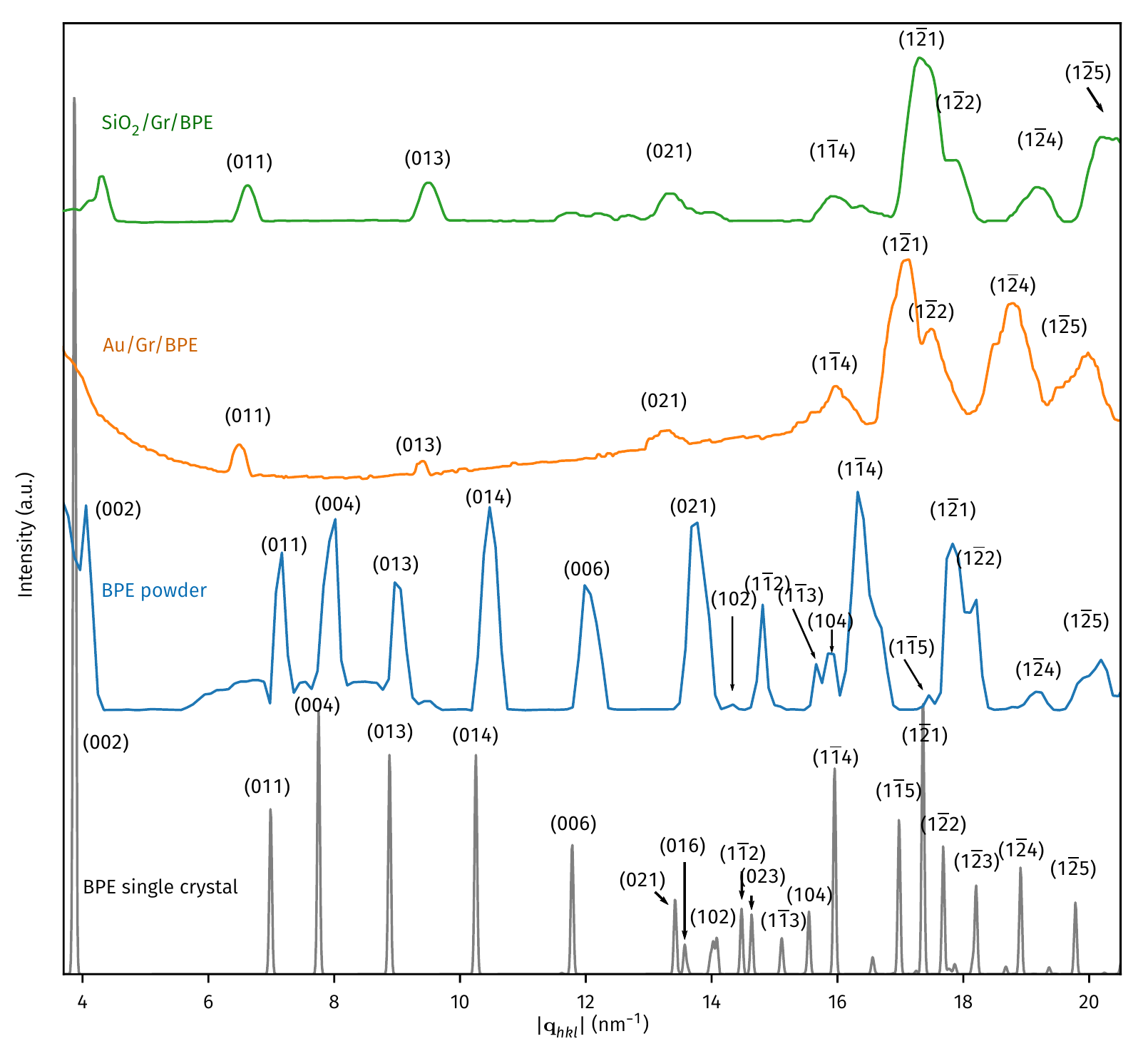}
  \caption{\textbf{1D X-ray diffraction patterns of BPE single
      crystal, BPE powder, SiO$_{2}$/Gr/BPE and Au/Gr/BPE}. The
    best-fitted $(hkl)$ indices are labeled. Diffraction of BPE single
    crystal is simulated using the RIETAN-FP package within the VESTA
    software\autocite{Momma_2008_vesta}.}
  \label{fig:bpe-1d-diffraction}
\end{figure}

Single crystal data of BPE molecule from
literature\autocite{Mizuguchi_1998_BPE} (CCDC DICNIM01) were
used as a reference for assigning the Miller indices for the lattice
planes. As shown in Fig. \ref{fig:bpe-lattice}a, the unit cell of BPE
single crystal is monoclinic with $\gamma \neq 90°$. The major
diffraction planes are shown in Fig. \ref{fig:bpe-lattice}b. The
high-$q$ (short interplanar distance) diffraction planes
$(1 \overline{1} 4)$, $(1 \overline{2} 1)$, $(1 \overline{2} 2)$,
$(1 \overline{2} 4)$ and $(1 \overline{2} 5)$ are associated with the
PTCDI basal plane of BPE molecule. The simulated 1D X-ray diffraction
profile from BPE single crystal data is shown in
Fig. \ref{fig:bpe-1d-diffraction} (black line).  In comparison,
Fig. \ref{fig:bpe-1d-diffraction} also show the experimental 1D X-ray
diffraction profiles for sublimed BPE powder (blue line),
SiO$_{2}$/Gr/BPE (orange line) and Au/Gr/BPE (green line).
The diffraction peaks in the experimental 1D X-ray diffraction
profiles were assigned as follows.
Assuming the BPE molecules in the experimental samples remained the
monoclinic lattice structure, we first guessed the Miller indices for
the most prominent peaks in each 1D profile using the reference data
from BPE single crystal. The lattice constants $a, b, c, \gamma$ were
then calculated using least-square fitting based on the equation of
interplanar spacing in a monoclinic unit
cell\autocite{Andrews_1967_EM_Book}:
\begin{equation}
  \label{eq:dhkl}
  \frac{1}{d^{2}_{hkl}}
  = \frac{1}{\sin^{2} \gamma}\left[
    \frac{h^{2}}{a^{1}} + \frac{k^{2}}{b^{2}} + \frac{l^{2} \sin^{2}\gamma}{c^{2}}
    - \frac{2h k \cos \gamma}{a b}
  \right]
\end{equation}
where $d_{\mathrm{hkl}} = 2 \pi (q_{hkl})^{-1}$ is the interplanar
distance.  The assignment of Miller indices was refined iteratively
until the mismatches between experimental and the fitted values
$d_{hkl}$ were minimized. The peak assignment for BPE powder,
SiO$_{2}$/Gr/BPE and Au/Gr/BPE are labeled in
Fig. \ref{fig:bpe-1d-diffraction}, with the fitted lattice constants
and $q_{hkl}$ values listed in Tables \ref{tab:bpe-best-fit-lattice}
and \ref{tab:bpe-fit-planes}, respectively.
Comparing with BPE powder, SiO$_{2}$/Gr/BPE and Au/Gr/BPE show
expansion in $b$-axis and shrinking in $c$-axis, possibly due to the
templating effect of graphene\autocite{Tian_2017_doping}. We find that
the best-fitted lattice constants $b$ and $c$ for Au/Gr/BPE has about
1\% expansion compared with that for SiO$_{2}$/Gr/BPE, corresponding
to the increasing of interplanar distance between the PTCDI basal
planes in Au/Gr/BPE.

\begin{table}[!htbp]
  \centering
  \caption{\textbf{Best fitted Lattice constants for BPE single
      crystal\autocite{Mizuguchi_1998_BPE},
      BPE powder sample, SiO$_{2}$/Gr/BPE and Au/Gr/BPE using monoclinic lattice model}.}
  \begin{tabular}[!htbp]{lllll}
  \hline
  \ & $a$ (Å) & $b$ (Å) & $c$ (Å) & $\gamma$ ($^{\circ}$) \\
  \hline
  BPE single crystal\autocite{Mizuguchi_1998_BPE} & 4.73 & 9.51 & 32.45 & 100.27 \\
  BPE powder & 4.64$\pm$0.03 & 9.34$\pm$0.04 & 31.09$\pm$0.08 & 99.71$\pm$0.84 \\
  SiO$_{2}$/Gr/BPE & 5.01$\pm$0.08 & 9.56$\pm$0.03 & 29.94$\pm$0.16 & 99.76$\pm$0.85 \\
  Au/Gr/BPE & 4.87$\pm$0.09 & 9.67$\pm$0.04 & 30.29$\pm$0.27 & 100.28$\pm$1.03 \\
  \hline
\end{tabular}
   \label{tab:bpe-best-fit-lattice}
\end{table}

\begin{table}[!htbp]
  \centering
  \caption{\textbf{Best fitted $|\mathbf{q}_{hkl}|$ values of major Bragg
      diffraction peaks $(hkl)$ for BPE single
      crystal\autocite{Mizuguchi_1998_BPE}, BPE powder,
      SiO$_{2}$/Gr/BPE and Au/Gr/BPE using monoclinic lattice
      model}. Blank fields indicate the corresponding diffraction
    peaks are weak in the sample.}
  ﻿\begin{tabular}[!htbp]{lrrrr}
  \hline
  \ & \multicolumn{4}{c}{$|\mathbf{q}_{hkl}|$ (nm$^{-1}$)} \\
  $(hkl)$ & BPE single crystal\autocite{Mizuguchi_1998_BPE} & BPE powder & SiO$_2$/Gr/BPE & Au/Gr/BPE \\
  \hline
  $(0 0 2)$ & 3.87 & 4.04 &  & \\
  $(0 1 1)$ & 6.99 & 7.12 & 7.26 & 7.00 \\
  $(0 0  4)$ & 7.75 & 8.09 &  & \\
  $(0 1  3)$ & 8.88 & 9.13 & 9.38 & 9.13 \\
  $(0 1  4)$ & 10.25 & 10.58 &  & \\
  $(0 0  6)$ & 11.62 & 12.13 &  & \\
  $(0 2  1)$ & 13.57 & 13.80 & 13.61 & 13.37 \\
  $(1 0  2)$ & 14.03 & 14.32 &  & \\
  $(1 \overline{1}  2)$ & 14.48 & 14.83 &  & \\
  $(1 \overline{1}  3)$ & 15.11 & 15.51 &  & \\
  $(1 0  4)$ & 15.55 & 15.94 &  & \\
  $(1 \overline{1}  4)$ & 15.95 & 16.40 & 16.02 & 15.99 \\
  $(1 \overline{1}  5)$ & 16.98 & 17.49 &  & \\
  $(1 \overline{2}  1)$ & 17.36 & 17.77 & 17.47 & 17.14 \\
  $(1 \overline{2}  2)$ & 17.68 & 18.11 & 17.84 & 17.51 \\
  $(1 \overline{2}  4)$ & 18.91 & 19.42 & 19.27 & 18.82 \\
  $(1 \overline{2} 5)$ & 19.78 & 20.34 & 20.27 & 19.93 \\
  \hline
\end{tabular}    \label{tab:bpe-fit-planes}
\end{table}

\subsection{Breakdown of wetting transparency theory}
\label{sec:breakd-wett-transp}

\begin{figure}[!htbp]
  \centering
  \includegraphics{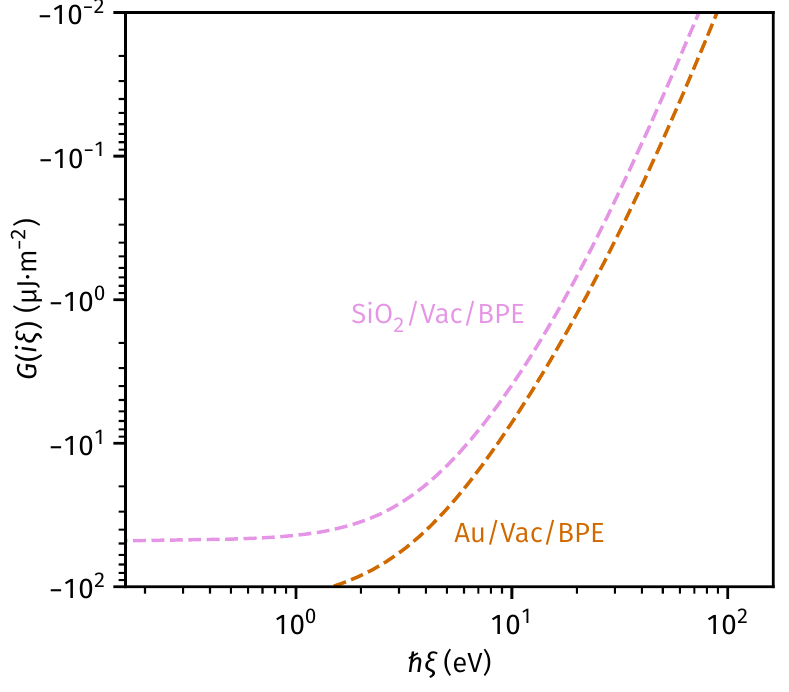}
  \caption{\textbf{Single-frequency interaction energy $G(i \xi)$ as
      function for $\hbar \xi$ of BPE/Vac/SiO$_{2}$ (violet) and
      BPE/Vac/Au systems.}
In contrast to the $G (i \xi)$ of graphene-mediated systems in
    Fig. \ref{main-fig:3}b, the interaction between BPE and bare Au
    substrate is significantly stronger (more negative interaction
    potential) than that on bare SiO$_{2}$ substrate over the entire
    frequency range.  }
\label{fig:bare-sub-att}
\end{figure}

An important evidence supporting the existence of repulsive vdW
interaction in Au/Gr/BPE system is the breakdown of wetting
transparency theory as observed from the lattice packing of BPE
molecules on bare substrates.
From the Lifshitz theory, the magnitude of attractive vdW interaction
between substrate and BPE layer over vacuum monotonically increases
when the dielectric responses of the substrate become
stronger. Therefore, when BPE molecules are evaporated on Au surface
(Au/Vac/BPE), the vdW interaction potential becomes more negative
compared with the SiO$_{2}$/Vac/BPE system, as shown in
Fig. \ref{fig:bare-sub-att}.

\begin{figure}[!htbp]
  \centering
  \includegraphics{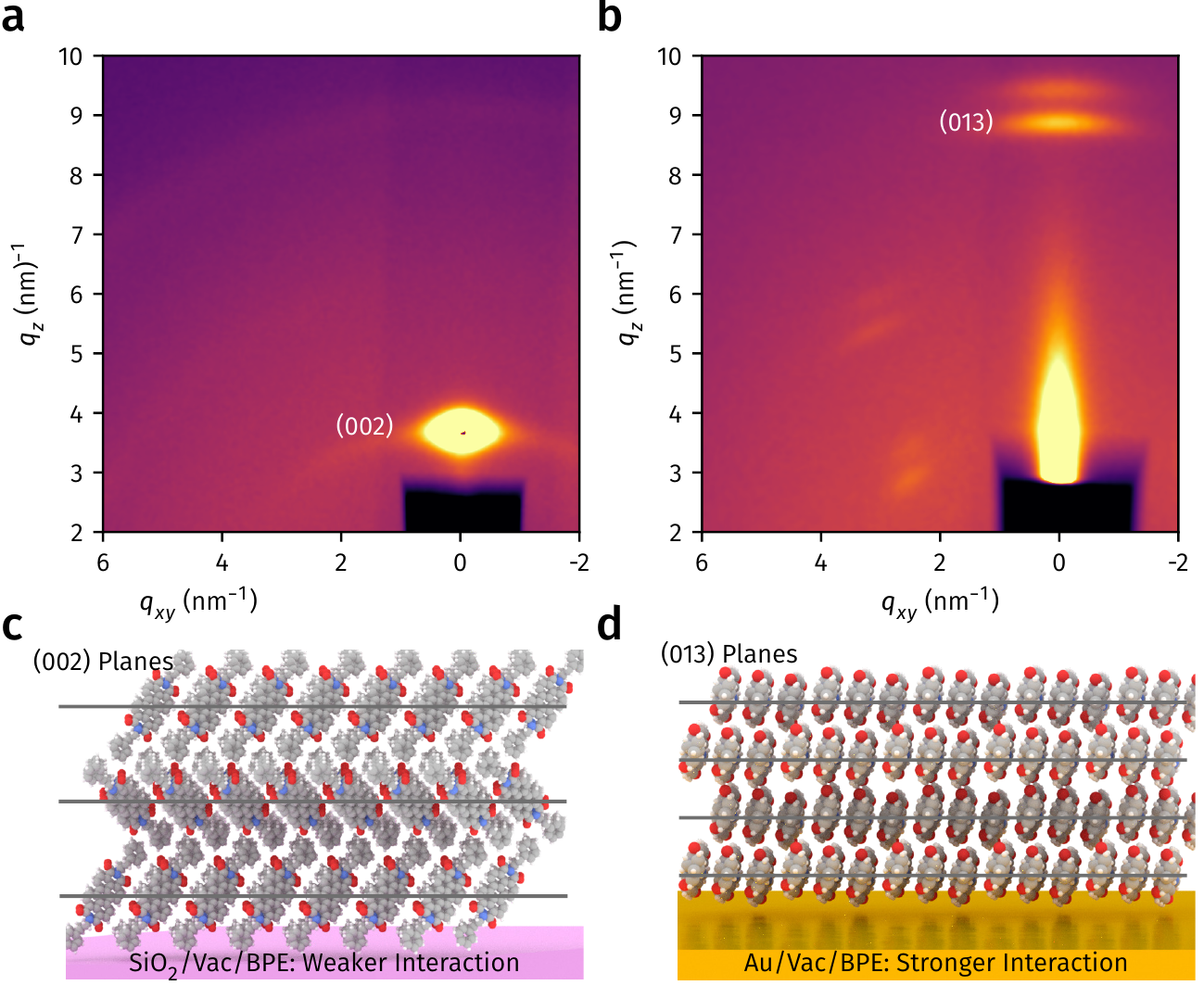}
  \caption{\textbf{GIWAXS spectra of BPE deposited on bare SiO2 a. and Au (B) substrates. }
The principal peaks in BPE/Gr/SiO$_{2}$ and BPE/Gr/Au are ($002$)
    and ($013$) lattice planes, respectively.
The larger distance of ($002$) compared with ($013$) lattice plane
    agrees with the stronger interaction in BPE/Vac/Au configuration.
\textbf{c} and \textbf{d} Schemes of ($002$) and ($013$) lattice
    planes corresponding to \textbf{a} and \textbf{b}, respectively.
  }
\label{fig:gixd-no-gr}
\end{figure}

The change of substrate-BPE interaction leads to significant
difference of BPE orientation between SiO$_{2}$/Vac/BPE and Au/Vac/BPE
as measured from 2D GIWAXS spectra. The diffraction plane
corresponding to the strongest Laue spot changes from $(002)$ in
SiO$_{2}$/Vac/BPE (Fig. \ref{fig:gixd-no-gr}a and
\ref{fig:gixd-no-gr}c) to $(013)$ in Au/Vac/BPE
(Fig. \ref{fig:gixd-no-gr}b and \ref{fig:gixd-no-gr}d). The PTCDI
planes are brought closer to the Au surface due to the stronger Au-BPE
interaction compared with that on the SiO$_{2}$-BPE interface.

The classical wetting transparency theory for 2D
materials\autocite{Rafiee_2012_trans,Shih_2012_prl} clearly breaks
down for the systems studied here. From the wetting transparency
theory based on additive model of vdW interactions, stronger
iterations between Au-BPE over vacuum will also lead to stronger vdW
interactions in Au/Gr/BPE. Our observation of lattice expansion in
Au/Gr/BPE clearly cannot be captured by classical
wetting transparency theory.

\FloatBarrier

\section{Additional Discussions}
\label{sec:addit-disc}

This section contains further discussions that are not directly
related to the results in the main text but may help understanding or
extending our current work.

\subsection{Comparison with remote epitaxy and wetting transparency}
\label{sec:effect-latt-transp}

In this section, we briefly compare our experimental observations with
other known 2D material-induced phenomena, including the remote
epitaxy\autocite{Kim_2017_remote,Kong_2018_pol,Lu_2018} (also known as
the lattice transparency\autocite{Chae_2017}) and wetting
transparency\autocite{Rafiee_2012_trans,Shih_2012_prl}. Both the
remote epitaxy and wetting transparency are examples demonstrating the
effect of substrate due to atomically-thin gap of 2D
materials. Although the concept seems similar to our system, we argue
that the underlying physics is substantially different.

\subsubsection*{Remote epitaxy}
\begin{enumerate}
\item \textbf{Substrate polarity} The lattice transparency and remote
  epitaxy phenomena are dominated by the polarity of oriented dangling
  bonds at bulk material surface (such as GaAs or
  ZnO)\autocite{Kong_2018_pol}, while the phenomena observed in our
  experiments, the polarity of material deposited (either Au or BPE
  molecule) are much smaller. Increasing the polarity of substrate
  generally enhances crystalline quality of epitaxy layer under the
  lattice transparency picture. On the contrary, the polarity
  dependency fails to explain the crystal growth we observed on
  freestanding graphene, where the substrate essentially has no
  polarity.
  
\item \textbf{Length scale} Remote epitaxy or lattice transparency can
  only occur when the gap separating the substrate and the deposited
  material is very thin (typically a few
  Å)\autocite{Kim_2017_remote,Chae_2017}, in order to allow
  redistribution of electron density induced by the substrate. On the
  other hand, our theoretical analysis (Section~\ref{sec:theory})
  shows that the many-body vdW interaction can penetrate much longer
  distance, and is essentially controlled by dispersion interactions.

\item \textbf{Crystallinity of epitaxy layer} For Au deposition on
  various graphene surfaces, the crystalline quality is best on
  freestanding graphene surface (due to ultrafast diffusion), while in
  lattice transparency experiments, better crystalline samples are
  observed when the deposited material and the substrate are of the
  same lattice structure.
\end{enumerate}

\subsubsection*{Wetting transparency}

In principle, the wetting transparency theory is based on classical
pair-wise vdW theory, which fails to capture any vdW repulsion. In
both of our experimental demonstrations, the observations contradicts
the predictions based on wetting transparency:

\paragraph{Wetting on freestanding graphene}

From the classical wetting transparency theory, the total interaction
energy $\Phi_{\mathrm{tot}}$ between gold and the entire surface
(\textit{S}) (i.e. graphene + the underlying substrate) is expressed
as: $\Phi_{\mathrm{tot}} = \Phi_{\mathrm{S/Au}} + \Phi_{\mathrm{G/Au}}$, where
$\Phi_{\mathrm{S/Au}}$ is the total vdW interaction between the substrate
and Au across vacuum, while $\Phi_{\mathrm{G/Au}}$ is the manybody vdW
interaction between monolayer graphene and Au over vacuum.  Note here
$\Phi_{\mathrm{S/Au}}$ is hypothetical "two-body" interaction energies,
i.e., independent of the existence of graphene between gold and the
substrate.  When \textit{S} becomes vacuum, one can assume
$\Phi_{\mathrm{S/Au}} = 0$. From the classical wetting transparency
model, $\Phi_{\mathrm{G/Au}}$ is estimated to be $\sim{}$84\% of that
between bulk graphite and Au\autocite{Shih_2012_prl}. In this sense,
one would expect the adhesion of Au on freestanding graphene to be
very close to that of Au on graphite. However, from our experimental
results and KMC simulations, surface diffusion of Au on freestanding
graphene is significantly faster than on graphite (Supplementary
Fig. \ref{fig:kmc-graphite}). In other words, making graphene
freestanding contributes to additional reduction in the total vdW
interaction, which we attributes to the many-body vdW repulsion in
such configuration.

\paragraph{Molecular packing on substrate-supported graphene}

From the GIWAXS data of BPE molecules deposited onto bare SiO2 and Au
substrates (Section \ref{sec:breakd-wett-transp},
Fig. \ref{fig:gixd-no-gr}), the PTCDI planes are brought closer to the
substrate on Au/Vac/BPE configuration, indicating a stronger vdW
interaction, which is also confirmed by our theoretical calculations
(Fig. \ref{fig:bare-sub-att}). If wetting transparency is the
underlying phenomenon, one would expect to see that lattice packing in
Au/Gr/BPE to be more closed-packed (if observable). However our GIWAXS
analysis in main text Fig. 3 showed the opposite, indicating that an
additional repulsive interaction exists in the system.

\subsection{Influence of bulk material bandgap}
\label{sec:infl-bulk-mater}

\begin{figure}[!htbp]
  \centering
  \includegraphics[width=0.95\linewidth]{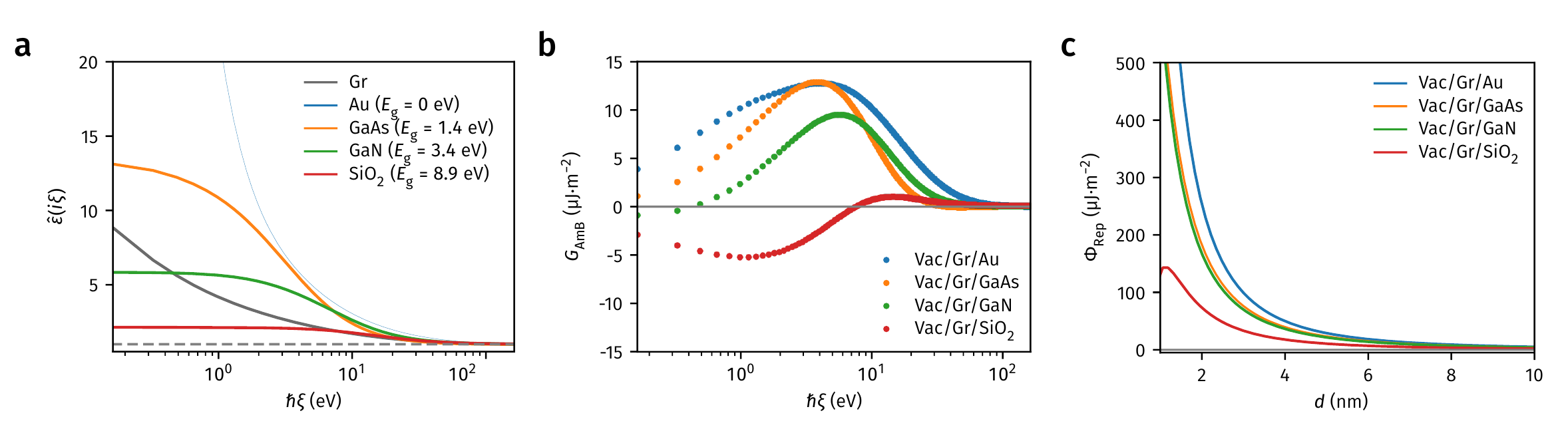}
  \caption{\textbf{Influence of bulk material bandgap on the interfacial vdW repulsion}
\textbf{a}. Dielectric functions for Au, GaAs, GaN and SiO$_{2}$
    compared with graphene at $d=1$ nm. \textbf{b}. Interaction
    spectra of Vac/Gr/B systems when B=Au, GaAs, GaN and SiO2,
    respectively. \textbf{c}. Distance-dependent vdW repulsion energy
    of corresponding system in \textbf{b}. }
\label{fig:vdw-rep-bulk-bandgap}
\end{figure}

In principle, the vdW repulsion may also be observed on freestanding
graphene even if the bulk material is semiconductor or insulator, as
long as the inequality (1) in main text holds. Here we theoretically
explore the influence of the bandgap of bulk material on the
interfacial vdW forces in a Vac/Gr/B system, where B is the bulk
material with varied bandgap.  Fig \ref{fig:vdw-rep-bulk-bandgap}a
compared the frequency-dependent dielectric functions for typical
metal (Au), semiconductors (GaAs, GaN) and insulator (SiO$_{2}$) as
compared with graphene at $d=1$ nm. As a general trend, the dielectric
response decreases for materials with larger bandgap.
As a result, the interaction spectra for Vac/Gr/B systems gradually
shift to attractive at lower frequencies, when the bandgap of B
increases (Fig.  \ref{fig:vdw-rep-bulk-bandgap}b and \ref{fig:vdw-rep-bulk-bandgap}c).
The results indicate that on freestanding graphene interface, it may
be easiest to observe the vdW repulsion via epitaxy of metal, compared
with semiconductors or insulators.

\subsection{Influence of 2D material layer number}
\label{sec:infl-2d-mater}

An intuitive way to control the degree of interfacial forces on 2D
materials is to control their layer numbers, which can also be
captured by our theoretical framework. We model the many-body vdW
interactions by extending the analysis in Section
\ref{sec:thickn-depend-repuls}. For a system of Vac/NL-Gr/Au where
NL-Gr represents $N$-layer graphene stacks, we assume that the
polarizability of NL-Gr $\alpha_{\mathrm{NL}}^{p}$ linearly scales with
$N$ \autocite{Tian_2019_nanolett}, such that
\begin{equation}
  \label{eq:linear-alpha}
  \alpha_{\mathrm{NL}}^{p} = N \alpha_{\mathrm{2D}}^{p}
\end{equation}
where $p$ is either in- or out-of-plane components and
$\alpha_{\mathrm{2D}}$ is the two-dimensional electronic
polarizability of monolayer graphene.

\begin{figure}[!htbp]
  \centering
  \includegraphics[width=0.60\linewidth]{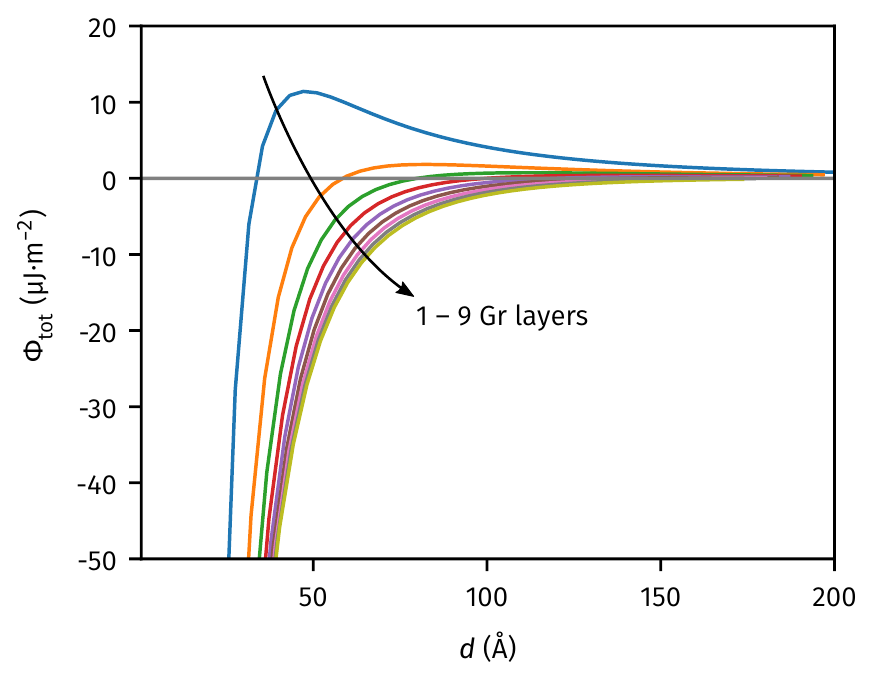}
  \caption{\textbf{Total vdW interaction energy for Vac/NL-Gr/Au system with various graphene layer numbers.} With increasing graphene layer thickness, the repulsive barrier diminishes.}
\label{fig:vdw-rep-NL}
\end{figure}

Using this model, we calculated the distance-dependent total vdW
interaction energy $\Phi_{\mathrm{tot}}$ for the Vac/NL-Gr/Au system
with varied layer numbers, as shown in Fig. \ref{fig:vdw-rep-NL}.
As expected, with increasing layer number of graphene, the
contribution of repulsive interactions decreases and the repulsive
barrier diminishes. We propose that such change of interaction can be
probed by observing metal epitaxy on freestanding graphene with
controlled numbers, which can be prepared using
mechanical exfoliation techniques.

\subsection{Choice of 2D material}
\label{sec:choice-2d-material}

\begin{figure}[!htbp]
  \centering
  \includegraphics[width=0.95\linewidth]{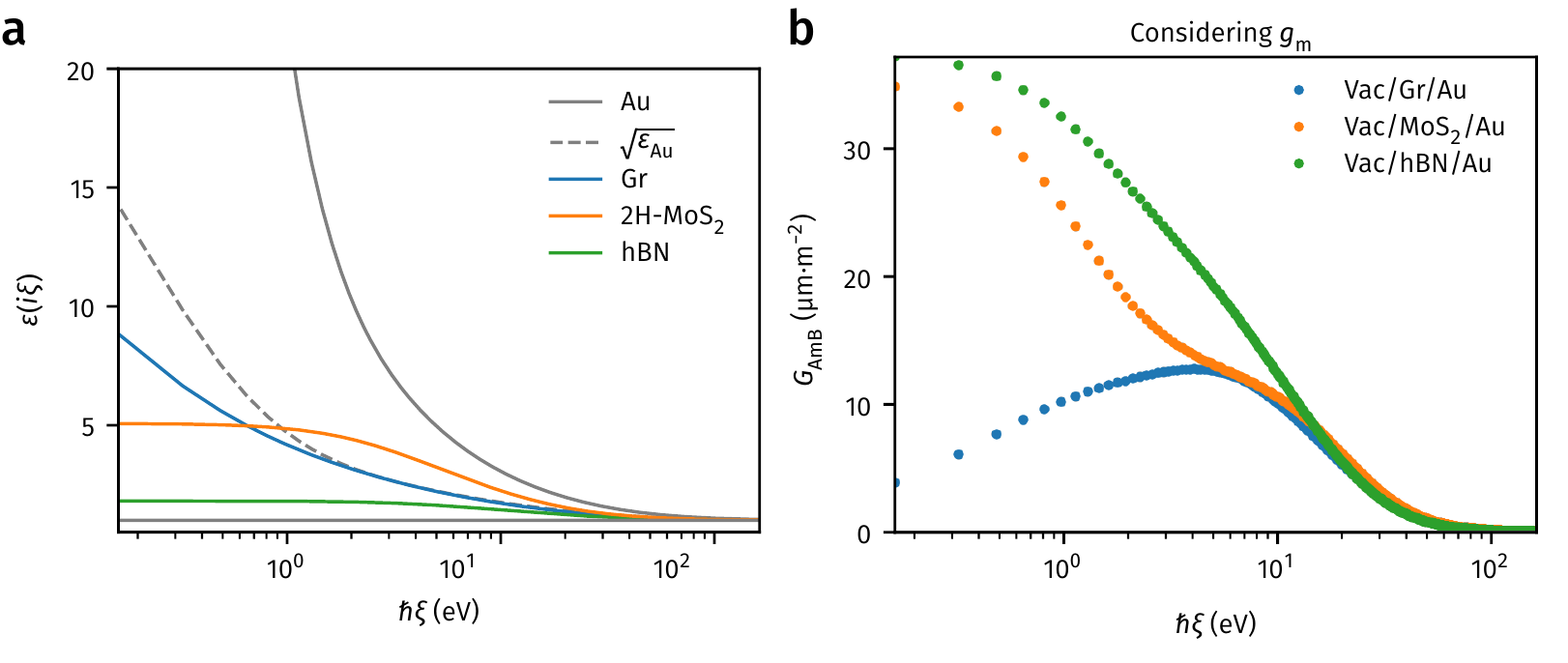}
  \caption{\textbf{Influence of choice of 2D material.}
    \textbf{a}. Frequency-dependent dielectric responses for graphene,
    MoS$_{2}$, hBN at $d=1$ nm. \textbf{c}. Interaction spectra in \textbf{b} when
    effect of $g_{\mathrm{m}}$ taken into account.}
\label{fig:max-2D-real-theory}
\end{figure}

Another degree of freedom for controlling the interfacial forces is to
change the kind of 2D material, in other words, replacing graphene
with MoS$_{2}$ or hBN. Simple analysis of
Eq. \eqref{eq:vdw-lifshitz-aniso-final} shows that
$\Phi_{\mathrm{AmB}}^{\mathrm{vdW}}$ depends on:
\begin{enumerate}
\item \textbf{Dielectric mismatch}: $\Delta_{\mathrm{Am}}\Delta_{\mathrm{Bm}}$
\item \textbf{Dielectric anisotropy of \textit{m}}: $g_{\mathrm{m}}$
\end{enumerate}
For isotropic medium \textit{m}, $g_{\mathrm{m}}=1$. The maximal vdW repulsion at certain frequency $\xi$ can be achieved when:
\begin{equation}
  \label{eq:max-isotropic-2D}
  \dfrac{\partial \Delta_{\mathrm{Am}}(\xi) \Delta_{\mathrm{Bm}}(\xi)}{\partial \hat{\varepsilon}_{\mathrm{m}}(\xi)} = 0
\end{equation}
by definition of $\Delta_{\mathrm{Am}}$ and $\Delta_{\mathrm{Bm}}$
this is equivalent to
$\hat{\varepsilon}_{m}(\xi) = \sqrt{\varepsilon_{\mathrm{A}}(\xi)
  \varepsilon_{\mathrm{B}}(\xi)}$.  However, using a 2D material as
medium, we have $g_{\mathrm{m}} < 1$, and $g_{\mathrm{m}}$ becomes
smaller when the bandgap of 2D material
decreases\autocite{Tian_2019_nanolett}. Therefore the situation is
more complex than homogeneous medium. Out theoretical analysis in
Fig. \ref{fig:max-2D-real-theory} shows that, when the effect of 2D
anisotropy is considered, hBN actually have stronger repulsive
interaction than graphene in the Vac/m/Au systems, due to less
screening at lower frequencies. The results indicate that similar
epitaxy morphology may also be observed on other freestanding 2D
materials, which we will investigate in future studies.

\subsection{Potential applications of repulsive 2D interface}
\label{sec:potent-appl}

The realization of repulsive vdW interactions at solid-state interface
by the existence of 2D material layers opens up opportunities for
practical applications, which we briefly discuss in the following
topics:

\paragraph{Two-dimensional epitaxy}

Following our demonstration of ultrathin metallic platelets grown on
freestanding graphene surface, the new 2D epitaxy mechanism can gain
more practical impact if:
\begin{enumerate}
\item \textbf{Precision control over nucleation site}:

  It is straightforward that if we can further control the location
  where the seed nuclei start to grow, such as via defect growth or
  ion beam drilling, it becomes possible to grow ultrathin and flat
  metallic thin films in a scalable manner.
  
\item \textbf{Extending to larger variety of metals}

  In principle the method does not limit the choice of metal, because
  basically all metals have higher dielectric response than graphene.
  It is noteworthy that the growth of ultrathin and flat metallic thin
  films has been very challenging due to the high surface energy. The
  state-of-the-art organometallic and colloidal chemistry only enables
  the synthesis of limited number of metallic platelet dispersions in
  solution, such as Au, Ag, and Al. The vdW-repulsion-induced 2D
  epitaxy of metal thin films is expected to be very impactful in the
  fields of plasmonics and optoelectronics.

\item \textbf{Growing ultrathin semiconductor layers}

  In addition, following our earlier discussions,
  Fig. \ref{fig:vdw-rep-bulk-bandgap} implies that the scenario of 2D
  epitaxy may also apply to the growth of ultrathin semiconductors,
  which may be also of interests for next-generation electronics.
  
\end{enumerate}

\paragraph{Molecular Sensors and Actuators}

The theoretical framework presented here allows us to predict the vdW
interactions between a number of given molecules and different
substrates through 2D monolayers. In principle, by coating a layer of
rationally chosen 2D material onto a designed substrate, one can
selectively make the vdW interactions being repulsive for a given
molecule. Accordingly, this would become a new sensing mechanism to
selectively adsorb analytes at a single-molecular level by designing a
2D material-coated substrate. An important advantage for this
mechanism is that one can in principle design and screen all analytes
and substrates \textit{in silico}.  Together with the freestanding 2D
materials systems, another interesting sensing platform one can
imagine is to first deposit ultrathin and flat 2D metal on one side of
2D monolayer, followed by sensing the adsorption events of molecular
analytes on the other side by the near-field Raman spectroscopy (for
example, in proximity to an AFM setup). The combination of
surface-enhanced Raman spectroscopy and Lifshitz-vdW interactions may
allow to detect molecular adsorption down to single molecular level.

\printbibliography{}